\documentclass[aps,reprint,prx,longbibliography]{revtex4-1}
\usepackage{graphicx,amsmath,subcaption}

\usepackage{amsmath}
\usepackage{epsf}
\usepackage{graphicx}

\usepackage{dcolumn}
\usepackage{bm}

\usepackage{amssymb}
\usepackage{array}
\usepackage{varioref}
\usepackage{color}
\usepackage[colorlinks, linkcolor=blue, citecolor=red]{hyperref}
\usepackage{soul}

\usepackage{float}
\restylefloat{table}

\providecommand{\nn}{\nonumber}

\providecommand{\pd}{\partial}

\providecommand{\bv}[1]{\bm{\mathrm{#1}}}

\providecommand{\w}{\omega}
\providecommand{\W}{\Omega}
\providecommand{\q}{\bv{q}}

\providecommand{\p}{\bv{p}}
\providecommand{\P}{\bv{P}}

\renewcommand{\k}{\bv{k}}

\providecommand{\hp}{\hat{p}}
\providecommand{\hq}{\hat{q}}
\providecommand{\hk}{\hat{k}}
\providecommand{\hz}{\hat{z}}

\providecommand{\cc}{\mbox{c.c.}}

\providecommand{\ve}{\varepsilon}
\providecommand{\ef}{\varepsilon_F}
\providecommand{\vf}{v_F}

\providecommand{\kf}{k_F}

\providecommand{\kf}{k_F}
\newcommand{\e}{\eta}
\renewcommand{\P}{\bv{\e}}
\renewcommand{\q}{\bv{q}}
\newcommand{\bvs}{\bv{\sigma}}

\providecommand{\gb}{\bar{g}}

\providecommand{\tp}{2\pi}
\providecommand{\tpp}{\left(2\pi\right)}

\providecommand{\Sg}{\Sigma}
\providecommand{\Sgt}{\tilde{\Sg}}
\providecommand{\sgn}{\mbox{sgn}}
\providecommand{\Trc}{\mbox{Tr}}
\newcommand{\spcr}{\qquad\qquad}

\newcommand{\QFM}{QFEM}
\renewcommand{\ef}{E_F}

\begin{document}

\title{A theory of criticality for quantum ferroelectric metals}

\author{Avraham Klein}
\affiliation{Physics Department, Ariel University, Ariel 40700, Israel}
\affiliation{Department of Chemical Physics, The Weizmann Institute of Science, Rehovot 76100, Israel}
\author{Vladyslav Kozii}
\affiliation{Department of Physics, Carnegie Mellon University, Pittsburgh, Pennsylvania 15213, USA}
\affiliation{Department of Physics, University of California, Berkeley, California 94720, USA}
\affiliation{Materials Sciences Division, Lawrence Berkeley National Laboratory, Berkeley, California 94720, USA}
\author{Jonathan Ruhman}
\affiliation{Department of Physics, Bar-Ilan University, 52900, Ramat Gan, Israel}
\affiliation{Center for Quantum Entanglement Science and Technology, Bar-Ilan University, 52900, Ramat Gan, Israel}
\author{Rafael M. Fernandes}
\affiliation{School of Physics and Astronomy, University of Minnesota, Minneapolis, Minnesota 55455, USA}

\begin{abstract}
  A variety of compounds, for example doped paraelectrics and polar metals,
  exhibit both ferroelectricity and correlated electronic phenomena such as low-density superconductivity and anomalous transport.
  Characterizing such properties is tied to understanding the quantum dynamics of inversion symmetry breaking in the presence of itinerant electrons. Here, we present a comprehensive analysis of the normal state properties of a metal near a quantum critical transition to a ferroelectric state, in both two and three dimensions. Starting from a minimal model of electrons coupled to a \emph{transverse} polar phonon via a
  Rashba-type spin-orbit interaction,
  we compute the dynamical response of both electrons and phonons. We find that the system can evince both Fermi and non-Fermi liquid phases, as well as enhanced pairing in
  both singlet and triplet channels. Furthermore, we systematically compute corrections to one-loop theory and find a tendency to quantum order-by-disorder, leading to a phase diagram that can include second order, first order, and
  finite-momentum phase transitions. Finally, we show that the entire phase diagram can be controlled via application of external strain, either compressive or volume-preserving. Our results provide a map of the dynamical and thermodynamical phase space of quantum ferroelectic metals, which can serve in characterizing existing materials and in seeking applications for quantum technologies.
\end{abstract}
\maketitle

\section{Introduction}
\label{sec:introduction}

Common wisdom holds that metallicity and ferroelectricity are mutually exclusive tendencies. The reason for this is straightforward: ferroelectricity is associated with a spontaneous formation and long-range ordering of dipole moments, whereas free carriers in a metal screen internal electric fields, including those associated with the ferroelectric polarization. In similar vein, one may argue that even in the rare situation where ferroelectricity  coexists with a metallic state, the electronic and lattice degrees of freedom should be only very weakly
coupled.
This intuition is born out by a variety of microscopic calculations (see e.g. Ref. \onlinecite{Gastiasoro2020} and references within).
%\cite{Gurevitch1962,Ruhman2016,Wolfle2018,Ruhman2019} % AK cite works on vanishing longitudinal coupling

In stark contrast to this view, in recent years dozens of materials have been identified which evince clear signatures of intertwined ferroelectricity and correlated electronic behavior \cite{Zhou2020}.
These compounds, which we shall call ``quantum ferroelectric metals'' ({\QFM}s), are typically doped semimetals and semiconductors. At low temperatures they undergo a phase transition to an inversion-symmetry broken state, which can be driven to zero temperature, i.e. to a putative quantum critical point (QCP), by various external parameters like pressure, strain, or doping. In the vicinity of the QCP, these systems exhibit enhanced superconducting $T_c$ as well as anomalous transport signatures. Furthermore, some of these compounds also have strong spin-orbit coupling leading to topological band structures, either due to or modified by the ferroelectric (FE) behavior. This rich behavior marks out {\QFM}s as prime targets for basic research, as well as very promising candidates to be material platforms for quantum technologies.

There are three families of compounds of particular interest, as they display both FE and metallic behavior. One type are quantum paraelectrics, such as SrTiO$_3$
or KTaO$_3$,
where quantum fluctuations prevent the undoped material from becoming polar~\cite{Barrett1952,Mueller1979}.
The 
transition into the FE phase is driven by strain or by chemical substitution \cite{Schlom2007,Salmani-Rezaie2020,Salmani-Rezaie2020a}, and the carrier density is controlled by doping or gating (in thin layers).
The second family are the IV-VI compounds such as PbTe, SnTe, or GeSe~\cite{bilz2006dynamical,yu2018unraveling}, which 
lie close to both FE and topological 
quantum critical points~\cite{hsieh2012topological}.
A third family are certain bilayer transition metal dicalchogenides (TMDs), which are both  metallic and polar by nature \cite{Dawson1987}, but can be tuned out of the polar state by doping and pressure \cite{Iqbal2019,Sakai2016,Vellinga1970}.
The latter two families host topological band structures \cite{Deng2016,Wu2016}, such as those of Weyl semimetals \cite{Jiang2017} and topological insulators~\cite{qian2014quantum}. In all types of materials there also appears superconductivity at low temperature \cite{Matsushita2006typeII,novak2013unusual,Koonce1967,Gastiasoro2020,Kamitani2017,Qi2016,Collignon2019,Ahadi2019,Stucky2016,Rischau2017,Tomioka2019,Enderlein2020,franklin2020giant,Ueno2011,Liu2022}. One important difference between these classes of systems is that, in the bilayer TMDs, the FE transition usually onsets at high temperatures, such that quantum FE fluctuations are unlikely to be driving their low-temperature properties. For this reason, we focus on quantum paraelectrics and IV-VI compounds as model {\QFM}s.  

\begin{figure*}
  \centering
  \begin{minipage}{0.49\hsize}
    \begin{subfigure}{0.75\hsize}
      \includegraphics[width=0.5\hsize]{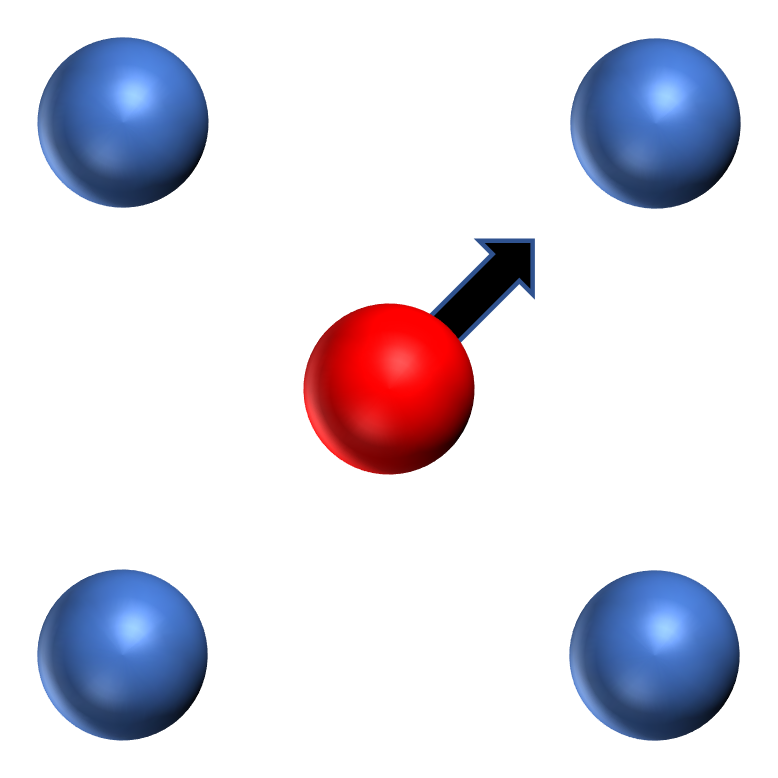}
      \caption{}
    \end{subfigure}
    \begin{subfigure}{\hsize}
      \includegraphics[width=0.5\hsize]{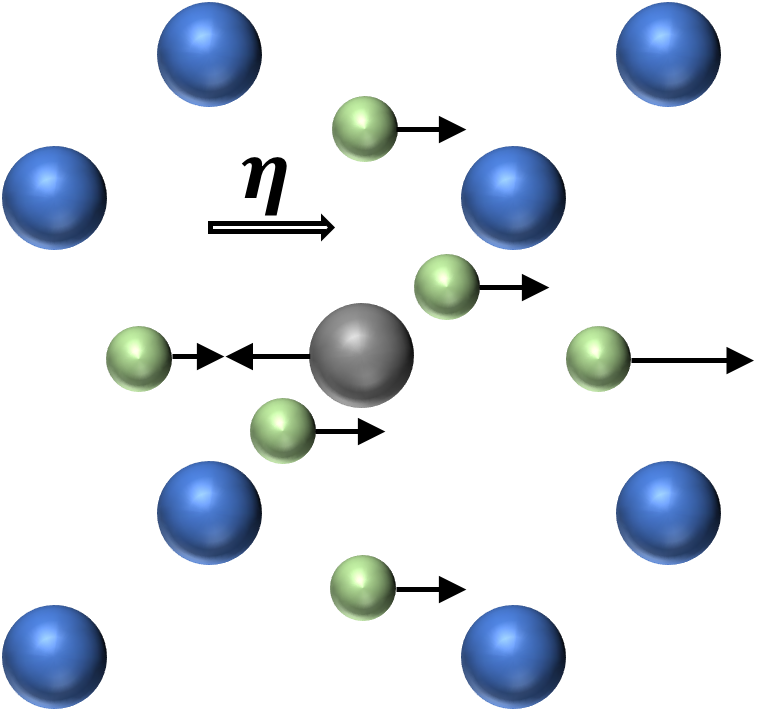}
      \caption{}
    \end{subfigure}
  \end{minipage}
  \begin{minipage}{0.49\hsize}
    \begin{subfigure}{\hsize}
      \includegraphics[width=\hsize]{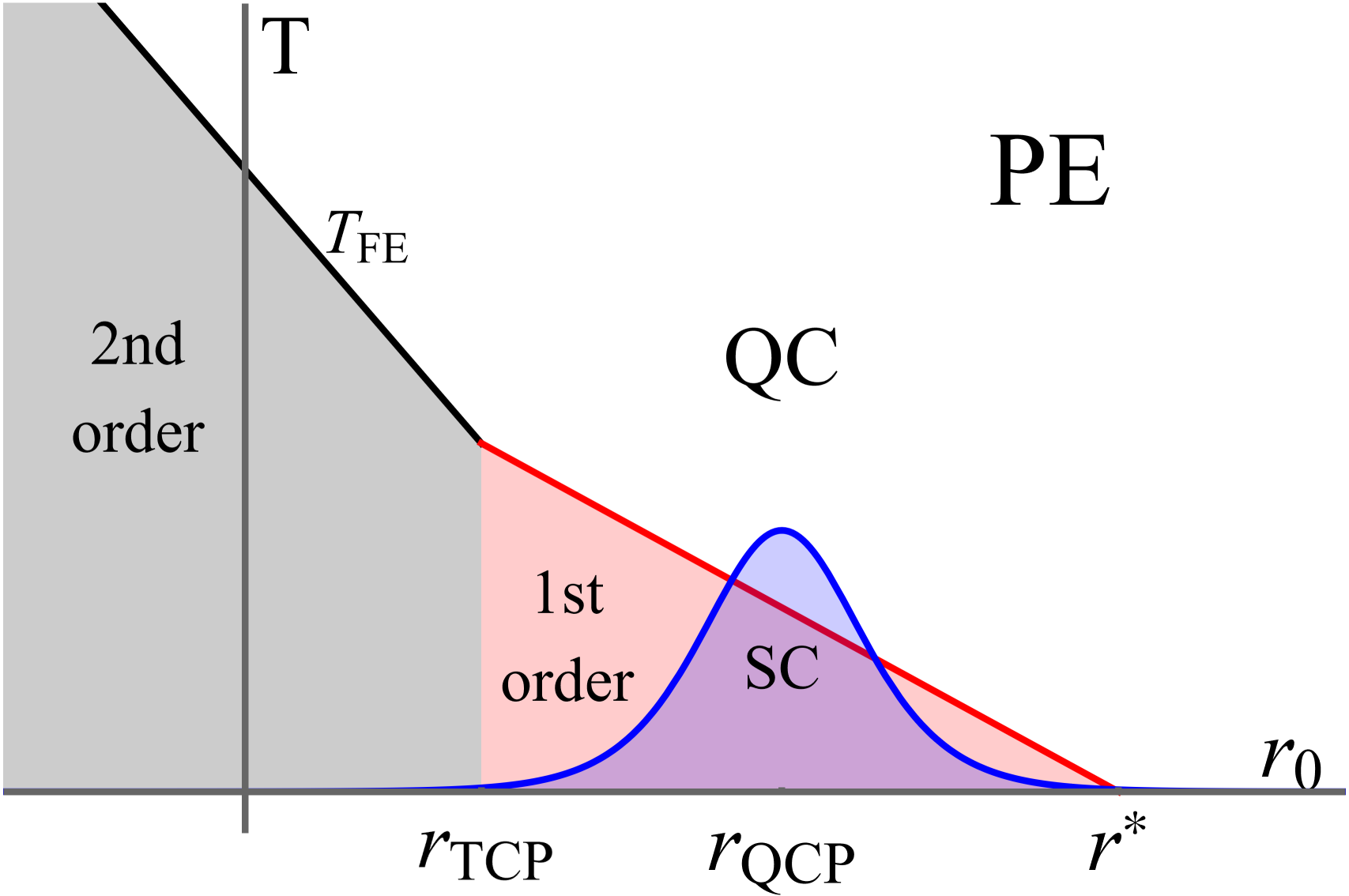}
      \caption{}
    \end{subfigure}
  \end{minipage}
  \caption{Schematic description of a \QFM. (a) Caricature of the FE deformation in a two dimensional ionic lattice. (b) Schematic of a polar deformation in a perovskite structure, e.g. SrTiO$_3$, where Sr, Ti, O are depicted as blue, gray, and green spheres. (c) Schematic phase diagram of a {\QFM} as a function of temperature and an external ``tuning'' parameter $r_0$, e.g. strain or doping, as calculated in this work. The phase diagram includes a paraelectric (PE) phase above a ferroelectric (FE) phase which spontaneously appears at a temperature $T_{FE}$ via a second-order transition. Within a mean-field picture, the FE phase transition line terminates at a QCP at $r_0=r_{QCP}$, where $r_{QCP}$ is a critical value of the tuning parameter. However, quantum fluctuation effects (specifically, QOBD) can cause a first-order transition that preempts the second-order one, resulting in a tricritical point $r_{TCP}$. The 1st-order transition line terminates at some $r^* \neq r_{QCP}$.
    Also within the mean-field picture, a superconducting (SC) dome rises in the vicinity of $r_{QCP}$,
    which
    can support both spin-singlet and spin-triplet superconducting states. We did not investigate a possible shifting of the SC dome due to QOBD in this work. The schematic is taken from one of the possible phases of a 2D {\QFM}  (see Sec.~\ref{sec:low-energy-theory}), but is qualitatively similar for a 3D \QFM, albeit with a much narrower first-order region.}
  \label{fig:schematic-intro}
\end{figure*}

In {\QFM}s the transition to the FE state typically occurs by a condensation of a polar
optical phonon that breaks inversion symmetry, see Figs. \ref{fig:schematic-intro}a and \ref{fig:schematic-intro}b.
This is the same type of transition that occurs in conventional insulating ferroelectrics and has been well studied in that context.
The transition is structural so that lattice effects are important, and may render the transition first order \cite{Cochran1960,Khmelnitski1973,Larkin1969}. More importantly, the strong lattice coupling makes the system sensitive to strain, which is a useful tuning parameter. The thermodynamics are well-captured by a vector field $\P$ that serves as an order parameter and which represents either the induced electric polarization or the atomic displacement associated with the soft polar phonon.
In the simplest case, it has a Ginzburg-Landau description of the form \cite{Rabe2007},
\begin{equation}
  \label{eq:GL-F}
  F \propto r |\P|^2 + u |\P|^4 - \kappa_{\e\varepsilon}\sum_{ij}\e_i \varepsilon_{ij} \e_j + \cdots,
\end{equation}
where $\varepsilon_{ij}$ is the elastic strain tensor, and $r,u,\kappa_{\eta\varepsilon}$ are functions of temperature,
stress, etc.
In many cases the transition temperature 
can be driven to low/zero temperatures, i.e. to a putative quantum critical point~\cite{Schlom2007,Gastiasoro2020,Sakai2016}. The quantum dynamics of these insulating systems has also been studied extensively~\cite{Roussev2003,Rowley2014,Narayan2019,Chandra2017}, 
and has a definite impact on the critical behavior~\cite{Khmelnitski1973}, captured by suitable corrections to Eq.
%Much of this knowledge can be summarized into a modified form of Eq.
~\eqref{eq:GL-F}.

In {\QFM}s the long-range dipole fields are screened by the itinerant electrons \cite{Benedek2016}, but the inversion-symmetry-breaking induced by the phonon remains and can be described by $\P$. Thus, the low energy dynamics are described by a coupled system of soft bosonic fluctuations of $\P$ and of itinerant electrons \cite{edgeQCorigin,Gastiasoro2020,Gastiasoro2020a,Gastiasoro2019}. Models of soft bosonic fluctuations mediating effective electronic interactions have long been recognized as paradigmatic examples of both unconventional superconductors and non-Fermi liquids (nFLs) \cite{Altshuler1994,Bonesteel1996}. Classic examples of these are the spin-fermion model \cite{Abanov2003} and the Ising-nematic model \cite{Metzner2003,Fernandes2014} used to describe spin-fluctuation and nematic-fluctuation mediated superconductivity, and with possible applications to the cuprate and iron based superconductors. Accordingly, soft FE fluctuations have been promoted in recent years as a possible mechanism behind {\QFM} phenomenology \cite{Rowley2014,Chandra2017,Enderlein2020,edgeQCorigin,Kedem2016,kumar2020quasiparticle,kiseliov2021theory,Gastiasoro2021}. 

However, in contrast with classical models of boson-fermion coupling,
in {\QFM}s even the simplest such coupling (which we will describe in detail later) 
is unique for several reasons.
The first reason is due to the structural properties of the parent ionic crystal.
The soft FE fluctuations are predominantly in the \emph{transverse} (TO) component of the polar phonon ~\cite{Wolfle2018,Ruhman2019}, such that the polarization and propagation direction of a fluctuating mode are perpendicular. This is a property inherited from the ``parent'' (insulating) ferroelectric, and is a result of the Coulomb interaction between dipoles, which splits the longitudinal (LO) mode off from the TO mode. In 3D this results in a hard gap \cite{lines2001principles}, and in 2D, while there is no gap at the Brillouin zone center \cite{Mele2002}, the dispersion curves are sharply separated \cite{SanchezPortal2002,Sohier2017}.  This effect persists even in the doped compounds, provided the density of itinerant electrons is not too high to fully  screen the Coulomb interaction~\cite{Kumar2021}.
The second reason is that the coupling is inherently odd under space inversion, and thus entangles spin and orbital degrees of freedom~\cite{Fu2015}. This is required 
to ensure the breaking of inversion symmetry while preserving the time-reversal symmetry, so that there is no creation of spontaneous currents \cite{Kiselev2017}. 
A proper treatment of the low energy properties of {\QFM}s therefore requires a comprehensive quantum theory of the coupled dynamics.

In this paper, we derive such a theory. We start from a microscopic model representing itinerant fermions coupled vectorially to the soft FE transverse phonon, and derive an effective Ginzburg-Landau action and phase diagram. We focus on systems with Fermi surfaces (FSs), relevant to describe quantum paraelectrics. For the IV-VI compounds, the coupling to Dirac electrons is a more appropriate starting point \cite{Kozii2019,Kozii2021}.
In analogy with e.g. the spin-fermion and Ising-nematic models, we take as a starting point Eq. \eqref{eq:GL-F}, describing a transverse phonon near a QCP, and systematically calculate the quantum corrections to the action from the coupling to electrons.

To obtain a qualitative picture of how a {\QFM} behaves, it is useful to think of another well-known model, namely a ferromagnetic spin-fermion model (see e.g. \cite{Belitz1997,Rech2006,Brando2016,Green2018} and references within). Such models have been used to describe itinerant ferromagnets such as the uranium superconductors (see e.g. \cite{Aoki2011,Mineev2017} and references within) and consist of a ferromagnetic bosonic field $\bv{S}$ minimally coupled to itinerant electrons. Similar to $\P$, the mode $\bv{S}$ is a vector, which however fluctuates purely in spin space. The phenomenology of $\bv{S}$ is by now well known, and is a result of (a) the strong fluctuations at the QCP, (b) the spin nature of the interaction, and (c) the vector nature of $\bv{S}$. At the critical point the fluctuations of $\bv{S}$ undergo strong Landau damping. In response, a nFL state arises, which also simultaneously mediates strong pairing fluctuations. Because the mode is in the spin sector, it contributes to spin-triplet rather than spin-singlet pairing. The vector nature gives rise to further complexity, because it implies the existence of soft Goldstone modes, which also interact with the itinerant electrons. The additional fluctuations modify both the normal state and the superconducting state. In the normal state they allow the system to avoid the QCP either via a preemptive first order transition \emph{or} via a transition to a finite momentum spin density wave state, a phenomenon often termed quantum order-by-disorder (QOBD). In the SC state, these additional modes
affect both $T_c$ and the nature of the
transition~\cite{ChubukovFink2003}.

The {\QFM} is deceptively similar to an itinerant ferromagnet, but the devil is in the details. On the one hand, it is described by a soft vector mode $\P$ near a QCP, like $\bv{S}$. On the other hand, this bosonic field acts \emph{both} in real and spin space and is transverse. Thus, it can in principle mediate both singlet and triplet superconductivity. Moreover, in contrast to $\bv{S}$, it has a restricted fluctuation space. Indeed, in an SU$(2)$ ferromagnet, via a slow spatial modulation, i.e. a Goldstone mode, the spin polarization can vary over the two-dimensional surface of a sphere of constant magnetization. 
For a {\QFM} the additional constraint of transverse polarization restricts the modulations to a one-dimensional circle.
Furthermore, because $\P$ also acts in real space, geometric considerations, such as real space momentum and energy conservation also play a role and further complicate the picture by restricting the phase space for FS scattering. At the same time, this entangling of spin and real space structure offers a convenient way to manipulate {\QFM}s: similarly to insulating FEs, external strain allows one
to tune the properties of {\QFM}s and evince their rich phase diagram. We will comment in detail on similarities and differences to ferromagnetic systems as we present our results.

Our central results can be summarized as follows. First (as expected) we find that the low-energy response of $\P$ is dominated by the coupling to electrons (Landau damping), which in turn leads to nFL behavior of the electrons as well as enhanced superconductivity. This information is encoded in the one-loop bosonic and fermionic self-energies $\Pi$ and $\Sg$. However, in contrast to the case of spin or Ising-nematic fluctuations, the {\QFM} bosonic response depends on the dimensionality: in 3D it is overdamped,
but in 2D, it has two separate modes, one overdamped and one underdamped.
Such behavior is more similar to XY nematics and is a result of the entangled spin and orbital (momentum) degrees of freedom.
Second, we find pairing instabilities to both spin-singlet and spin-triplet states, which have enhanced pairing temperatures compared to BCS theory, 
i.e. compared to the classic exponential dependence on inverse coupling strength. Here too  there is a dependence on dimensionality. In 2D, the singlet and triplet instabilities are almost degenerate. In 3D, the singlet dominates,
but applied strain can make the two instabilities almost degenerate.
Third, we find that {\QFM}s have a low energy tendency to form preemptive states either by a first-order transition to a homogeneous phase or by formation of a finite-momentum state, i.e. a Ferroelectric Density Wave (FDW).
This is a manifestation of QOBD as described above.
These orders modify, but do not prevent, the enhancement of superconductivity arising from proximity to the QCP. 
Fourth, we show that coupling of FE order to strain controls all of the above properties, so that by applying external compressive or tensile strain one may control the normal state order, the fermionic behavior, and the dominant pairing instability. We provide a schematic phase diagram in Fig.~\ref{fig:schematic-intro}, and several more detailed phase diagrams in the remainder of the paper.

We derive these results by performing a field-theoretical analysis of
the model within the Eliashberg formalism and analyzing the low-energy
bosonic and fermionic self-energies. Then, we go beyond the
Eliashberg theory to identify the relevant quantum processes for
quantum order by disorder. Finally, we account for the modification of the results by
external strain. Thus, our main contribution is to apply established techniques of field-theoretical diagrammatic calculations to a relatively unexplored quantum critical system. At almost every step, we find features, some rather surprising, due to the interesting nature of the space-odd transverse fluctuations that characterize the FE QCP. Therefore, our results provide at least a semi-quantitative picture of how the {\QFM}  properties are expected to behave. 

The paper is organized as follows. In Sec.~\ref{sec:minimal-model-qfm}, we present our minimal model for a {\QFM}. In Sec.~\ref{sec:low-energy-theory}, we study the two-dimensional problem. We calculate the bosonic and fermionic self-energies in the normal state  and the pairing instabilities. Then we identify relevant soft fluctuations driving QOBD. We consider the impact of external strain and generalize the results to finite temperatures, thus constructing the schematic phase diagrams of Figs.~\ref{fig:phaseDiag2d} and \ref{fig:strain1}. In Sec.~\ref{sec:theory-3d-qfm}, we present the results for the three-dimensional case. We end with a discussion of the broader implications of our work in Sec.~\ref{sec:discussion}.

\section{Minimal model for a {\QFM}}
\label{sec:minimal-model-qfm}

A description of a {\QFM} requires three ingredients. First, the structural transition leading to a FE state is described by a boson mode that softens at the transition. Because it is a property of the phonons, such a description is valid regardless of whether the compound is insulating or metallic.  In order to construct a universal theory, we will assume from the start an isotropic, rotationally invariant system, leaving a discussion of actual lattice effects to Sec. \ref{sec:lattice-properties}. Second, the low-energy itinerant fermions are described by a Fermi liquid (FL). Third, the two systems are coupled by an appropriate interaction. The complete system is described by the field theory with the dynamical action,
\begin{equation}
  \label{eq:action-1}
\resizebox{1.\hsize}{!}{$\mathcal{S} = \int d\tau \left[\sum\limits_{\q}\mathcal{L}_\e(\q) + \sum\limits_{\p}\mathcal{L}_{FL}(\p) +\left(\frac{a}{L}\right)^d\sum\limits_{\p,\q}  \mathcal{L}_{int}(\p,\q)\right]$.}
\end{equation}
Here, $\tau$ is imaginary time in units with $\hbar =1$; $a,L$ are respectively the lattice constant and the size of the system; $d$ is the dimensionality; and $\p,\q$ denote respectively momenta of fermionic and bosonic degrees of freedom. We write out all of the relevant degrees of freedom directly in momentum space, summing over the first Brillouin zone. The various Lagrangian densities are described below. 

In this work, we assume that the transition to the FE state is a displacive transition, i.e. driven by the softening of the transverse optical phonon discussed in the introduction.
The other main mechanism for ferroelectricity is an ``order-disorder'' one, which is reminiscent of an Ising or Heisenberg ferromagnetic transition. The order-disorder scenario describes systems whose unit cells are deformed along certain lattice-preferred orientations, creating a series of ``pseudospin'' FE moments. The FE transition is governed by the angular fluctuations of these moments, rather than by a softening of their amplitude.
Since the displacive transition is more common at low temperatures \cite{lines2001principles,Chandra2017} we focus on it here.
The phonon mode is described by the Lagrangian density
\begin{equation}
\mathcal{L}_{\eta} = \e_i(\q)D_0^{-1}a^2\left(|\q|^2 - c^{-2}\pd_\tau^2 + c^{-2}\w_T^2 \right)\e_i(-\q),\label{eq:L-b}
\end{equation}
where 
\begin{equation}
  \label{eq:P-def}
  \e_i(\q) = \mathcal{P}_{ij}(\hq)u_j(\q), \qquad \mathcal{P}_{ij}(\hat q) = \delta_{ij}-\hq_i\hq_j
\end{equation}
describes a transverse component of a dimensionless phonon displacement $u_j$, with $\mathcal{P}$ a projection operator 
onto the transverse sector, i.e., $\hq\cdot\P = 0$. $D_0$ is a constant with units of inverse energy,
% $k_a$ is a reciprocal
% $a$ is a lattice constant,
$c$ is the transverse phonon velocity, and $\w_T$ is the transverse phonon excitation energy, inversely proportional to the correlation length, which softens to zero at the QCP. The model of Eq. \eqref{eq:L-b} gives rise to a Ginzburg-Landau free energy similar to that of Eq. \eqref{eq:GL-F}, with
%e.g.
\begin{equation}
  \label{eq:r-basic}
  r = (\w_T a/c)^2.
\end{equation}
Note that while $r$ measures how close the phonon system is to the QCP compared to, e.g. another non-critical insulator, it does not contain information about the criticality of the electrons, as evidenced by the fact that the Fermi momentum does not appear in its definition. This is a consequence of the fact that $\P$ represents phonons, which are microscopically an independent degree of freedom from the electrons. In many other models of quantum criticality, the bosons represent an ordered state of the electrons themselves, and have the same fundamental energy scales as these electrons do. The relevant energy scales for the coupled system show up in the renormalization of $r$, and  in the electronic and bosonic self-energies, as described later.

We describe the FL by the Lagrangian density
\begin{equation}
  \label{eq:L-f}
  \mathcal{L}_{FL} = \psi^\dagger_{\alpha}(\p)(\pd_\tau + \epsilon(\p))\psi_{\alpha}(\p),
\end{equation}
where the repeated index $\alpha$ denotes implicit summation over spin indices. $\epsilon(\p)$ can describe
any rotationally invariant dispersion. For our purposes, to keep the discussion general, we linearize the dispersion near the Fermi surface (FS), $\epsilon(\p) \approx \vf (|\p|-\kf)$, where $\kf$ is the Fermi wave-vector and $\vf$ is the FL effective velocity.

We now turn to the interaction term. In general we expect the condensation of $\P$ to break inversion symmetry for the fermions as well. This is accomplished by a linear coupling term which must also be polar from symmetry considerations. In a FL such coupling can be in either spin or charge sector. However, ordering in the charge sector in a model with orbitals that have the same parity would imply creation of spontaneous currents, thus breaking time reversal symmetry, and would also require fine-tuning to avoid Bloch's theorem \cite{Bohm1949}. Hence, it is natural to expect a coupling in the spin channel.  

In systems with strong spin-orbit coupling, these couplings were shown~\cite{Fu2015,Martin2017,Gastiasoro2020a} to have the form $\psi_{\alpha}^\dagger(\p) \hat F_{\alpha\beta}[\p]\psi_\beta(\p)$, where $\hat F$ encodes a type of spin-orbit  coupling.  Based on the transformation properties under mutual rotation of spin and momentum, it may take the form $\hat F = \bvs\cdot\p$, $\bvs \times \p$, $\bvs\otimes\p,...$, where $\bvs$ is a vector of Pauli matrices, corresponding to scalar, vector, tensor, etc. couplings. We note that strictly speaking $\bvs$ here is not spin, since it is not a good quantum number in spin-orbit coupled systems. However, in the presence of both time-reversal and inversion symmetries (which is the case in this paper), the Bloch states remain doubly degenerate at each crystal momentum $\bf k$, thus allowing to introduce a ``pseudospin'' basis $\bvs$, which we refer to as ``spin'' for simplicity hereafter. The most natural coupling is the vector one, since it  couples linearly to the phonon displacement vector $\boldsymbol \eta$ (for any other type of coupling, we need to introduce either a nonlinearity or break the symmetry explicitly to couple to a vector). Thus we have
% . , i.e.
\begin{equation}
  \label{eq:L-i}
  \mathcal{L}_{int} = \frac{\lambda}{k_I} %\sum_{\q}
  \e_i(\q)\psi^\dagger_\alpha(\p+\q/2)(\p\times\bvs_{\alpha\beta})_i\psi_{\beta}(\p-\q/2),
\end{equation}
where $\lambda$ has units of energy and $k_I$ is a parameter introduced for convenience to rescale the interaction constant. It is important to note that the typical interaction strength in Eq. \eqref{eq:L-i} goes down if we decrease the Fermi momentum, since $|\p|\sim \kf$. However, since this distinction will not be important in this paper, we henceforth set $k_I = \kf$ for simplicity.
We note that the scalar form of coupling is what is expected for an Ising-type transition, e.g. an order-disorder one, and that the tensor form is just the spin-nematic from a traditional FL that has been studied previously \cite{Kirkpatrick2011,Klein2019a}.

Equations~\eqref{eq:L-b}-\eqref{eq:L-i} form a complete model for a {\QFM}. We chose this model both for universality and for simplicity. It can be readily checked that our conclusions from the study of this model generalize to more realistic forms of interactions, band structures, lattices and so forth. For instance, the microscopic origin of the coupling constant $\lambda$ in the quantum paraelectric STO has been recently discussed in Ref \cite{Gastiasoro2021}. The model makes sense in any dimension where a cross-product can be defined. We now proceed to study its dynamics. As we shall see, the unique properties of {\QFM}s are most transparently seen in 2D, and it is also easier to study the model analytically in 2D than in 3D. For this reason, we will next concentrate on the 2D effective low-energy theory.

\section{Low-energy theory of a 2D {\QFM}}
\label{sec:low-energy-theory}

\begin{figure*}
  \centering
  \begin{subfigure}{0.3\hsize}
    \includegraphics[width=\hsize,clip,trim=0 0 0 0]{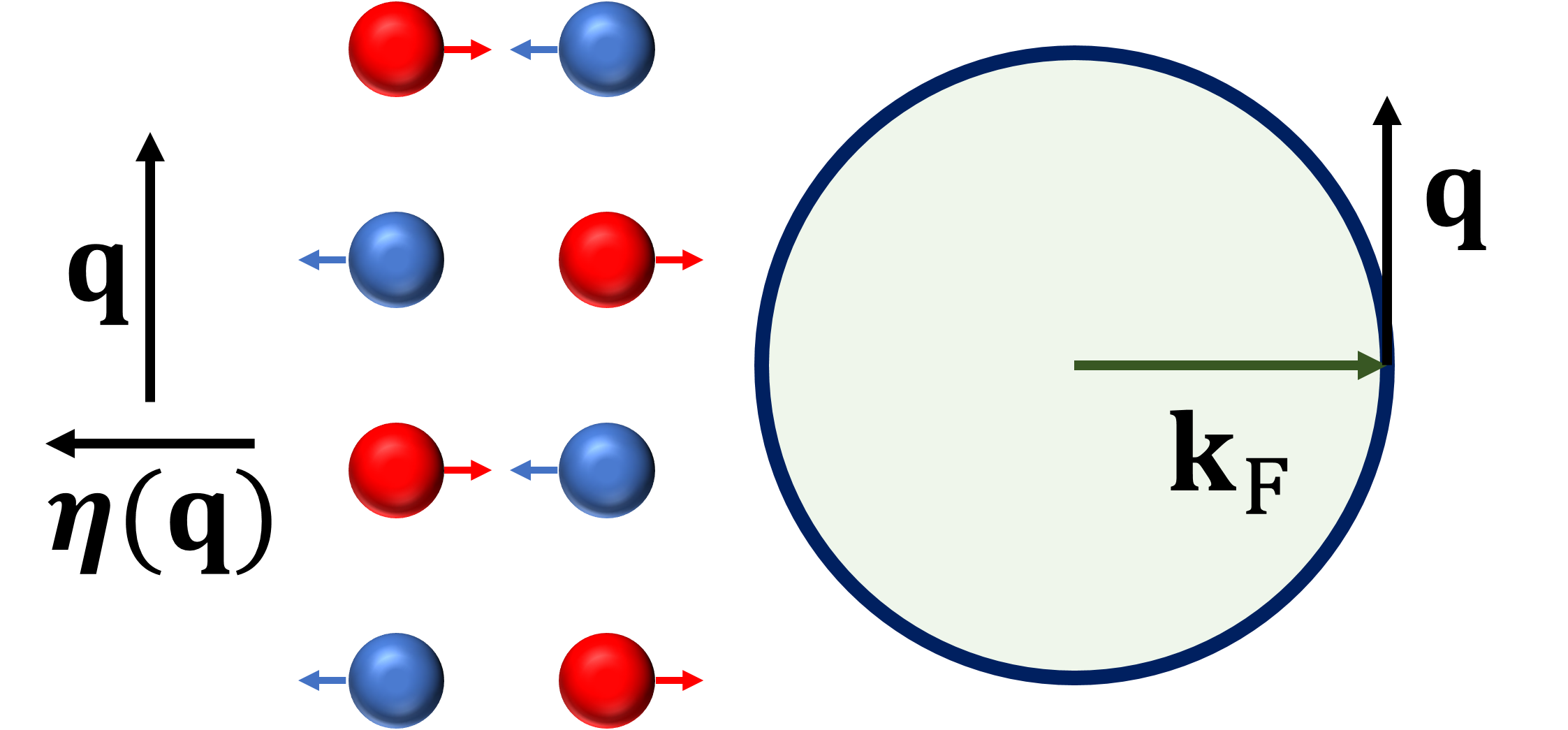}
    \caption{\label{fig:qfmDist}}
  \end{subfigure}\hfill
  \begin{subfigure}{0.3\hsize}
    \includegraphics[width=\hsize]{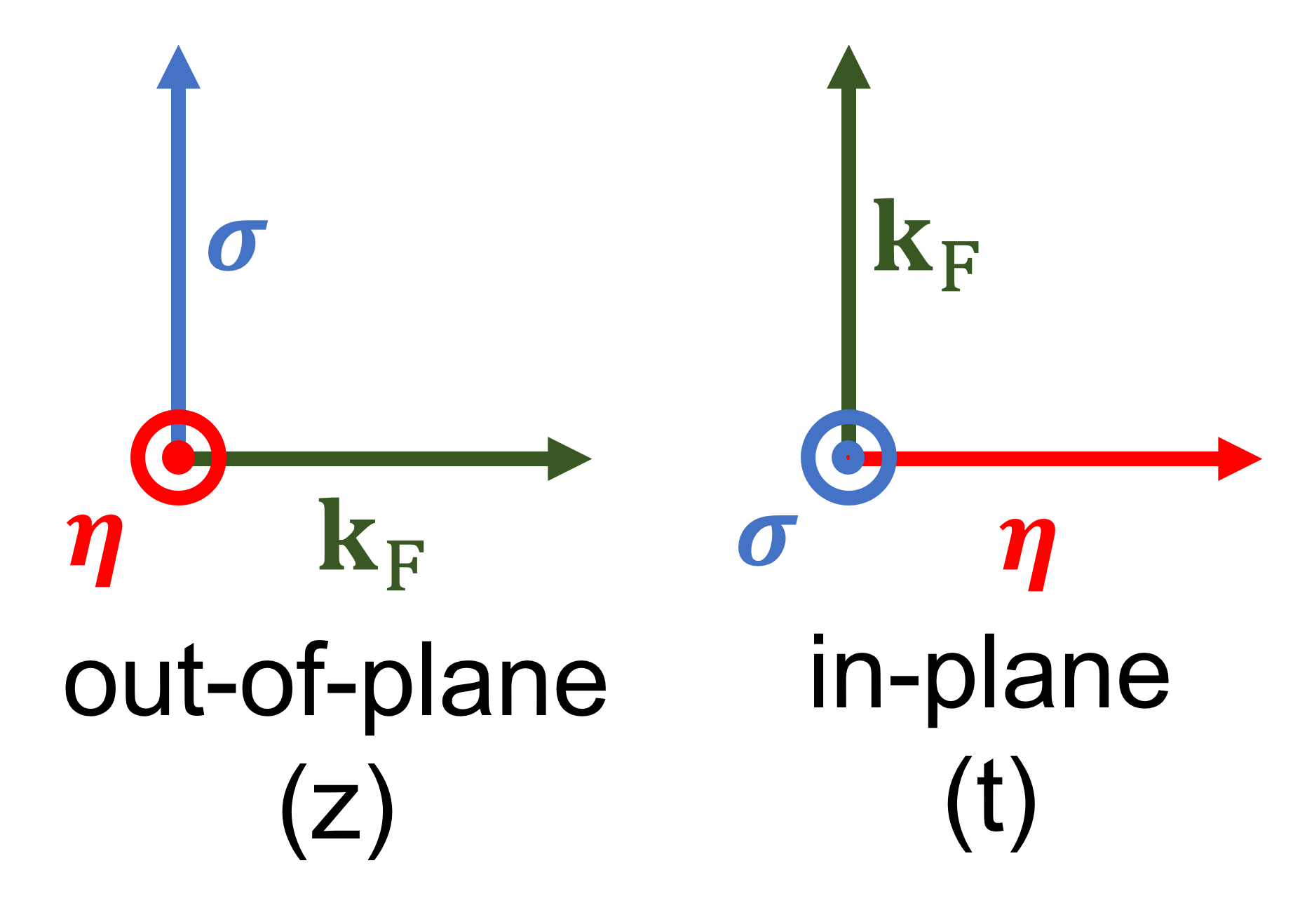}
    \caption{\label{fig:qfmStat}}
  \end{subfigure}\hfill
  \begin{subfigure}{0.3\hsize}
    \includegraphics[width=\hsize]{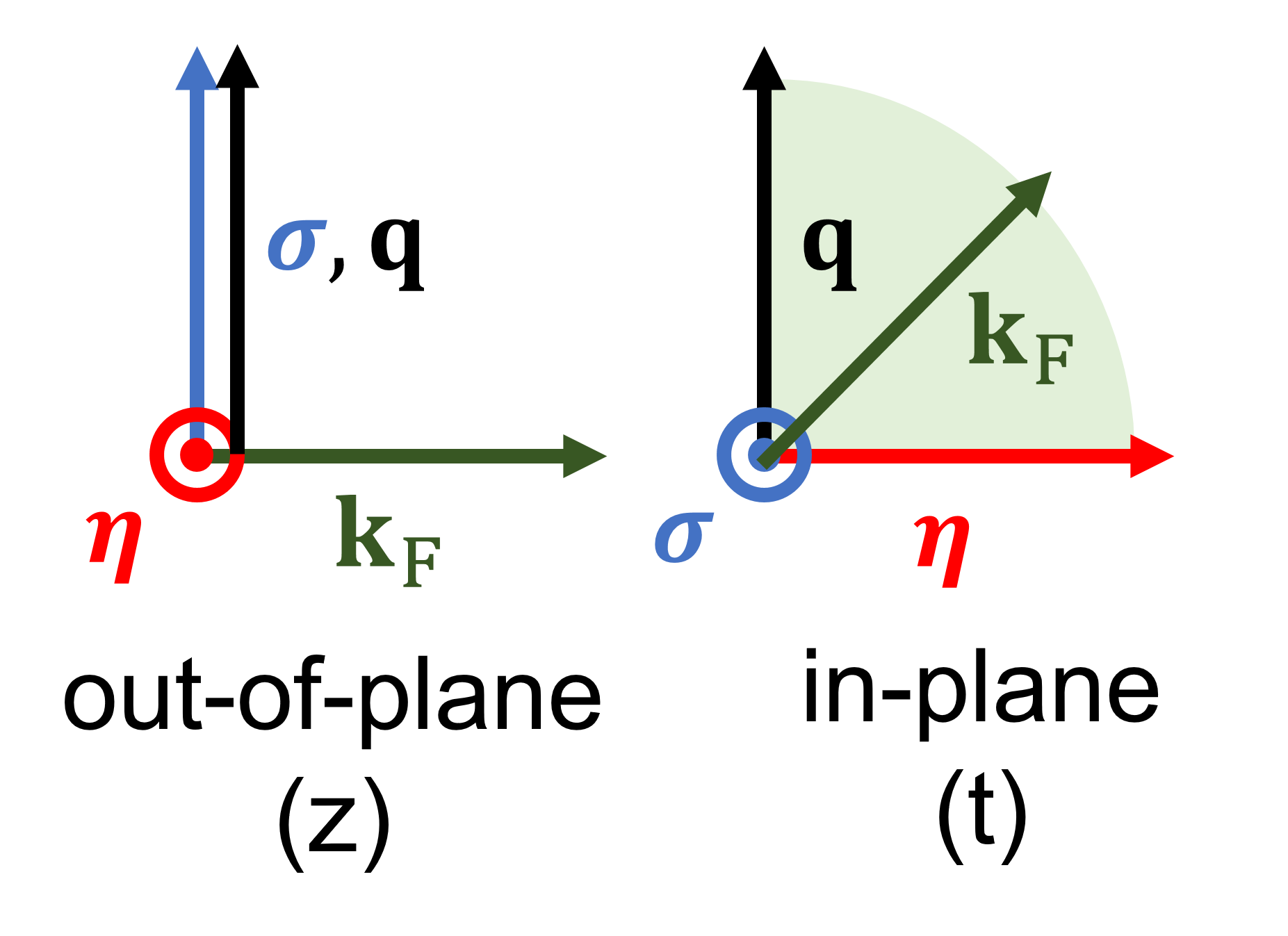}
    \caption{\label{fig:qfmDyn}}
  \end{subfigure}
  \caption{The {\QFM} model in 2D and its geometrical constraints. (a) A schematic of the distortion caused by a transverse phonon mode. The red and blue balls represent a toy model of an ionic lattice. The transverse phonon propagates in direction $\q$ and induces a distortion, creating local dipole moments, in the direction of $\P(\q)$. $\P$ scatters electrons most strongly when their Fermi vector $\k_F\perp \q$. In the limit $|\q|\to 0$, $\bv{\eta}$ becomes  the homogeneous polar distortion depicted in Fig. \ref{fig:schematic-intro}a. (b) The two possible polarizations for $\P$, and the associated spin-orbit splitting of the electronic dispersion parametrized by $(\k_F \times\bvs)$. In the $z$ polarization, the spin is in-plane, while in the $t$ polarization, the spin is out-of-plane. (c) The geometric constraints on scattering of electrons by the phonon modes. The $t$ mode has suppressed scattering, because it is not possible for $\P$ to efficiently scatter electrons parallel to the Fermi surface, see Eq. \eqref{eq:ff-t-2d}.}
  \label{fig:intro1}
\end{figure*}

In this section we perform a comprehensive analysis of a 2D {\QFM}.
Before proceeding, we need to define precisely what is meant by a 2D system, since the interaction term in Eq. \eqref{eq:L-i} is inherently three-dimensional. In this work, we will assume that the 2D system is a thin film of a material, which we take to be aligned with the xy plane.
This configuration splits the transverse phonon into an Ising-like out-of-plane mode and an XY-like in-plane mode, corresponding to out-of-plane or in-plane polarization. It is convenient to assign to every vector $\k$ in the 2D plane the three orthonormal vectors
% To see explicitly the momentum structure in the polarization, it is convenient to work with the right-hand trio
\begin{equation}
  \label{eq:trio-def}
  \hat k = \left(\begin{array}{c} \hat k_x \\ \hat k_y \\ 0\end{array} \right), \, \hat k_t = \hat z \times \hat k = \left(\begin{array}{c} -\hat k_y \\ \hat k_x \\ 0\end{array} \right), \,  \hat z = \left(\begin{array}{c} 0 \\ 0\\ 1\end{array} \right),
\end{equation}
such that the components transverse to $\hat k$ are $\hat k_t$ and $\hat z$. Similarly, we can decompose the spin degrees of freedom into
% Although it is straightforward, it is worthwhile to pay some attention to the summation in Eq. \eqref{eq:Sig-dis-1loop}. For every vector $\hat k$ the Pauli matrix vector can be decomposed into a parallel/transverse basis,
\begin{equation}
  \label{eq:sig-decomp}
  \sigma_{k;l}= \hat k \cdot \bvs = \hat k_x \sigma_x + \hat k_y \sigma_y, \,  \sigma_{k;t} = \hk_t \cdot \bvs = \hat k_x \sigma_y - \hat k_y \sigma_x, \, \sigma_z,
\end{equation}
where henceforth we will for clarity suppress the $k$ subscript. Then, the interaction splits into
\begin{equation}
  \label{eq:k-cross-decomp}
  % \hat k \times \bvs = \hat z\sigma_{t} + \hat k_t \sigma_z.
  \hat k \times \bvs = \hat z\sigma_{t} - \hat k_t \sigma_z.
\end{equation}

The transverse phonon,denoted in the previous section by $\P$, creates a structural distortion in the lattice, depicted schematically in Fig.~\ref{fig:qfmDist}.
When $\P$ condenses, i.e. when it acquires a nonzero static expectation value, the maximum energy gain is in a configuration where the polarization direction $\hat{\eta}$,
% \vk{ (how P is related to $\eta$ or $u$?)},
the fermionic ordering vector $\k$, and the associated Zeeman splitting $\bvs$, are all perpendicular to one another, see Eq. \eqref{eq:L-i}. This means that, if the phonon condensation results in an out-of-plane displacement, it will result in an in-plane spin-orbit splitting, whereas if the displacement is in-plane, the spin-orbit splitting will be out-of-plane. We denote these two modes the ``$z$'' and ``$t$'' modes, and depict these geometric constraints in Fig.~\ref{fig:qfmStat}. Since the 3D rotational symmetry is explicitly broken, the distance to the QCP of these two modes will be different, and we denote them by $r_z,r_t$. In other words, at the QCP, only one of the modes will be soft.

In what concerns static properties, these geometric constraints do not play an important role. However, once we consider dynamical properties arising from the scattering of electrons by the phonon modes, the situation changes.
The reason is that $\P$ couples most strongly to low-energy particle-hole excitations parallel to the FS (Fig.~\ref{fig:qfmDist}), which generates Landau damping. Thus, the momentum transfer vector $\q$ must be perpendicular to both $\k_F$, to maximize scattering, and to $\P$, since the phonon mode is transverse.
As long as $\P \propto \hz$ (the $z$ mode), this condition can be satisfied. In contrast, for the $t$ mode, it is impossible to place three vectors in a plane that are all perpendicular to each other, as shown in Fig.~\ref{fig:qfmDyn}. Consequently, scattering is suppressed in the $t$ channel, which is the source of most of the unique properties of {\QFM}s as compared to, say, a quantum ferromagnetic metal.

To see this more clearly, we rewrite the interaction term, Eq.~\eqref{eq:L-i}, in terms of the decomposition into the two modes,
\begin{align}
  \label{eq:L-i-2d}
  &\mathcal{L}_{int}^{(d=2)} = \frac{\lambda}{k_I} \left[\e_z(\q) \psi^\dagger(\p+\q/2)\sigma_t\psi(\p-\q/2) \right.\nn\\
                       &\qquad+ \left.(\e_xp_y-\e_yp_x) \psi^\dagger(\p+\q/2)\sigma_z\psi(\p-\q/2)\right],
\end{align}
% where we denoted the (scalar) $z$-component of $\P$ as $\eta_z$, and the (2D vector) in-plane component of $\P$ as $\P_t$.
In principle, the coupling constant $\lambda$ can take different values for the $z$ and $t$ sectors, but for simplicity we neglect this difference here. The angular form-factor in the $t$ sector can be rewritten as
\begin{equation}
  \label{eq:ff-t-2d}
  k_I^{-1}(\e_xp_y-\e_yp_x) \approx \pm \sqrt{\eta_x^2+\eta_y^2}\cos(\theta_p-\theta_q),
\end{equation}
where we assumed that the fermionic momentum $|\p| = \kf$ resides on the FS and denoted the in-plane angles of $\q,\p$ as $\theta_q,\theta_p$. We also used the fact that the angle between $\q$ and $\P_t$ must be $\pm \pi/2$. We see that the interaction term in the $z$ channel is similar to that of an Ising ferromagnet, in the sense that there is no explicit momentum dependence of the coupling constant.
However, the coupling constant in the $t$ channel depends on $\cos(\theta_q-\theta_p)$, which vanishes precisely at the angles $\pi/2$ where we expect electron-phonon scattering to be the strongest.

The ramifications of this result for the quantum critical (QC) behavior of the system are as follows. Consider first the $z$ mode. Since there is no geometric constraint on the scattering, it behaves rather similar to other itinerant quantum critical systems: it evinces strong Landau damping of $\P$, which in turn generates strong non-FL self-energy for the fermions. Then, most of the behavior of the system is determined by the competition between strong scattering, leading to e.g. strong pairing, and strong damping. On the other hand, the $t$ mode has a different behavior. The Landau damping is suppressed, resulting in an underdamped boson. Consequently, the fermions remain coherent even near the QCP. The result is that correlation effects like pairing are still enhanced, but the mechanism is completely different from the standard QC scenario. In the next sections, we perform a quantitative analysis that corroborates the expectations from this qualitative assessment.

\subsection{Disordered phase: self-energies to one-loop order}
\label{sec:self-energies-one}

We begin with analyzing the dynamics in the disordered phase and calculate the bosonic and fermionic self-energies to one-loop order.
The bosonic self-energy is obtained from the one-loop fermionic bubble, which from Eq. \eqref{eq:L-i} has the form,
\begin{align}
  \label{eq:bubble-bare}
  \resizebox{1.\hsize}{!}{$\Pi_0^{lk}(q) =\frac{\gb T}{k_F^2} \Trc  \sum\limits_{p}  (\p \times \bvs)_l   G(p-q/2) G(p+q/2) (\p\times \bvs)_k.$}
\end{align}
Here the trace is over spin indices and $G(k)$ are fermionic propagators. Here and henceforth, $\sum_p\equiv \left(\frac{a}{L}\right)^2\sum_{\p}\sum_{p_0}$ denotes a sum over the three-vector $p = (p_0,\p)$, where $p_0$ is a Matsubara frequency and $\p$ is a 2D vector (we leave the frequency normalization out of the sum to make the dimensional analysis more transparent.) Finally, $\gb = \lambda^2D_0$ is the effective fermion-boson vertex, which we take to be small, $\gb\nu_F \ll 1$, in order to control our calculations. Here, $\nu_F$ is the 2D FS local density of states summed over both spins,
\begin{equation}
  \label{eq:nuF-2d-def-main}
  \nu_F = \frac{\kf a^2}{\pi\vf},
\end{equation}
which for later convenience we define with a factor of $a^2$ to obtain a quantity with units of inverse energy.

We replace the sum over Matsubara frequencies and momenta by its infinite-system, zero-temperature limit
$T\sum_{p} \to a^2\int d^3p/\tpp^3$, 
except where we explicitly treat finite $T$ effects.
We then assume that the integral over fermionic momentum can be separated into an integral transverse to and parallel to the FS, which is the so-called Eliashberg approximation employed in many previous studies \cite{Abanov2003}. Then, the transverse momenta are restricted to the vicinity of the FS, which results in (see Appendix~\ref{sec:one-loop-self}):
\begin{align}
  \label{eq:Pi0-dis-final}
  \hat \Pi_0(q) = \gb\nu_F \int \frac{d\theta_p}{\tp} \frac{\vf q \cos(\theta_p-\theta_q)}{i q_0 - \vf q \cos(\theta_p-\theta_q)}\hat{\mathcal P}(\hat p).
\end{align}
Here and henceforth we present matrices in momentum (not spin) space with a $\hat{\cdots}$ symbol.
The projection matrices arise from the spin trace and the momentum dependence of the interaction. 
Performing the various summations and integrations we obtain,
\begin{align}
  \label{eq:Pi0-dis-form-1}
  &\hat\Pi_0(q) = -\hat z\hat z \left[ \delta r_z - \delta\Pi_z(q_0/\vf|\q|)\right] \\
    &\quad- \hat q_t\hat q_t\left[ \delta r_{t} - \delta\Pi_{t}(q_0/\vf|\q|)\right] - \hat q\hat q\left[ \delta r_{t} - \delta\Pi_{l}(q_0/\vf|\q|)\right] \nn
\end{align}
where we explicitly wrote the nonzero components of $\hat \Pi_0$ in unit-vector form, with $\hq_t$ defined in Eq. \eqref{eq:trio-def}. In Eq. \eqref{eq:Pi0-dis-form-1}, $\delta r$ are the static corrections to the energies of the phonon modes (or, equivalently, to the distance to the QCP),
\begin{equation}
  \label{eq:delta-rs}
  \delta r_z = \gb \nu_F = 2\delta r_{t},
\end{equation}
and $\delta\Pi$ are given by
\begin{align}
  \label{eq:delta-pi-defs}
  \delta\Pi_z(x) &= \gb\nu_F |x| l_0(x),
  % \frac{|x|}{\sqrt{1+x^2}}
                    \nn\\
  \delta\Pi_{t}(x) &= \gb\nu_F |x| l_1(x), \nn\\
  % \gb\nu_Fx^2\left(1-\frac{|x|}{\sqrt{1+x^2}}\right),\nn\\
  \delta\Pi_{l}(x) &= \delta\Pi_z(x)-\delta\Pi_{t}(x).
\end{align}
corresponding respectively to the dynamical contributions to the polarization in the out-of-plane z sector, the planar transverse t sector, and the planar longitudinal l sector. Here $l_0(x) = (1+x^2)^{-1/2}$ and  $l_1(x) = |x|(1-|x| l_0(x))$ are well known from the Lindhard functions of a 2D FL~\cite{Klein2019a}.

We emphasize that the static interaction renormalizes the energy of the phonon modes according to
\begin{equation}
  \label{eq:masses}
  r_z \to r_z - \delta r_z,r_t \to r_t - \delta r_t,
\end{equation}
so that the $z$ and $t$ sectors are split even if in the absence of electronic interactions their original splitting is negligible. In that case, the 
$z$ sector reaches the QCP first, since the gap in the $t$ sector remains finite,
\begin{equation}
  \label{eq:splitting}
  r_z = 0 \Rightarrow r_{t} = \gb\nu_F /2.
\end{equation}
In practice, though, as we discussed above, lattice effects provide their own splitting between the z and t modes, and can change the order of the transition. Moreover, we will show later that external strain tunes the energies of the two phonon modes in a controllable way. For this reason, we treat $r_z$ and $r_t$ as parameters and continue our analysis for both the case when the $z$ mode goes critical first and for the case when the $t$ mode goes critical first.

Importantly, because $l_0(x\rightarrow 0) \rightarrow 1$ whereas $l_1(x\rightarrow 0) \rightarrow |x|$, the $t$ phonon mode is underdamped while the $z$ phonon mode is overdamped. This can be seen by continuing to the real $\w$ axis, in which case the $z$ phonon has a classic Landau overdamped $\Gamma_z \sim i\w/\vf |\q|$ behavior while the planar mode is almost ballistic. This should be compared to the usual FL case \cite{Klein2019a}, where it is known that in the $p-$wave channel the longitudinal mode is underdamped and the transverse mode is overdamped.

To obtain the bosonic self-energy, we project the particle-hole bubble $\hat \Pi_0$ onto the transverse component,
\begin{align}
  \label{eq:Pi-1loop-0}
  \varPi^{ij}(q) &=  -\mathcal P^{ik}(\hat q) \Pi_0^{lk}(q)\mathcal P^{jl}(\hat q).
\end{align}
yielding the renormalized bosonic propagator,
%\vk{ (there's some mess with the factor $D_0$)}
\begin{align}
  \label{eq:chi-def}
  \hat D^{-1}(q) &= D_0^{-1}\left(\hat z\hat z D^{-1}_z(q) + \hat q_t\hat q_t  D^{-1}_t(q)\right), \\
  % D_z^{-1}(q) &= r_z + k_a^{-2}q^2 + \w_c^{-2}q_0^2+ \delta\Pi_z, \\
  % D_t^{-1}(q) &= r_t + k_a^{-2}q^2 + \w_c^{-2}q_0^2+ \delta\Pi_t.
                  D_z^{-1}(q) &= r_z + (|\q| a)^2 + (q_0a/c)^{2}+ \delta\Pi_z,\label{eq:Dz-def} \\
  D_t^{-1}(q) &= r_t + (|\q| a)^2 + (q_0a/c)^{2}+ \delta\Pi_t. \label{eq:Dt-def}
\end{align}
% where $r_z,r_t$ are the renormalized inverse correlation lengths squared.
Next, we calculate the fermionic self-energy $\Sg(k)$, which has a different form depending on which phonon sector becomes critical first. It has the form,
\begin{align}
  \label{eq:Sig-dis-1loop}
  \Sigma_{\alpha\beta}(k) &= \sigma_{0,\alpha\beta}\Sg(k) \nn\\
                          &\approx
                          % \frac{\gb}{k_a^2D_0}
                            \frac{\gb a^2}{D_0\kf^2}
                            \int \frac{d^3p}{\tpp^3} (\k \times \bvs_{\alpha\gamma})_i \nn\\
  &\qquad\qquad \times G(k-p)D^{ij}(p)(\k \times \bvs_{\gamma\beta})_j.
\end{align}
The splitting of the interaction, as in Eq.~\eqref{eq:k-cross-decomp}, also splits the contributions to the self-energy. Performing the summations we find
\begin{equation}
  \label{eq:sig-2-dis}
  \resizebox{1.\hsize}{!}{$\Sg(k) \approx
  % \frac{\gb}{k_a^2}
  \gb a^2
  \int \frac{d^3p}{\tpp^3} G(k-p)\left(D_z(p) + \cos^2(\theta_k-\theta_p)D_t(p)\right).$}
\end{equation}
The $\cos^2(\theta_k-\theta_p)$ term arises from the projection of $\hat k_t$ onto $\hat p_t$. This is very different from a transverse boson interacting via a conventional current-type coupling, in which case we would get a $\sin^2(\theta_k-\theta_p)$ term \cite{Wu2007,Klein2019a}. This difference is important because, when the fermions are restricted to the FS, the condition $\theta_k-\theta_p \approx \pm \pi/2$ is enforced. Thus, in contrast to the usual current-like coupling, the one-loop contribution to the self-energy from $D_t$ is greatly suppressed, as it comes from
% the \emph{incoherent} tail of the scattered fermions with a finite angle between the scattering vector and the fermion's Fermi vector
scattering of fermions on the FS to states away from the FS (and vice versa). As a result, the two sectors yield different fermionic behaviors.
In the $z$ sector, directly at the QCP ($r_z = 0$), the system displays nFL behavior with the self-energy %\vk{ (add $\text{sign}(k_0)$?)}
  \begin{equation}
    \label{eq:SE-z-QCP}
    \Sigma_z(k_0) = -i \w_z^{1/3} |k_0|^{2/3}\sgn(k_0),
  \end{equation}
where
\begin{equation}
  \label{eq:w-z-form}
  \w_z = \frac{\gb^2}{24 \sqrt{3}\pi^2\vf \kf} \sim \gb^2/E_F
\end{equation}
characterizes the typical energy scale at the QCP. Away from the QCP, where $r_z$ is finite but small, the self-energy will only have the characteristic QC form at frequencies $k_0 \gg \w_{r,z}$, where
\begin{equation}
\omega_{r,z} = r_z^{3/2}\vf k_a/(\gb\nu_F) \propto r_z^{3/2}(E_F^2/\gb),
\end{equation}
At lower frequencies $k_0 \ll \w_{r,z}$ the self-energy has the standard FL form
\begin{equation}
\label{eq:sig-z-fl}
    \Sg_z = -i \frac{\gb}{4\pi\vf k_a\sqrt{r_z}} k_0.
\end{equation}
where we defined
\begin{equation}
  \label{eq:ka-def-ms}
  k_a = \frac{1}{a}
\end{equation}
to make the units more transparent
(see Appendix~\ref{sec:one-loop-self} for details). 

In contrast, the self-energy at the $t$ channel QCP ($r_t = 0$) is given by
% , for $r_t = 0$, has the form, \vk{ (CHECK PREFACTORS)}
\begin{equation}
 \label{eq:Sg-t-qcp}
  \Sg_t(k_0) = -i \frac{\w_t}{8 \kf\vf} k_0 \log\left(\frac{\w_t}{|k_0|}Z_{UV}^2\right),
  %\mathcal{U}_t(\w_{r,t}/k_0)\right)
  %\log \frac{\w_{t}}{k_0}
  % \mathcal{U}_t(\w_{r;t}/k_0)
\end{equation}
where
\begin{equation}
  \label{eq:w-t-def}
  \w_{t} = \sqrt{\gb\nu_F}k_a v_F \propto \sqrt{\gb E_F}
\end{equation}
characterizes the typical energy scale at the QCP. Here, $Z_{UV} = \mbox{min}(1,c/\vf)$ denotes whether the high-energy cutoff in the system is given by the bare bosonic speed of sound or by the Landau damping. For finite but small $r_t$, i.e. away from the $t$-channel QCP, Eq. \eqref{eq:Sg-t-qcp} is only correct for $k_0 \gg \w_{r,t}$, with 
\begin{equation}
    \omega_{r,t} = r_t k_a\vf/\sqrt{\gb\nu_F} \propto r_t(E_F^2/\gb)^{1/2}.
\end{equation}
%$\omega_{r,t} = r_t k_a\vf/\sqrt{\gb\nu_F} \propto r_t(E_F^2/\gb)^{1/2}$ and $\mathcal{U}_t(x\to 0) = 1,\mathcal{U}_t(x\to \infty) \propto x^{-1}$.
For low frequencies, $k_0 \ll \w_{r,t}$ the self-energy is linear and obeys the usual FL behavior
\begin{equation}
\label{eq:Sg-t-rt}
    \Sg_t(k_0) = -i\frac{\w_t}{8\vf \kf}k_0 \log\left(\frac{\w_t}{\w_{r,t}}  Z_{UV}^2 \right).
\end{equation} 
% At the critical point, $r_t = 0$, the expression simplifies to
% \begin{equation}
%     \label{eq:Sg-t-mnFL}
%     % \Sg_t(k_0) = \frac{\sqrt{\gb\nu_F}k_a}{8 k_F} k_0 \log\left|\frac{\w_{t}}{k_0}\right|\sgn(k_0)
%     \Sg_t(k_0) = -i \frac{\w_t}{8 \vf \kf} k_0 \log\left|Z_{UV}^2\frac{\w_{t}}{k_0}\right|.
% \end{equation}
% where 
% \vk{ (This factor appears under the log only. Also, I have smth a little different for $U_t$.).
% \begin{equation}
% \Sigma_t(k_0) = \frac{k_0 \sqrt{\bar g}}{4 \pi v_F \sqrt{\nu_F k_a^2}} \ln \left( \min\left\{ \frac{\sqrt{\bar g \nu_F k_a^2 c^2}}{\omega};  \sqrt{\frac{\bar g \nu_F c^2}{r_t v_F^2}}    \right\}   \right).
% \end{equation}
% } 
Equation~\eqref{eq:Sg-t-qcp} represents a marginal FL. We note that Eqs.~\eqref{eq:Sg-t-qcp} and~\eqref{eq:Sg-t-rt} are valid as long as the argument of the log is large, i.e. $\max\{k_0, \omega_{r,t}\} \ll \omega_t Z_{UV}^2.$ %At finite $r_t$, the logarithm is replaced by
%\begin{equation}
%  \label{eq:Sg-t-r}
  %\log\left(\frac{\w_t\mathcal{U}_t(\w_{r,t}/k_0)}{k_0}\right),
%\end{equation}
%where $\omega_{r,t} = r_t k_a\vf/\sqrt{\gb\nu_F} \propto r_t\gb^{-1/2}E_F^{3/2}$ and $\mathcal{U}_t(x\to 0) = 1,\mathcal{U}_t(x\to \infty) \propto x^{-1}$. 

The total electronic self-energy is thus the sum of the contributions from the two sectors,
\begin{equation}
  \label{eq:Sg-sum}
  \Sg(k_0) = \Sg_z(k_0,r_z) + \Sg_t(k_0, r_t).
\end{equation}
 
The consequences of Eq. \eqref{eq:Sg-sum} are as follows. Without fine-tuning, only one of the sectors can become fully critical, while the other one retains a finite $r$.
% mass-squared.
If $r_z \to 0$, the system displays nFL behavior due to the contribution from the $z$-sector, analogous to that of an Ising ferromagnet, since the contribution from the $t$ sector just renormalizes the FL parameters. On the other hand, if $r_{t}$ goes to zero, then the system is a marginal FL, since the nFL contribution from $\Sg_z$ is cut off by the finite value of $r_z$, leaving just a linear FL-like contribution.

Our results were obtained under the simplifying assumption of a single fermion-boson coupling constant $\gb$, see Eq. \eqref{eq:ff-t-2d} and the discussion after Eq. \eqref{eq:Pi0-dis-final}. Similarly to $r_z,r_t$, lattice effects can also split the coupling, which will just modify somewhat the prefactors to the various self-energies. We neglect all such effects in our work. We also explicitly computed vertex corrections within the model to check that they do not qualitatively modify the one-loop results. In Appendix \ref{sec:vertex-corrections} we show that while vertex corrections can be divergent (as is the case in several other boson-fermion models \cite{Chubukov2005a}), they do not invalidate the results in this section.

\subsection{Pairing in 2D {\QFM}s}
\label{sec:pairing-2d-qfms}

We now investigate
% Finally, we turn to study
superconductivity arising from QC FE fluctuations. In this paper, we will not make a comprehensive study of the superconducting phase of a {\QFM}, since our focus is on the normal state quantum-critical properties.
% since our focus till now has been on normal state properties in the QC regime.
Instead, we will study the pairing instabilities via a linearized gap equation to determine which properties are unique to {\QFM}s and which ones are similar to other QC unconventional superconductors, like the ferromagnetic and nematic ones. In doing this, we go beyond the FL regime, which has been at least partially studied previously, see e.g. Refs.~\cite{Kozii2015,Wang2016,Gastiasoro2020a,Gastiasoro2021}.
As in Sec.~\ref{sec:self-energies-one}, we assume that both $r_z$ and  $r_t$ are parameters that can be tuned to criticality independently.

%The pairing diagram is shown in Fig. (...).
The pairing equation is given by 
\begin{widetext}
\begin{equation}
  \label{eq:pairing-eq-1}
  \Phi_{\alpha\beta}(k) = \frac{\gb T}{\kf^2D_0} \sum_{p} \gamma_{\nu\alpha}^{i}\left(\frac{-\k-\p}{2}\right)G_{\nu\kappa}(-p)\Phi_{\kappa\mu}(p)G_{\mu\sigma}(p)\gamma^{j}_{\sigma\beta}\left(\frac{\k+\p}{2}\right)D^{ij}(k-p),
\end{equation}
\end{widetext}
where
\begin{equation}
  \label{eq:gamma-def}
  \gamma^i_{\alpha\beta}(\k) = (\k\times\bvs)_{\alpha\beta}^i\,.
\end{equation}
At first sight, the pairing interaction seems to be repulsive in the singlet channel, since $\gamma_{\nu\alpha}^{i}\left(\frac{-\k-\p}{2}\right)\gamma^{j}_{\sigma\beta}\left(\frac{\k+\p}{2}\right)= -\gamma_{\nu\alpha}^{i}\left(\frac{\k+\p}{2}\right)\gamma^{j}_{\sigma\beta}\left(\frac{\k+\p}{2}\right)$. This is nothing but the well-known statement that a current-like interaction (i.e. with form-factor $\k$) is repulsive. However, the spin summation in Eq.~\eqref{eq:pairing-eq-1} gives another $-1$ factor, reflecting the fact that a magnetic-mediated pairing interaction is generally repulsive. Thus, the total pairing interaction is attractive in the singlet channel precisely because of the spin-charge mixing, even though each component by itself would be repulsive.

\begin {table}[tp]
    \caption{Table of the first four irreducible representations of a rotationally invariant model with inversion symmetry. The last two columns denote whether the pairing channel is attractive (+) or repulsive (--).}
    \label{tab:SC_decomp}
    \begin{tabular}{| l | c |  c |c |c |}
    \hline
    irrep  &  Matrix form $F_{nj}(\hat \k)$  & Inv. symmetry  & $z$  & $t$   \\ \hline
     $n=0$  &   $1$  &   even  &  + &  + \\ \hline
     $n=1$  &  $ \hat \k \cdot \boldsymbol \sigma $&  odd & -- & -- \\ \hline
     $n=2 $ &   $\hat k_x \sigma^y - \hat k_y \sigma^x$  &  odd  & + & --\\ \hline
       $n=3 $ &   $\{k_y\sigma^z,k_x\sigma^z \}$  &  odd  & -- & + \\ \hline
     
    \end{tabular}
\end {table}

It is convenient to decompose the pairing function into irreducible representations
\begin{align}
\label{eq:rep-phi-2d}
\Phi(k) = i\sigma^y\sum_{nj} \phi_{nj}(k_0) F_{n}^j(\hat \k),
\end{align}
where $F_{n}^j(\hat \k)$ is a 2$\times$2 matrix function encoding the $j$th member of representation $n$ (see Table~\ref{tab:SC_decomp} for the first few representations in the case of a fully rotationally and inversion symmetric system)~\cite{Kozii2015,Kozii2019}. In this notation, each representation has its own transition temperature, which is obtained from the gap equation, and $T_c$ is set by the highest one. Table~\ref{tab:SC_decomp} reveals that the $z$ mode is attractive in the $n=0$ singlet channel and the $n=2$ nodeless triplet channel, which is a superposition of the $m_z=\pm 1$ spin-triplet channels in the $\ell=1$ spin  sector of the Cooper pair. The $t$ mode is attractive in the singlet channel and in the $n=3$ doublet, which is the $m_z=0$ spin-triplet channel. These results follow qualitatively from the ``double repulsive" nature of the pairing interaction discussed above. Since the $t$ mode has a $\sigma_z$ spin dependence, it is attractive in the channels for which a $z$-axis Ising spin mode would be repulsive, namely spin-singlet and $t$ spin-triplet, but repulsive in the channels for which the spin mode would be attractive, namely spin-polarized channels. Conversely, the $z$ mode is attractive only in the singlet and in the $z$-axis spin-polarized channels, since it has $t$ spin polarization.

To discuss the superconducting transition temperature $T_c$ resulting from Eq.~\eqref{eq:pairing-eq-1}, we consider two scenarios $r_z \gg r_t \to 0$ and $ r_t \gg r_z \to 0$ separately.
As we will show, there is a qualitative difference between the pairing promoted by the $z$ and $t$ modes at the QCP. In the case of the out-of-plane $z$ mode, QC pairing arises from the standard interplay of a singular interaction  with reduced fermionic coherence from the non-FL self-energy. The case of the $t$ mode is different because of the $\cos^2$ term in the effective interaction. In what concerns the normal-state properties, as we showed in the previous section, this angular-dependent term suppresses both the interaction strength and the fermionic incoherence. As for the pairing instability, the angular term implies that there is no pairing between two fermions exactly on the FS, removing the weak-coupling FS instability towards pairing. On the other hand, the bosonic mode is underdamped, again because of the reduced phase-space for scattering. This enhances the pairing attraction strength and gives rise to a logarithmic divergence, similar to the Cooper instability, but stemming from the \emph{bosonic} degrees of freedom. Pairing is strong only near the QCP, and completely vanishes away from it.

We start with the case where $r_z \to 0$, and consider only the attractive channels, $n=0,2$. In agreement with Refs. \onlinecite{Kozii2015,Wang2016} we find that the two channels  are degenerate. 
To estimate  $T_c$ we assume that the gap equation is purely local in momentum space on the FS,  neglect all nonsingular  $\p$ dependence in the gap equation and integrate over $\p$.
For the $z$ mode we find, after the angular integration,
\begin{widetext}
  \begin{align}
    % \phi_{nj}(k_0,|\k|) &= \frac{\pi \gb T \nu_F}{\kf} \sum_{k_0\neq p_0}\int_0^\infty \frac{d|\p|}{2\pi}\frac{1}{|p_0|+|\Sg(p_0)|} \frac{\phi_{nj}(p_0,|\p|)}{r_z + |\p|^2/k_a^2 + \gb\nu_F|p_0-k_0|/(\vf |\p|)}.
                            \label{eq:pairing-z-2}
                          \phi_{nj}(k_0) &= \frac{\pi \gb T \nu_F}{\kf} \sum_{p_0\neq k_0}\int_0^\infty \frac{dp}{2\pi}\frac{1}{|p_0|+|\Sg(p_0)|} \frac{\phi_{nj}(p_0)}{r_z + p^2/k_a^2 + \gb\nu_F|p_0-k_0|/(\vf p)}.
\end{align}
\end{widetext}
Equation~\eqref{eq:pairing-z-2} has the same form as the linearized gap equation of other itinerant QC systems, e.g. ferromagnets and nematics.
The pairing instability results from the $(|p_0|+|\Sigma|)^{-1}$ term arising from the fermionic Green's function, which in the FL regime yield the Cooper logarithm. Note that we have removed the diagonal $k_0=p_0$ term representing thermal fluctuations; we will comment on this shortly. Integrating over $p$ results in a gap equation local in $\theta_k$ and with a purely frequency dependent effective interaction $U_{\mbox{eff}}\propto |p_0-k_0|^{-1/3}$ at the QCP. The solution is well known and obeys \cite{Moon2010,Wang2016a,Metlitski2015,Klein2019}, 
\begin{align}
  \label{eq:Tc-estimates}
  T_{c,z} = 
  \left\{\begin{array}{ll}
           a_z \frac{\gb^2}{\kf\vf}
           % \frac{\}{\vf\kf}
           & r_z \ll \w_z/\w_{r,z} \\
    b_z\frac{r_z^{3/2}}{\gb\nu_F}\vf k_a\exp\left(-\frac{4\kf\sqrt{r_z}}{\gb\nu_F k_a}\right) &r_z \gg \w_z/\w_{r,z}    % correct the factors
  \end{array}.\right. 
\end{align}
The parameters $a_z,b_z$ are $O(1)$ and are discussed in more detail in Appendix \ref{sec:pairing-near-z}. To obtain Eq. \eqref{eq:Tc-estimates} and also the estimates appearing later in this section, we assumed for simplicity that the upper cutoff for the pairing logarithm, when it exists, is given by the interplay between momentum and the Landau damping induced polarization, and that the lower cutoff is just $2\pi T_c$. We also note that some of the numerical coefficients we presented are obtained while neglecting the frequency dependence of $\phi_{nz}$, which is not justified at the critical point.

Let us now turn to the case of $r_t\to 0$, again restricting to the attractive $n=0,3$ channels which are approximately degenerate at the critical point.
% As apposed to the out of plane mode, the $t$-mode mediates an attractive interaction in the $n = 0$ and $n=3$ channels (see Table~\ref{tab:SC_decomp}).
Now, however, the angular integration yields
\begin{widetext}
  \begin{align}
  \phi_{nj}(k_0) = \frac{\pi \gb T \nu_F}{\vf\kf} \sum_{p_0\neq k_0}\int_0^\infty \frac{dp}{2\pi p }~l_1\left(\frac{p_0}{\vf p}\right) \frac{\phi_{nj}(p_0)}{r_t + p^2/k_a^2 + \gb\nu_F(k_0-p_0)^2/(\vf p)^2},\label{eq:pairing-t-2}
\end{align}
\end{widetext}
where $l_1(x)$ is defined below Eq.~\eqref{eq:delta-pi-defs}. Note the disappearance of the Cooper instability, which is replaced by $p^{-1}l_1(p_0/\vf p)$, which in turn is non-singular at small $p$ or small $p_0$. The reason for this is that for small angle scattering, the interaction is proportional to $\cos^2(\theta_k-\theta_p)$ and vanishes exactly when the fermion is scattered parallel to the FS in the small angle scattering limit. Since the QC contribution to the Cooper instability in the vicinity of the QCP arises precisely from this regime of scattering, it is suppressed. 
Consequently, away from the critical point there is no logarithmic divergence and the pairing instability is absent. At criticality, however, the singular nature of the interaction balances  this vanishing factor and causes a logarithmic divergence, with  $T_c$  given by   
\begin{equation}
  \label{eq:Tt-p}
  T_{c,t} = a_tZ_{UV}^2\vf k_a \sqrt{\gb\nu_F}\exp\left(-\frac{8\kf}{k_a \sqrt{\gb \nu_F}}\right),
  % \frac{\vf }{} a_t v_F k_a \sqrt{\gb \nu_F} \exp(-1/\sqrt{\gb\nu_F}).
\end{equation}
where $a_t$ is detailed in Appendix \ref{sec:pairing-near-t}  and $Z_{UV}$ was defined after Eq. \eqref{eq:w-t-def}. Note that the same caveats specified after Eq. \eqref{eq:Tc-estimates} apply here as well.

Thus, superconductivity is enhanced compared to the naive BCS type $T_c \sim \exp(-1/V_0)$, where $V_0 \sim \gb \nu_F /r$, but only in an exponentially narrow region around the QCP. At finite but small $r_t$ we find $T_{c,t} \to T_{c,t}-\delta T_{c,t}$, where $\delta T_{c,t} \sim r_t\frac{\vf k_a}{\sqrt{\gb\nu_F}}$.
Thus the pairing vanishes when $r_t \sim Z_{UV}^2\gb \nu_F \exp(-8\kf/(k_a\sqrt{\gb\nu_F}))$.

We finish this section by commenting on the dropping of the diagonal $k_0 = p_0$ term in the gap equations, representing thermal fluctuations. They are formally divergent, since $D_z,D_t$ both diverge at the QCP at zero frequency. It can be shown that for singlet pairing, these terms drop out of the gap equation, as they are cancelled by similar fluctuations renormalizing the self-energy $\Sigma$ \cite{Chubukov2003,Chubukov2005}. This effect is just a manifestation of Anderson's theorem, since static thermal fluctuations can be considered a form of nonmagnetic disorder. However, in triplet channels this exact cancellation does not take place, and can result in a reduction of $T_c$ or in a first order transition~\cite{Chubukov2003}. We verified, by calculating corrections to the self-energy, that in our model there is an approximate cancellation as long as $r_z,r_t$ are sufficiently separated (see Appendix \ref{sec:vertex-corrections}), which justifies dropping the diagonal terms. We did not study in detail additional instabilities that may arise within the superconducting state due to this non-exact cancellation. We similarly leave a detailed investigation of the effect of $r_z \sim r_t$ on the superconducting state for a later work.

\subsection{Quantum order-by-disorder phases}
\label{sec:quant-order-disord}

Our analysis so far has concentrated on establishing the properties of {\QFM}s within a one-loop approximation. This is justified in 2D when there is a large splitting between the out-of-plane ($z$) sector and the in-plane ($t$) sector, since there is little feedback between the two channels. However, if $r_z \sim r_t$, feedback effects may not be neglected. There are three reasons to study this regime in detail. First, one may expect that if the lattice energy scale $\w_T$ is smaller than or of order of $\ef$, the splitting between the sectors will not be large in comparison with the typical energy scale associated with electronic fluctuations, i.e. $\gb$. Second, as we show later, the splitting between sectors can be tuned by strain, so even a system with $\w_T \gg \ef$ may be brought into a state with $r_t \sim r_z$. Finally, in 3D (e.g. for a cubic lattice) there is no splitting between the sectors at all, so that it is helpful to study the degenerate case in 2D as a warm-up for the 3D problem. We will therefore now turn to the case where the splitting is comparable with the electronic scale, and for simplicity will study the fully degenerate case $r_t = r_z$.

It is known that for a metallic system near a QCP, the expected nFL behavior may not be realized for several reasons. First, there is the issue of pairing, which may preempt the nFL region \cite{Metlitski2015,chubukov2020interplay}.
Second, unless the ordered phase breaks a discrete  symmetry, the soft Goldstone modes that accompany the second order QCP have their own dynamics, which when coupled to the fermions can give rise to additional orders, in a mechanism known as quantum-order-by-disorder (QOBD) \cite{Green2018}. The idea is that since fluctuations diverge near the QCP, spontaneously breaking the symmetry introduces a finite cutoff to the fluctuations, which reduces their energy cost. This can happen either via a first-order transition or by shifting the wave-vector of the instability to a non-zero value.

QOBD has been extensively studied for magnetic systems \cite{Belitz1997,Kirkpatrick2011,Brando2016,Kirkpatrick2020,Maslov2009,Rech2006,Chubukov2003,Green2018}. Its main signature is the emergence of nonanalytic terms in the magnetic correlations  generated by soft particle-hole excitations that are cut-off by the preemptive order. In a two-dimensional ferromagnetic system, these generate respectively a $-|\mathbf{M}|^{2+a}$ term in the magnetic free energy (where $\mathbf{M}$ is the magnetic order parameter) and a $-|\q|^a$ term in the (inverse) magnetic correlation function, where $a=1$ at higher temperatures and $a=3/2$ in the nFL region, weakening the transition. In three dimensions the nonanalytic behavior is logarithmic and therefore much weaker (see Sec.~\ref{sec:quant-order-disord-1}).

The situation for a {\QFM} is more complex than the ferromagnetic one.
First and foremost, as we discussed in Sec.~\ref{sec:self-energies-one}, the $z$ and $t$ sectors are split due to the spin-momentum mixing, see Eq. \eqref{eq:splitting}. This is in contrast with a magnetic system, where in the absence of an explicit magneto-elastic coupling, reducing the dimensionality of the lattice does not break the SU$(2)$ spin symmetry.  Thus, in a magnetic system, the soft fluctuation space is three-dimensional, and upon condensing at the QCP, there are still 2D soft Goldstone fluctuations. For {\QFM}s, the transverse phonon is constrained to only two soft directions, similar to an XY magnet, such that soft fluctuations are one-dimensional.
In ferromagnets, even a 1D soft fluctuation is enough to trigger QOBD. In {\QFM}s, though, fluctuations along the remaining dimension are also gapped out, as noted above, due to the $z$/$t$ splitting from the electronic polarization, see Eq. \eqref{eq:delta-rs}. This provides an intrinsic IR (infrared) cutoff to the fluctuations, which however can be of the order of the typical electronic scale $\gb\nu_F$. 
%\vk{ (Why is it gapped out? Where did we show it?)}. 
On the other hand, we already saw that a {\QFM}  can remain a FL down to the QCP, which gives rise to stronger quantum fluctuations than in a nFL with strongly damped fermions.
%\vk{ (Landau damping suppresses quantum fluctuations?)}. 
% AK Landau damping is suppresses bosons not fermions. incoherence suppresses quantum fluctuations because it reduces FS spectral weight.
As we shall show, the end result is that the system \emph{does} in general enter a QOBD phase, but that it is easy to tune the system (e.g. via strain) out of this phase.

\begin{figure}
  \centering
  \includegraphics[width=\hsize]{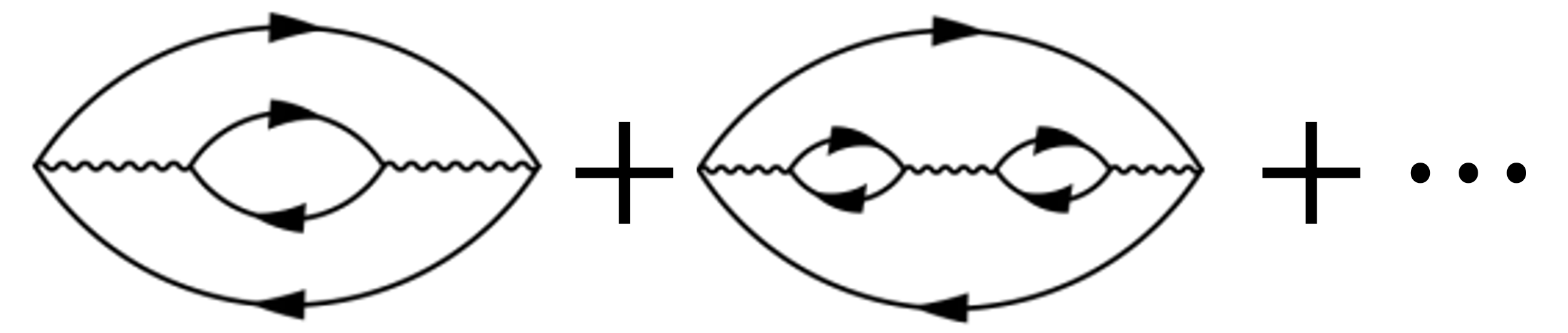}
  \caption{The ladder diagrams yielding nonanalytic contributions to the free energy in the ordered state, which promote a tendency to a first-order transition.}
  \label{fig:ladderQOBD}
\end{figure}
\begin{figure}
  \centering
  \includegraphics[width=\hsize]{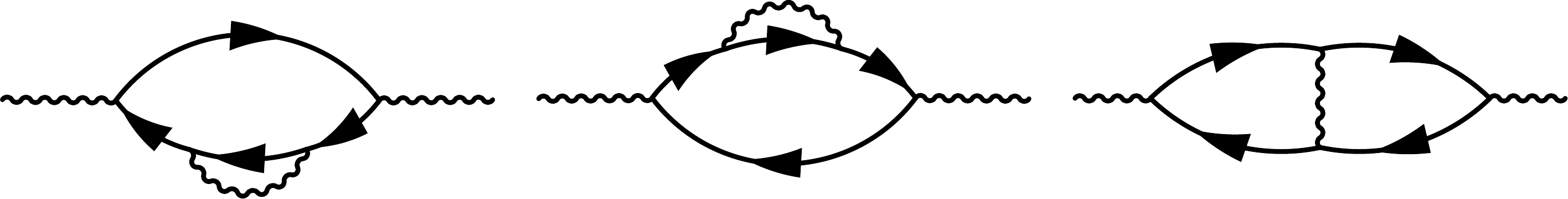}
  \caption{The lowest-order diagrams in the disordered phase that contribute a negative term proportional to $-|\q|$ in the bosonic inverse propagator, promoting an instability towards a FDW (i.e. finite-$q$) state.  %\vk{ Do we have understanding why exactly these diagrams are so important? Same question to Fig. 3.}
  }
  \label{fig:finiteQOBD}
\end{figure}

To perform our detailed calculations we will follow the methods of Ref. \cite{Maslov2009}. We will seek for both a first-order instability to a homogeneous state and a finite-wavevector transition to a Ferroelectric Density Wave (FDW) state (which remains second order within our approach).
To identify the propensity to a first-order transition we shall compute nonanalytic corrections to the free energy by self-consistently generating an effective action~\cite{Karahasanovic2012} in a ladder approximation (see Fig.~\ref{fig:ladderQOBD}). In order to identify finite-$q$ FDW instabilities, we will calculate the leading order corrections to the one loop calculations from Sec.~\ref{sec:self-energies-one} in the disordered phase. The reason for this is that we were unable to generalize the method of effective action to the finite-$q$ {\QFM} scenario. The effective-action method has been used in the magnetic case to study finite-$q$ behavior, but the method utilizes the spin/charge decoupling, so it cannot be applied to the {\QFM} case.

\subsubsection{Instability towards a first-order transition}
\label{sec:inst-1st-order}

We consider the possibility of a preemptive first-order transition by calculating the effective free energy of the coupled phonon-fermion system in the presence of static FE order.
% and compute the free energy in the presence of this order. 
A first-order transition will occur if the dynamical fluctuations generate terms that drive the free energy negative even in the presence of a finite mass term. We integrate out the fermions to generate an effective bosonic action for the FE modes, and then integrate out FE fluctuations to obtain an effective action for the static FE order parameter. This procedure is equivalent to summing up the series of diagrams in Fig.~\ref{fig:ladderQOBD} (the first-order diagram was already included in Sec.~\ref{sec:self-energies-one}). To make the calculations more transparent we assume that the fermions are in a FL state. As discussed above, this description is correct down to the QCP for the $t$ mode but not for the $z$ mode. However, the distinction turns out to be not  very important, and we will comment on the nFL situation later.

To proceed, we introduce a static FE order parameter
%\vk{ (not well defined at zero momentum)}
\begin{equation}
  \label{eq:Delta-def}
  \bv{\Delta}_j = \lambda \bv{\eta}_j(|\q| = 0,\W=0),
\end{equation}
where $j=z,t$.
% We start by introducing a static order-parameter
% \begin{equation}
%   \label{eq:Delta-def}
%   \Delta_i = \lambda P_i(\q=0,\W=0),
% \end{equation}
Equation~\eqref{eq:Delta-def} should be understood as the term obtained by first taking the static $\W\to 0$ limit and then the uniform $|\q|\to 0$ limit, such that the transverse $\hq \cdot \P_j = 0$ nature of the phonon is obeyed. Note that the procedure is well defined since there is only one state with $\q=0$, corresponding to the 
%static 
$q=0$ 
term in the interaction, Eq.~\eqref{eq:L-i}.
%\vk{ (How do we even define $\bv{\eta}_j(\q=0,\W=0)$?)}.
This term modifies the effective fermionic action by inducing a Rashba splitting,
\begin{align}
  \label{eq:Lfl-mod}
  \mathcal{L}_{FL} &= \psi^\dagger_\alpha(p)\left[(i p_0 - \vf (|\p|-\kf))\delta_{\alpha\beta}\right. \nn\\
  &\qquad\qquad-\left.\bv\Delta_j \cdot \hp\times\bvs_{\alpha\beta}\right]\psi_\beta(p),
\end{align}
where for convenience we already wrote the FL form of the bare propagator linearized near the FS. We then integrate out the fermions to obtain the effective bosonic action.
It has two parts. First, a static free energy for~$\Delta$,
\begin{equation}
  \label{eq:F-delta-bare}
  F_{\Delta} = \frac{r_j}{\gb}|\bv\Delta_j|^2 + \frac{u_jD_0}{2\gb^2}|\bv\Delta_j|^4,
\end{equation}
where again $j = z,t$ depending on the mode, and we also phenomenologically added a quartic term to ensure the stability of the free energy. The $\gb$ factors arise because we incorporated factors of $\lambda$ into the definition of $\Delta$. The second part is a dynamical effective action for the phonons,
\begin{align}
  \label{eq:LP-mod}
\mathcal{L}_{P} &= \eta_i(q) D_0^{-1}\left[\hz \hz D_z^{-1}(q,\bv{\Delta}_j)\right. \nn\\
&\qquad\left. + \hq_t\hq_tD_t^{-1}(q,\bv{\Delta}_j)\right]_{ij}\eta_j(-q).
\end{align}
%\vk{ (Split the equation into 2 lines.)} 
% \vk{ (I thought we stuck to $q$ for bosons.)}
There are no $\hz \hq_t,\hq_t\hz$ cross-terms in the action in the 2D case, as we show explicitly later. Here,
\begin{align}
  \label{eq:D-def-mod}
   D_z^{-1}(q,\bv\Delta_j) &\approx r_z + k_a^{-2}|\q|^2 + \delta\Pi_{\Delta;z}\left(\frac{q_0}{\vf|\q|},\frac{|\bv\Delta_j|}{\vf|\q|}\right), \\
  D_t^{-1}(q,\bv\Delta_j) &\approx r_t + k_a^{-2}|\q|^2 + \delta\Pi_{\Delta;t}\left(\frac{q_0}{\vf|\q|},\frac{|\bv\Delta_j|}{\vf|\q|}\right) 
\end{align}
are the renormalized propagators in the presence of static FE order, where $k_a$ is a reciprocal of the lattice spacing, see Eq. \eqref{eq:ka-def-ms}. For simplicity we removed the quadratic frequency terms which are irrelevant in the low-energy regime. The correction to the free energy from the presence of static FE order is obtained by tracing out the bosonic action. Recalling that
$\eta_i(q) = \hat{\mathcal{P}}_{ij}(\hq)u_j(q)$
is just a phonon already projected onto its transverse component,
%we could just as easily have written out Eq. \eqref{eq:LP-mod} replacing $P_i\to u_i$ 
%\vk{ (Do you mean $\eta_i \to u_i$? And why can we do this?)}.
% AK Because there is already a projection in that equation. Will calrify the language
%Thus, 
the effective free energy density is the sum of the bare energy \eqref{eq:F-delta-bare} and a correction from the trace-log of the propagator in $\mathcal{L}_P$, namely
\begin{equation}
  \label{eq:free-1st-order-def}
  \delta F(\bv\Delta_j) =
                        % \bv{\Delta} \cdot (\hz\hz r_z + \hat x\hat xr_t) \cdot \bv{\Delta} \nn\\
                      % &
                        T \sum_{q}\left[\log D_z^{-1}(q,\bv{\Delta}_j) + \log D_t^{-1}(q,\bv{\Delta}_j)\right] - F_0,
\end{equation}
where $F_0$ is the bare energy.

The free energy correction will have a different form depending on whether $\bv\Delta$ condenses in the $t$ or $z$ configuration. Without loss of generality we will pick
\begin{equation}
  \label{eq:Delta-directions}
  \bv\Delta_z = \hat z \Delta \qquad \mbox{or} \qquad \bv\Delta_t = \hat x \Delta,
\end{equation}
where $\Delta > 0$, depending on which type of transition we consider. We then calculate Eqs. \eqref{eq:LP-mod}-\eqref{eq:free-1st-order-def} for both cases, thus checking whether one order has a stronger tendency to an instability than the other (we will find that the $z$ mode is typically more unstable). 

\paragraph{The free energy for an out-of-plane FE transition.}

Let us begin with the simpler case of $\bv\Delta_z = \hz \Delta$. To gain insight into this problem, it is convenient to consider the ferromagnetic analogue of this situation, in which case the coupling of $\Delta$ to the fermions has a form factor $\bvs$ instead of $(\p\times \bvs)$ in Eq.~\eqref{eq:Lfl-mod}. In the ordered phase, when $\Delta > 0$, the magnetic response splits into two sectors - an out-of-plane (longitudinal magnetic) sector representing intraband excitations that is independent of $\Delta$, and an in-plane (transverse magnetic) sector representing interband spin-flips. The transverse response is nothing but the Goldstone mode in the ordered state, and hence remains gapless but with a nontrivial functional dependence on $\Delta$. To obtain the contributions to the free energy from the magnetic fluctuations it is enough to log-trace out the magnetic inverse susceptibilities. When this is done, one finds that the contributions from the transverse sector generate nonanalytic $\Delta$-dependent terms in $F$. Importantly, the contributions arise from fluctuations with $\vf|\q| \sim \W \sim \Delta$, which is a different regime than the fluctuations giving rise to quantum critical behavior and pairing instabilities, in which typically $\vf |\q| \gg \W$ \cite{Maslov2009}. In Fig. \ref{fig:ladderQOBD} we depict the diagrams which are summed up in the ladder approximation as we discussed in the introduction to this section. 
%\vk{ (Not sure if it's clearly written here.)}{\jr[I agree. This is a very nice concept, but the common reader will not be able to follow. If i understood correctly, what happens here is that the FSs are split and the Goldestone mode connects them, so there it does not couple strongly in the $q\to 0$ limit? Also there is a paper by Watanabe and Vsihwanath about a criterion for when there is strong coupling and when not. This can also be cited here... ]}

We have already shown that when the coupling to electrons drives the $z$ mode to the critical point $r_z=0$, $r_t$ remains finite but small, even if the bare $z$ and $t$ masses were equal.
%, see Eq.~\eqref{eq:splitting} \vk{ (What if $r_z$ and $r_t$ are very different to start with and, say, $t$ transition takes place first?)}. 
This implies that any generated nonanalytic terms are always sensitive to the finite correlation length in the $t$ sector. Let us demonstrate how this happens in practice. The first step is to compute $\varPi^{ij}_\Delta$, the polarization bubble projected onto the transverse sector (see Eq.~\eqref{eq:Pi-1loop-0}), in the presence of the finite FE order parameter. The bubble, before projection, has a nontrivial spin texture,
\begin{flalign}
  \label{eq:bubble-Delta}
  \Pi_0^{lk}(q) =\frac{\gb T}{k_F^2} \Trc  \sum_{p}   (\p\times \bvs)_l G_\Delta(p-q/2) (\p \times \bvs)_k \cdot \nn\\
  G_\Delta(p+q/2) ,
\end{flalign}
where $G_\Delta$ represents the fermionic propagator in Eq.~\eqref{eq:Lfl-mod}. To proceed, we diagonalize the propagators via the $p-$dependent transformation $\psi \to U\psi$, where
\begin{align}
  \label{eq:U-def}
  U &= e^{-\frac{i}{2} \sigma_z (\theta_p+\pi/2)} e^{-\frac{i}{2}\sigma_y (\pi/2)} \\
  \Rightarrow G_\Delta^{-1} &= (ip_0 - \vf(|\p|-\kf)) \sigma_0 - \Delta \sigma_z.\label{eq:G-delta-def}
\end{align}
As a result the interaction form factor changes to
\begin{equation}
  \label{eq:ff-change}
  (\hat p \times \bvs) \to U^\dagger (\hat p \times \bvs) U = \hp_t \sigma_x + \hz \sigma_z.
\end{equation}
Thus, the polarization splits into in-plane and out-of-plane sectors, as was the case for the disordered phase calculation, see Eq. \eqref{eq:k-cross-decomp}, with the $z$ sector behaving as an effective Ising spin. Furthermore, since spin-flip processes occur only for the $\sigma_x$ form-factor, the $z$ and $t$ sectors are analogous to the longitudinal and transverse components of the magnetic system. Performing the various summations we find Eqs.~\eqref{eq:LP-mod} and~\eqref{eq:D-def-mod} (see Appendix~\ref{sec:1st-ord-transition}) with
\begin{align}
  \label{eq:delta-pi-Deltas}
  \delta\Pi_{\Delta;z}(x,y) &= \gb\nu_F |x|l_0(x), \nn\\
                               % \delta\Pi_z\left(\frac{q_0}{\vf|\q|}\right), \nn \\
  \delta\Pi_{\Delta;t}(x,y) &= \frac{\gb\nu_F}{2} |x|l_1(x - 2iy) + \cc,
\end{align}
where $l_i(x)$ are the same functions given after Eq.~\eqref{eq:delta-pi-defs}. 

Now we can compute the corrections to the free  energy in  Eq. \eqref{eq:free-1st-order-def}. Clearly, the contribution from $D_z$ is zero, as it does not depend on $\Delta$ even in the ordered state. All that is left is the contribution from $D_t$, which has the form
\begin{align}
  \label{eq:f-oop-1}
  \delta F &= T\sum_q \left(\log D_{t}^{-1}(q,\Delta\hat z)- \log D_{t}^{-1}(q,0) \right) \nn\\
           &= k_a^{-2} \int \frac{d^3q}{\tpp^3} \log \left[\frac{r_{t} + |\q|^2 + \delta\Pi_{\Delta;t}}
             % \left(\frac{1}{2}\delta\Pi_t\left(\frac{q_0 + 2i \Delta}{\vf |\q|}\right) + \cc\right)}
             {r_{t} + |\q|^2 +\delta\Pi_{0;t}}\right]
             % \delta\Pi_t\left(\frac{q_0}{\vf |\q|}\right)} \right]
\end{align}
where $k_a$ was defined in Eq. \eqref{eq:ka-def-ms}. There are several features to note here. First, an expansion of the integrand in powers of $\Delta$ yields even powers $\Delta^2,\Delta^4,\cdots$. Nonanalytic terms can be generated if there is nonanalytic behavior related to the lower limit of integration. Second, $F$ has a $\Delta^2$ term that arises from the UV (ultraviolet) limit of the momentum integration, which can be checked by expanding in powers of $\Delta/\vf|\q|$. This term can be incorporated into $r_z$. Third, the remainder of the integral is convergent and peaked at $q_0 \sim \vf|\q| \sim \Delta$, so that for small enough $\Delta$ we may neglect the analytic $|\q|^2$ terms in the propagator in comparison with the $q_0/\vf|\q|$ terms in $\delta\Pi$. Finally, as we discussed in the opening statements to this section, we are assuming that before coupling to electrons $r_z = r_t$, so that after the coupling is included, when $r_z = 0$ then $r_{t}  = \gb \nu_F/2$, see Eq. \eqref{eq:splitting}. This means the integral is completely dimensionless, since both $r_t$ and $\delta\Pi_{\Delta;t}$ have the same $\gb/\nu_F$ prefactor (this remains true so long as $|r_z-r_t| \ll \gb\nu_F$ before the coupling to electrons). After appropriate rescaling we find,
\begin{align}
  \label{eq:delta-f-final}
  &\delta F_z(\Delta) = 
    \frac{\Delta^3}{\vf^2k_a^2} \int_0^\infty \frac{x^2 dx dz}{2\pi^2} \times \nn\\
  &\quad\log \left[\frac{1 +
    % \left(\frac{1}{2}\delta\pi_t\left(z+2ix^{-1}\right) + \cc\right)}
    z\left(l_1(z+2ix^{-1})+l_0(z-2ix^{-1})\right)}
    {1 + 2zl_1(z)}\right]\nn\\
    % \delta\pi_t\left(z\right)} \right] \nn\\
  &\approx -0.18 \frac{\Delta^3}{\vf^2k_a^2}, % ka^2 = (ka^2/kF^2*\pi vF kF)*kF^2/(pi vF kF) = 1/nu_F * kF^2/(pi kF vF) ==> -3.3 D^3 / vF^2 / (kF^2/(pi vF kF)/(nu_F)) = -3.3 D^3 pi nu_F/(v_F k F)
\end{align}
% where we set $\delta\pi_t = \delta\Pi_t / (\gb \nu_F)$.
where the prefactor was computed numerically, see Appendix \ref{sec:1st-ord-transition}. The final form of the free energy is
\begin{equation}
  \label{eq:F-oop-final}
  F \approx \frac{1}{\gb}\left(r_z\Delta^2 - 0.56 \frac{\gb \nu_F}{\vf \kf}\Delta^3 + \frac{u_zD_0}{2\gb}\Delta^4\right).
\end{equation}
The cubic term with a negative coefficient implies a first-order transition. Therefore, fluctuations in the ordered state drive a preemptive first-order transition before the QCP is reached.

\paragraph{The free energy for an in-plane FE transition.}

We now calculate what happens when $r_t = 0$ but $r_z$ remains finite and small. For concreteness we take $r_z = \gb\nu_F/2$, i.e. exactly the opposite limit to what we assumed in the previous calculation. The result does not change significantly as long as $r_z \sim \gb\nu_F$. The calculation proceeds in a similar manner as for the out-of-plane transition. The splitting introduced by the static order is
\begin{equation}
  \label{eq:zeeman-ip}
  (\Delta \hat x) \cdot \hp \times \bvs = \sin\theta_p \sigma_z.
\end{equation}
The fermionic Green's function is already diagonal, and the form-factor for the interaction is given by Eq. \eqref{eq:k-cross-decomp}. Clearly, in this case the role of ``longitudinal'' and ``transverse''  between the $z$ and $t$ sectors is reversed. Performing the calculation we find (see Appendix \ref{sec:1st-ord-transition})
\begin{align}
  \label{eq:Pi-delta-ip}
  \delta\Pi_{\Delta;t}(x,y) &= \gb\nu_F |x| l_1(x), \nn\\
                            % \Pi_t(q), \nn\\
  \delta\Pi_{\Delta;z}(x,y,\theta_q) &= \gb\nu_F\frac{|x|}{2}l_0(x-2 i y\cos\theta_q) + \cc.
                            % \delta\Pi_z \left(\frac{q_0 + 2i\Delta\cos\theta_q}{\vf|\q|}\right) + \cc.
\end{align}
 The appearance of a renormalized Zeeman field $\Delta\cos\theta_q$ is just the result of the suppression of forward scattering in the $t$ sector discussed previously. However, since there is a significant angular phase space where $
\cos\theta_q = O(1)$, the $\Delta^3$ term is still generated. The free energy correction is now
\begin{align}
  \label{eq:f-ip-1}
  \delta F_t(\Delta) &= T\sum_q \left(\log D_z^{-1}(q,\Delta\hat x)-\log  D_z^{-1}(q,0) \right) \nn\\
    &\approx 0.09 \, \delta F_z (\Delta).
\end{align}
Consequently, we obtain the nonanalytic cubic term, like in the $z$ sector case, but with a significantly reduced numerical prefactor (Eq. \eqref{eq:F-oop-final}). This is due to both the different forms of $l_0, l_1$ appearing in $\delta\Pi$ and the averaging over $\theta_q$ in Eq.~\eqref{eq:Pi-delta-ip} (see Appendix~\ref{sec:1st-ord-transition}).

\subsubsection{Instability towards a ferroelectric density-wave state}
\label{sec:instability-finite-q}

We now determine whether an instability to a FDW (i.e. finite-$q$) state is driven by fluctuations. This is done by computing the leading order diagrams in the disordered state. 
As we discussed in the beginning of this section, the instability to a finite-$q$ state arises from a similar mechanism to the one that gives rise to the first-order transition. Namely, if finite-$q$ order exists in the system, it cuts off the IR divergence of fluctuations near the QCP. At a technical level, the effect is manifested by the generation of negative nonanalytic momentum-dependent terms in the inverse boson propagator. Therefore, to study this instability it is not necessary to introduce FE order, but to compute the leading order correction to the RPA susceptibility at $\W = 0,|\q|>0$. The relevant diagrams are shown in Fig. \ref{fig:finiteQOBD}. 
%\vk{ (What about Aslamazov-Larkin correction? And I still don’t understand how very different diagrams lead to
%essentially the same or similar physics.)} 
Note that the so-called Aslamazov-Larkin diagrams which are important for the ferromagnetic QCP case \cite{Rech2006} are omitted from Fig.~\ref{fig:finiteQOBD} as they give zero due to the form-factor of the coupling term. As in the previous section, if the diagrams in Fig. \ref{fig:finiteQOBD}
% \vk{(who ``they''?)}
contribute a negative, nonanalytic term at finite $|\q|$, an instability towards FDW order preempts the FE-QCP. Because we have already given a detailed account of the process for the case of a first-order transition, here we only provide the main steps of the calculation in the cases of both $z$ and $t$ transitions.

The diagrams in Fig. \ref{fig:finiteQOBD} acquire the following forms, for small external momenta $\q$ and near the QCP, 
\begin{widetext}
\begin{align}
  \label{eq:delta-D-def}
  \Pi_q^{ij}(\q,0) &= d^{ij}_1(\q,0) + d^{ij}_2(\q,0),\\
  d^{ij}_1(q=(\q,0)) &= 2\frac{\gb^2T^2}{\kf^4D_0} \sum_{p,k} G(k-q)G(k-p)G^2(k) D^{mn}(p)\mathcal{F}^{imnj}_0(\k), \\
  d^{ij}_2(q=(\q,0)) &= \frac{\gb^2T^2}{\kf^4D_0} \sum_{p,k} G(k-q)G(k-p)G(k)G(k-p-q) D^{mn}(p)\mathcal{F}_0^{imjn}(\k).
\end{align}
Here, we introduced the form-factors for the self-energy and vertex diagrams,
\begin{align}
  \label{eq:form-factors}
  \mathcal{F}_0^{imnj} = \Trc(\k\times\bvs)^i(\k\times\bvs)^m(\k\times\bvs)^n(\k\times\bvs)^j.
  % , \\
  % \mathcal{F}_2^{nijm} = \Trc(\k\times\bvs)^n(\k\times\bvs)^k(\k\times\bvs)^j(\k\times\bvs)^m. \\
\end{align}
Although $d_1$ and $d_2$ seem to have different propagator dependence on momentum and frequency, they can be recast into a more compact form. This is a result of a Ward identity for the density-density correlation function, which imposes a certain analytic structure on the three diagrams even though the correlator we compute is not the density-density one (see e.g.~\cite{Chubukov2018}).
After algebraic manipulation (see Appendix \ref{sec:finite-transition-1}), we find
\begin{align}
  \label{eq:delta-D-form}
  \Pi_q^{ij}(\q,0) &= 2\frac{\gb^2T^2}{\kf^4D_0} \sum_{p,k} G(k-q)G(k-p)G^2(k) D^{mn}(p)\mathcal{F}^{imnj}(\k), \\
  \mathcal{F}^{imnj}(\k) &= \Trc(\k\times\bvs)^i(\k\times\bvs)^m\left[(\k\times\bvs)^n,(\k\times\bvs)^j\right].
\end{align}
%\vk{ (Some $($ is missing?
If we replaced $\k\times\bvs \to \sigma_0$, then we would obtain zero, since the commutator would vanishes. This turns out to be a requirement of the Ward identity mentioned above. Performing the spin traces we find
\begin{equation}
  \label{eq:Pi-q-form-2}
  \hat \Pi_q = 8\gb^2T^2 \sum_{p,k}G(k-q)G(k-p)G^2(k) \left[\hz\hz \cos^2(\theta_p-\theta_k) D_t(p) + \hk_t\hk_t D_z(p)\right]
\end{equation}
\end{widetext}
where we assumed that $|\k| \approx \kf$. Although the diagrams contributing to the finite-$q$ instability are different from those that we evaluated to obtain the first-order instability in the previous section, the physics is qualitatively the same. First, the fluctuation contribution to the out-of-plane mode comes from the in-plane fluctuations (and vice-versa). Second, note that the expression in Eq. \eqref{eq:delta-D-form} has the form of a self-energy correction to the polarization bubble. It involves a secondary scattering of an excited electron-hole pair with momentum $\q$, which serves as an IR cutoff for the fluctuations.
% Add comment on not near FS dynamics?
Making use of these insights, the evaluation is straightforward but tedious, yielding (see Appendix \ref{sec:finite-transition-1}),
\begin{equation}
  \label{eq:Pi-q-final}
  \hat\Pi_q = -\gb\nu_F\frac{|\q|}{\kf}\left(0.03\hz\hz+0.003\hq_t\hq_t\right),
  % \frac{\gb}{\pi^3\vf \kf}\frac{q}{k_a}\left(0.4\hz\hz +0.1\hq_t\hq_t\right).
\end{equation}
where the prefactors were evaluated numerically. Thus, the static propagators for the two modes have the form
\begin{align}
  \label{eq:D-q-final}
  D_z^{-1} = r_z + k_a^{-2}q^2 -0.03 \gb\nu_F\kf^{-1} q,
  % \frac{\gb}{\vf\kf}\frac{q}{k_a}
  \nn\\
  D_t^{-1} = r_t + k_a^{-2}q^2 -0.003\gb\nu_F\kf^{-1} q.
  % \frac{\gb}{\vf\kf}\frac{q}{k_a} 
  \end{align}

  Minimization with respect to the momentum results in a non-zero wave-vector, corresponding to a preemptive FDW transition before the QCP is reached. Note that this effect is stronger for the $z$ mode than for the $t$ mode. 

In the derivation above, we assumed a FL form for the fermionic self-energy, which is not justified at the QCP. Indeed, for the ferromagnetic QCP case, it has been shown \cite{Rech2006,Maslov2009} that both the first-order and finite-$q$ transitions are modified at the QCP, with the nonanalytic terms scaling as $\Delta^{7/2},|\q|^{3/2}$ due to nFL contributions. We now argue that this is not the case for the FE QCP. The reason is that the change in power law for the ferromagnetic QCP case can be traced to the singular form of the polarization bubble, see Eq.~\eqref{eq:delta-pi-Deltas} for the case of a first-order transition. In the FL  regime, the typical frequency and momentum scales are $\W_n  \sim \vf |\q| \sim \Delta$. By power counting this gives rise to a $\Delta^3$ contribution to the free energy. In the nFL regime \cite{Maslov2009}, the scaling changes to $\Sigma(\W_n)  \sim \vf |\q| \sim \Delta$, and since $\Sigma\sim\W_n^{2/3}$, this changes the free-energy contribution to $\Delta^{7/2}$. 

However, such a scaling analysis neglects vertex corrections, which are necessary to maintain both spin and charge conservation and can cancel out self-energy contributions. It has been shown that for a ferromagnetic QCP, vertex corrections do not restore the FL form of the nonanalytic terms \cite{Rech2006,Chubukov2005a,Maslov2009}. The reason for this is that within a spin-fermion model, the spin associated with ferromagnetic order is not conserved independently, but only in combination with the fermionic spin. In the FE case, the $t$ and $z$ modes have different behaviors near the QCP. Because the $t$ mode remains FL all the way down to the QCP, the polarization bubble retains its qualitative form, up to logarithmic factors that can be neglected. The $z$ mode does give rise to nFL behavior. However, this mode behaves qualitatively like an Ising degree of freedom, which to a first approximation \emph{is} conserved separately from the fermionic spin-orbit moment. Thus, vertex corrections ensure that the FL form of the polarization bubble remains approximately the same (see Appendix \ref{sec:vertex-corrections}), and that the nonanalytic terms retain their form. In this sense, the FE and the ferromagnetic QCPs are qualitatively different.

\subsection{The phase diagram}
\label{sec:phase-diagram-finite-T}

\begin{figure*}
  \centering
  \begin{subfigure}{0.45\hsize}
    \includegraphics[width=\hsize]{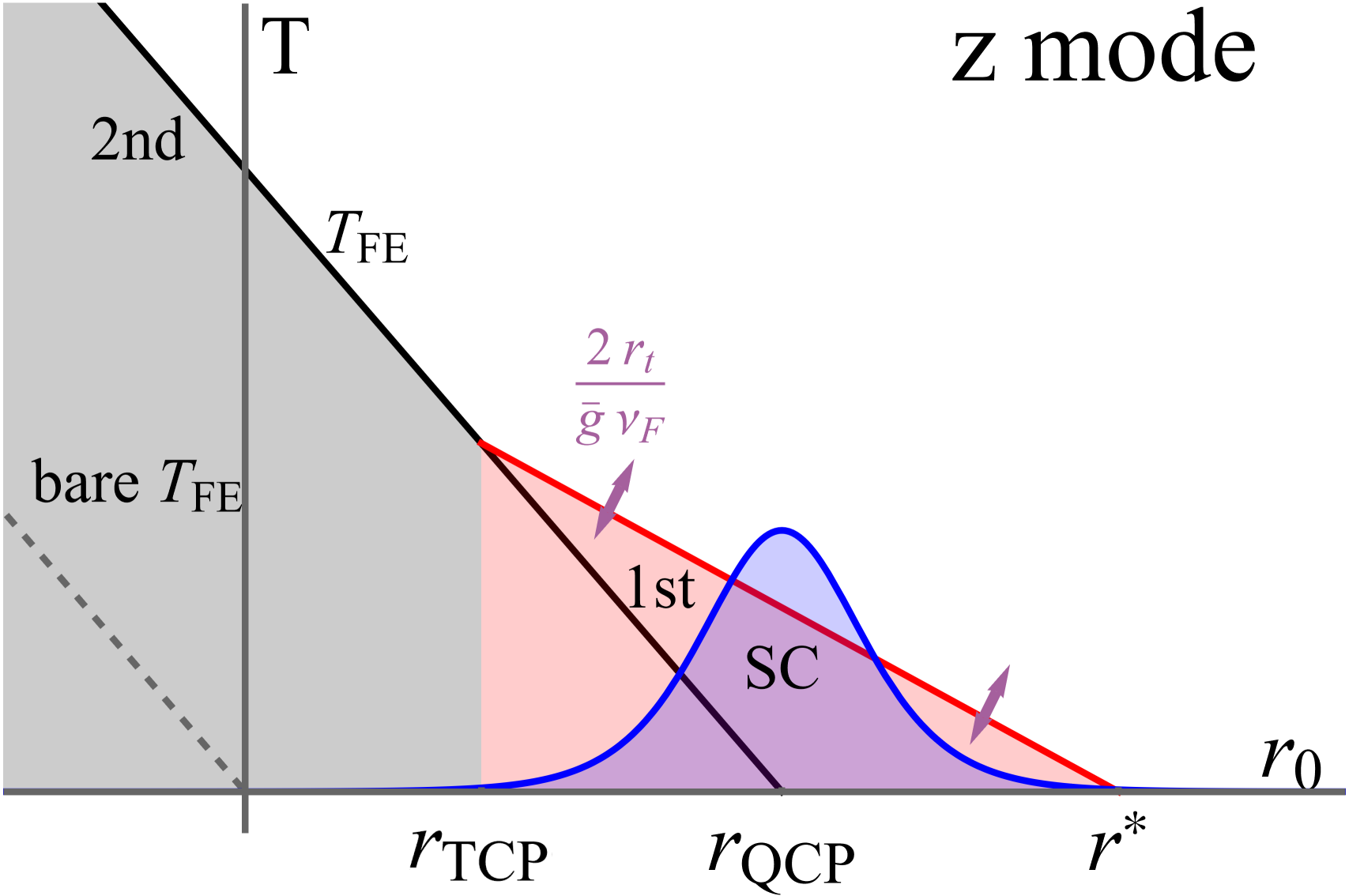}
    \caption{}
  \end{subfigure}\hfill
  \begin{subfigure}{0.45\hsize}
    \includegraphics[width=\hsize]{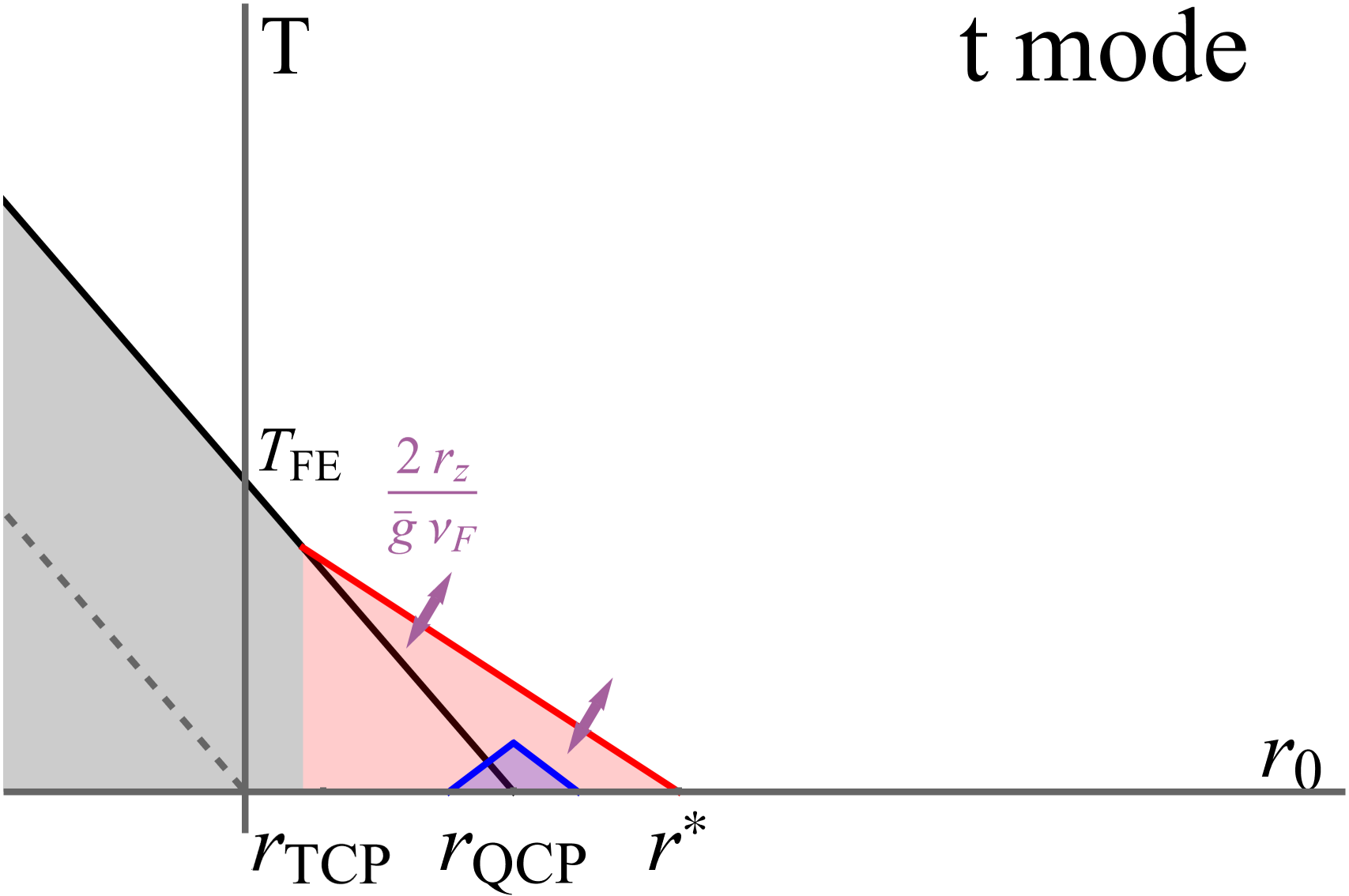}
    \caption{}
  \end{subfigure}
  \caption{
  The schematic phase diagram for the 2D QFEM. (a) The phase diagram for the $z$-mode transition, corresponding to out-of-plane FE, tuned by the parameter $r_{0,z}$, see Eq. \eqref{eq:T-j-def}. The black line and gray shading denote a second-order transition to the FE state, terminating at the putative QCP, while the red line and shading denote the first-order transition, which extends the transition line to $r^*$. A SC dome rises above the first-order transition line. (b) The phase diagram for the $t$-mode transition, corresponding to in-plane FE, tuned by the parameter $r_{0,t}$. The shading and labels are the same as for panel (a). For the $t$ mode, the termination point  is shifted to a lower value compared to the $z$ mode, and the SC dome, if present, is buried in the ordered state. For both figures, the purple arrows show the direction along which the first-order transition line moves when the ratio $r_{t/z}/\gb\nu_F$ is changed, where $r_{t/z}$ define the mass of the \emph{non-}critical mode. For $r_{t/z}/\gb\nu_F \gg 1$, the first-order line is suppressed almost entirely.  }
  \label{fig:phaseDiag2d}
\end{figure*}

Based on the results of this section, we can now construct the phase diagram of a 2D QFEM, which consists of a normal paraelectric and ferroelectric state, and a superconducting dome. In addition, the second-order FE transition may be preempted by a first-order one, or by a transition to a finite-$q$ FDW state. As usual, the phase diagram for either $z$ or $t$ modes depends on the mass terms $r_{z}$ and $r_t$ in the action, see e.g. Eqs.~\eqref{eq:chi-def}--\eqref{eq:Dt-def}. 
Up to now, we treated $r_z$ and $r_t$ as independent parameters which can be tuned by e.g. doping or pressure. To extend the analysis to finite temperatures, we include a phenomenological temperature dependence, which for concreteness we assume to have the usual Curie-Weiss form,
%\vk{ (Where does this come from? I thought that all the previous analysis was purely of $T=0$ and $r_j$ were independent parameters controlled externally (doping, pressure, etc.)}
\begin{align}
  \label{eq:r-T-dep}
  r_j =\alpha(T-T_j),
\end{align}
where $T_j$ is the transition temperature to the FE state after renormalization due to the coupling to electrons.
At $T=0$, $r_j = -\alpha T_j$ is just the tuning parameter used in the previous sections, such that when $T_j < 0$ the system is in the disordered phase and when $T_j > 0$ the system is in the ordered phase. Explicitly, $T_j$ is given by,
%Its connection to the zero-temperature tuning parameter $r$ is given by
\begin{equation}
  \label{eq:T-j-def}
  \alpha T_j = -r_{0,j} + \delta r_j, \quad r_{0,j} = \left(\frac{\omega_{T,j}}{k_a c}\right)^2,
\end{equation}
where the sound velocity $c$ was defined in Sec.~\ref{sec:minimal-model-qfm}, see Eq.~\eqref{eq:r-basic}, and $\omega_{T,j}$ are the transverse phonon optical frequencies at $T=0$, which in our previous treatment we considered to be equal for simplicity $\w_{T,z}=\w_{T,r}=\w_T$, see Eq.~\eqref{eq:r-basic}. $\delta r_j$ 
%are the contributions to $r_j$ arising from the coupling to the electrons, see 
were defined in Eq.~\eqref{eq:masses}. Thus, $\alpha$ has units of inverse energy and $r_{0,j}$ is a ``tuning parameter'' towards the QCP as determined by e.g. pressure or doping. While the temperature dependence in Eq.~\eqref{eq:r-T-dep} may not capture the actual behavior near the FE critical point \cite{Rowley2014}, our results are not qualitatively changed by assuming another $T$ dependence.

While the phase diagrams for the $z$ or $t$ modes are similar overall, there are some qualitative differences. First, $\delta r_z = 2\delta r_t$, so even when the purely bosonic system has $r_{0,z}=r_{0,t}$, as we assumed for the analysis above, the originally degenerate transitions are shifted to two different critical points.
% two systems are shifted to a different critical point
Second, the superconducting dome for the $t$ mode occupies an exponentially smaller area than the $z$ mode dome, due to the unconventional form of $T_{c,t}$, see Eq. \eqref{eq:Tt-p}. Except for that, however, both modes display the ``standard'' QCP picture of an ordered state tapering down to a SC dome.

To understand how the presence of QOBD modifies the phase diagram, we assume that the back action of SC on QOBD is weak enough that we can analyze the two tendencies independently. Consider first $T=0$ in the normal state. Starting from the case where the bare masses are identical, $r_{t,0} = r_{z,0}$, the changes in the renormalized masses due to the coupling to electrons are such that the $z$ mode reaches the quantum critical point first. %\vk{ (I thought that $r_t = r_z$ means that the transitions occur simultaneously.)}. 
The first-order transition occurs when $F=0,\pd_\Delta F = 0$, and the FDW transition occurs when 
$D_z^{-1}=0, \pd_q D_z^{-1} = 0$. Performing these calculations, we find that the first-order transition occurs at
\begin{equation}
  \label{eq:1st-order-trans}
  \Delta_j^* = a_j (\gb\nu_F)^2 E_0, \quad r_j^* =  \frac{a_j^2}{2} \gb^3E_0\nu_F^4\frac{k_a^2}{\kf^2},
\end{equation}
%\vk{ (Have we defined $u$?)}
where $E_0 = \frac{k_a^2}{u D_0 \kf^2}$ is an UV scale of the system, $u$ is the quartic coefficient defined in Eq. \eqref{eq:LP-mod}, and $a_z = 1.76, a_t = 0.14$. The FDW transition occurs for
\begin{equation}
  \label{eq:fdw-trans}
  \tilde q_j = b_j (\gb\nu_F) \frac{k_a^2}{\kf}, \quad \tilde r_j =  b_j^2 (\gb\nu_F)^2\frac{k_a^2}{\kf^2}
\end{equation}
where $b_z = 0.015, b_t = 0.0015$.
% , which we show in Fig. \ref{fig:phaseNormal}.

From Eqs. \eqref{eq:1st-order-trans} and \eqref{eq:fdw-trans} it appears that because $\tilde{r}_j$ is proportional to a lower power of $\gb\nu_F$ than $r_j^*$,
%, 
the transition 
%$r$ 
%\vk{ (What does ``the transition $r$'' mean?)} 
to the FDW state is 
%of lower order in $\gb\nu_F$ and should be 
preferred. However, for typical cases where $E_0\nu_F =O(1)$, $\gb\nu_F = O(1)$, the numerical prefactors $b_j$ in $\tilde r_j$ render it very small, favoring the first-order transition. To complete the picture, we extend our $T=0$ analysis of the first-order transition to finite temperatures. We show in Appendix~\ref{sec:1st-ord-transition} that for finite temperatures, the nonalytic term in the free energy scales as $\Delta^3f_T(T/\Delta)$, with $f_T(0) = 1$ and $f_T(x) \propto 1/x$ for $x\gg 1$. Thus, the tendency to a first-order transition weakens with increasing temperature. This implies that if 
%, so that for positive enough 
$T_j^0 = \alpha^{-1} r_{0,j}$ (see Eq. \eqref{eq:T-j-def}) is positive and large enough, the transition occurs at a high enough temperature that the cubic term is absent from the free energy and
%that the nonanalytic term is negligible compared to the quartic term in the free energy, 
the transition is 
%becomes 
second-order.
%\vk{ (What's ``positive enough''? What's $T_j^0?$)}. 
The tricritical point occurs at a temperature (see Appendix~\ref{sec:1st-ord-transition})
\begin{equation}
  \label{eq:T-TCP}
  T_{\mbox{\tiny TCP},j} \approx \frac{a_j}{4\pi} (\gb\nu_F)^2E_0.
\end{equation}
Interestingly, this temperature is of order of the superconducting $T_c$ for the $z$ mode. Thus, we can expect that the SC dome will rise above the first-order transition line for the case we just analyzed, i.e. when the $z$ mode is unstable. For the case where the $t$ mode is unstable, we may expect the SC phase to be buried inside the ordered FE state. The two schematic phase diagrams are depicted in Fig.~\ref{fig:phaseDiag2d}, of which a simplified version appeared already in the introduction. We emphasize that these phase diagrams are constructed without taking into account the feedback between the ordered FE state and superconductivity.

\subsection{Lattice properties and strain effects}
\label{sec:lattice-properties}

\begin{figure}
  \centering
  \includegraphics[width=\hsize]{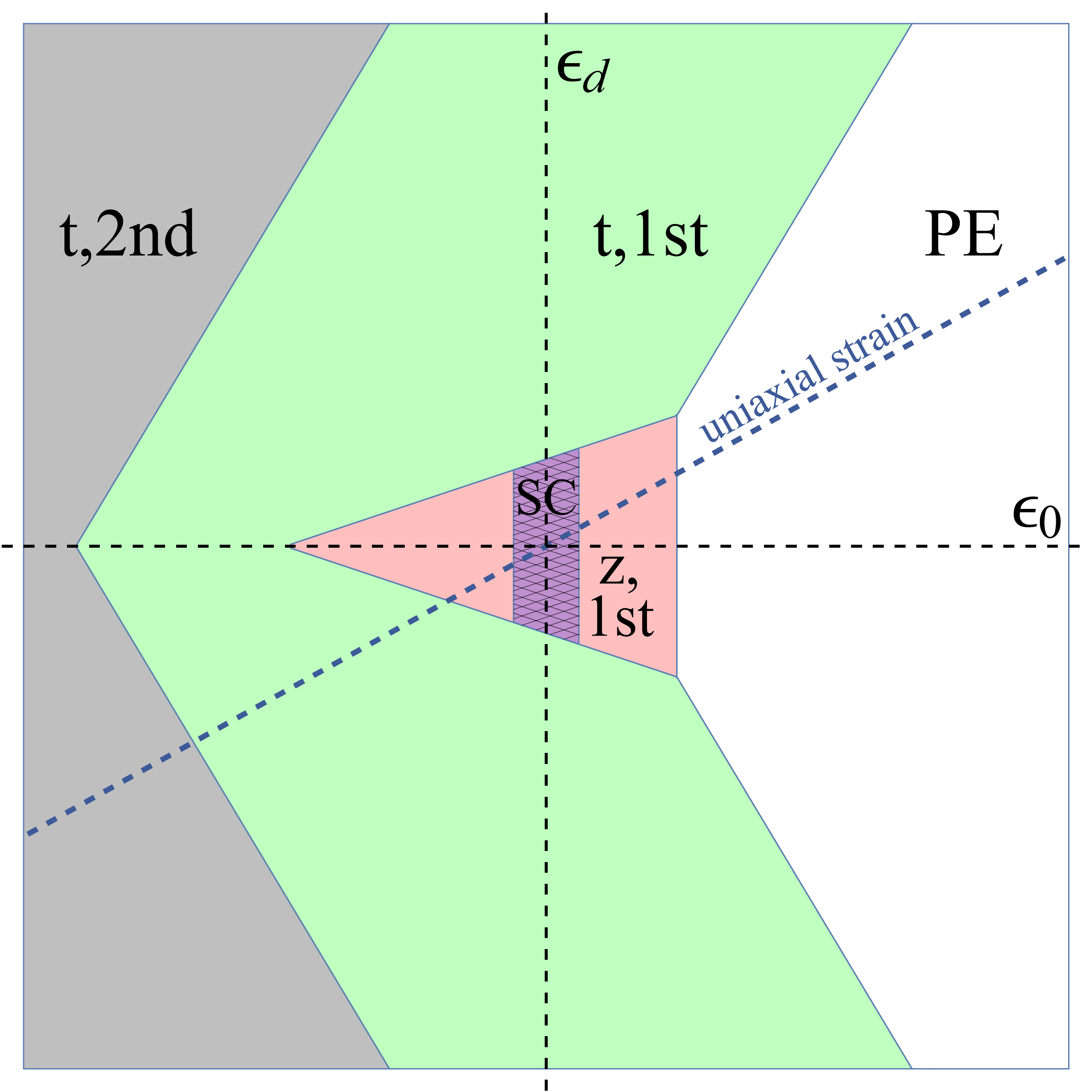}
  \caption{ 
  Schematic phase diagram of a 2D {\QFM} in the presence of external strain parametrized by a volume-changing symmetry-preserving component $\ve_0$ and a volume-preserving symmetry-breaking component $\ve_d$. At zero strain, the system is assumed to be at a $z$ mode FE+SC state, as depicted in Fig. \ref{fig:phaseDiag2d}(a), with $r_z \approx 0$, $r_t \approx \gb\nu_F/2$. Upon applying compressive (tensile) $\ve_0$ strain, $r_z$ and $r_t$ are both shifted positively (negatively), but the magnitude of the negative shift for $r_t$ is greater, thus driving a $z$ to $t$ transition for compressive strain. Upon applying $\ve_d$ strain, $r_t$ is shifted negatively in a preferred direction, see Eq. \eqref{eq:rt-strain}. The dashed line depicts the response to uniaxial strain.
  See Appendix \ref{sec:app-num} for numerical parameters.}
  \label{fig:strain1}
\end{figure}
We now discuss how the
% normal state
phase diagram for the {\QFM} derived in the previous section is modified by the lattice degrees of freedom.
%We will discuss the relevance of this picture for real materials, e.g. STO, at the end of the paper.
The existence of an underlying 2D crystal has two main implications.  First, the phonon polar modes couple nonlinearly
% have a nonlinear coupling to acoustic modes, and in particular
to the strain tensor, so that
% . This means that
applying external stress also modifies the phonon propagator. We shall see that this property allows one to control the phase diagram of a {\QFM}.
% , and we shall see that this allows for changing the behavior of both t and z modes.
Second, the rotational symmetry is broken down to a discrete one, implying
% . This changes both the form of the polar phonon propagator,
a mixing between the longitudinal and transverse modes away from high-symmetry directions. In what follows, we will assume that the lattice only weakly perturbs the rotationally invariant modes,
% and that
so that we only need to calculate the effect of the lattice anisotropies projected onto the transverse mode. We will also neglect additional instabilities arising from the coupling between the polar phonon modes and elastic fluctuations, such as
% the strong mixing of acoustic and polar modes,
the Larkin-Pikin \cite{Larkin1969,Larkin1969a} instability, as these effects are unrelated to the coupling to electrons.
Instead, our focus will be on the impact of externally applied uniform strain on the QC behavior
% impact that the strain has on the quantum critical behavior
of the coupled system.
% and furthermore limit ourselves to uniform strains.

The coupling to strain
% is nonlinear due to symmetry constraints, but
can
% still
be treated within a continuum theory. It is convenient to start from a 3D isotropic crystal, which is characterized by two elastic constants, namely, the bulk modulus and the shear modulus. Strain is defined as $\varepsilon_{ij} = (\partial_i \tilde{u}_j + \partial_j \tilde{u}_i)/2$, where $\tilde{\mathbf{u}}$ is the lattice displacement vector (not to be confused with the displacement $\mathbf{u}$ associated with the polar mode). The coupling between the FE order parameter and the strain tensor in an isotropic 3D crystal is then given by:
\begin{equation}
  \label{eq:L-strain}
  \mathcal{L}_{\ve,3D} = u_i(\q)D_0^{-1}\left[\lambda_0\ve_0\delta_{ij} + \lambda_1\left(\ve_{ij}-\frac{\ve_0}{3}\delta_{ij}\right)\right]u_j(-\q).
\end{equation}
Here, $\ve_0 = \Trc\ve_{ij}$ is the symmetry-preserving longitudinal strain and $\ve_{ij}-\frac{\ve_0}{3}\delta_{ij}$ is the $d$-wave rank-2 traceless tensor corresponding to symmetry-breaking shear strain. Note that $u_j$ is the polar mode before being projected onto the transverse and longitudinal components. Moreover, $\lambda_0$ and $\lambda_1$ are coupling constants. 

To go to the 2D limit, we need to establish the boundary conditions for the out of plane direction $\hz$. One possibility is a thin film that is clamped on one face
% \vk{ (Is it called ``face'' or ``side''?)}
% AK face
and free on the other. In that case, the strain tensor has  nonzero  $z$ components  that can be replaced by their value averaged over the $z$ direction, e.g. $\langle\ve_{zx}\rangle = \langle\ve_{yz}\rangle = 0, \langle \ve_{zz}\rangle \neq 0$, assuming no shear stresses are present.
% \vk{ (What exactly is meant here? Averaging over what?)}
 Another possibility is to have both faces clamped, such that we can set $\langle\ve_{zi}\rangle = 0$. For simplicity, we will assume the latter scenario. Upon taking the 2D limit, the $d$-wave tensor splits into a symmetry-preserving $r^2-3z^2$ term and a symmetry-breaking $xy$, $x^2-y^2$ doublet. Similarly, we expect $\lambda_1$ to split in two terms, $\lambda_{1z}$ and $\lambda_{1t}$. Then, decomposing
% going through the various steps \vk{ (What steps? Either specify or make a reference to an appendix.)}, decomposing
$u_j$ into its components, and 
%dropping the longitudinal part, 
projecting on the transverse sector, we get,
\begin{align}
  \label{eq:L-strain-2D}
  \mathcal{L}_{\ve,2D} &= \eta_i(\q)D_0^{-1}\left[\left(\lambda_{0t}\ve_0  +\frac{\lambda_{1t}}{2} \hq_t\cdot\hat{\ve}_{d}\cdot\hq_t \right)\hq_t\hq_t
                         \right.\nn\\
                       &\spcr\spcr \left. +\lambda_{0z}\ve_0\hz\hz
                        \right]_{ij}\eta_j(-\q),
\end{align}
where the 2D ``$d$-wave'' strain tensor is $\hat{\ve}_{d} = \left((\hat x\hat x-\hat y\hat y) (\ve_{xx}-\ve_{yy})+2(\hat x\hat y +\hat y\hat x) \ve_{xy}\right)$. Clearly, longitudinal compressive or tensile strain $\ve_0 \neq 0$ shifts the gaps of both modes but by different amounts, which allows for external control of the phases. On the other hand, shear strain only affects the $t$ mode.

To see these effects, we analyze a simple case where none of the couplings in 2D are modified from their 3D values. In that case we find $\lambda_{0t} = \lambda_0 + \lambda_1/6$, $\lambda_{0z} = \lambda_0-\lambda_1/3$, $\lambda_{1t} = \lambda_1$.
Symmetry-preserving volume-changing strain, which corresponds to
% $\ve_0 \neq 0, \ve_d = 0$ \vk{ (Does it imply
  $\ve_0 = \ve_{xx}+\ve_{yy} \neq 0$, $\ve_{xx} - \ve_{yy} = \ve_{xy} = 0$,
  % $\hat \ve_d$ is defined as a tensor.)}
  shifts the phonon mode gaps such that
\begin{align}
  \label{eq:strain-shifts}
  \overline{r} &\to \overline{r} +\ve_0(\lambda_0 - \lambda_1 /
  12)\nn\\
  \Delta r &\to \Delta r - \ve_0\lambda_1 / 2,
\end{align}
where $2\overline{r} = r_z+r_t$, $\Delta r = r_z-r_t$.
% We see right away, that applying a compressive strain (positive or negative) shifts the masses of both modes. Most importantly, they are generally shifted by different amounts. To see this clearly,
The effect is most pronounced when
% let's assume that $\lambda_i$ are not significantly changed by creating a thin film and that $|\lambda_1|\gg |\lambda_0|$ and that \
$|\lambda_1|\gg |\lambda_0|$, in which case the $t$ mode is favored by compressive strain and the $z$ mode, by tensile strain, assuming $\lambda_1 > 0$ (for $\lambda_1 < 0$, the role of tensile and compressive strains switch). For concreteness, we consider $\lambda_1 \gg \lambda_0 >0$ and the situation where the mass terms obey $r_z \ll r_t \approx \gb\nu_F/2$, as it would be the case if the splitting between the modes was dominated by the effects of the electronic renormalization.
% \vk{ (Isn't this a fine-tuning of two independent parameters ($r_z$ and $r_t$)?)}
% AK It is, but we already discussed this extensively in the previous sections
Applying compressive $\ve_0<0$, the transition for the $t$ mode  is triggered when $\lambda_1|\ve_0|/6 \approx \gb\nu_F/2$.
% \vk{ (I keep getting confused. Are $r_t$ and $r_z$ bare masses or those renormalized by the polarization operator?)}.
For this strain value, the $z$ mass will be $r_z \approx \lambda_1|\ve_0|/3 \approx \gb\nu_F$. This means that $r_z$ is still small in the sense discussed in the previous section, resulting in a first-order transition due to QOBD mechanism promoted by the soft fluctuations associated with the $z$ mode. Upon further increasing $|\ve_0|$, the $z$ mode fluctuations are no longer soft, and the $t$ mode transition becomes second-order. In the opposite limit, $\lambda_0 \gg |\lambda_1|$, volume-changing symmetry-preserving strain $\ve_0$ can drive the system in and out of the FE state i.e. change the sign of $\bar{r}$), but will not in general affect the hierarchy of the phonon gaps (i.e. the sign of $\Delta r$). 
% \vk{ (Where does this conclusion come from? Do we neglect QOBD of the $t$ mode because it's numerically smaller? Also, what's the role of $\lambda_{1t}$ term?)}
% AK It is just the conclusion of the previous section - we don't need a true Goldstone mode, just a small enough gap.
%and the prefactors are the same as those found in Sec.~\ref{sec:quant-order-disord}, up to factors of order 1.

If, instead, we apply a volume-preserving symmetry-breaking strain, e.g. $\ve_{xx}=-\ve_{yy} = \ve_d/2$, $\ve_{xy}=0$, the $t$ mode acquires a preferred direction and is always favored over the $z$ mode, whose gap remains unchanged. To see this, we rewrite $\hat\ve_d$ in the $\hq$, $\hq_t$ basis to get
\begin{equation}
  \label{eq:ve-2d-pbasis}
  \hat\ve_{d} = \ve_{d}\cos2\theta_q\left(\hq\hq - \hq_t\hq_t \right)
\end{equation}
%where $(\cdots)$ represent longitudinal components. 
The implication is that the $t$ mode has a directional mass,
\begin{equation}
  \label{eq:rt-strain}
  r_t^\ve = r_t - \frac{\lambda_{1t}\ve_d}{2}\cos 2\theta_q.
\end{equation}
The system will condense  % \vk{ (Do you mean $\lambda_{1t}$?)}
% AK I am using a specific choice where \lambda_{1t} = \lambda_1, see above.
in the $\hq = \pm \hat x$ or $\hq = \pm \hat y$
% , \hat p_t = \pm \hat y$
configuration when $|\lambda_1\ve_d| = 2r_t$,
% \vk{ ($\hat p$ and $\hat p_t$ are not indipendent in that sense. The only thing that condenses is the polarization vector.)}
 where the choice of axis depends on the sign of $\lambda_{1t}\ve_d$.
% \vk{ (It seems that if $\lambda_{1t}\ve_d$ is negative then $\hat p = \pm \hat y$.)}.
% AK correct
Interestingly, the tendency to QOBD is not modified by the volume-preserving symmetry-breaking strain, since
% Now, we may ask what happens to the first order transition, which naively we would expect to be suppressed now that the t mode is constrained. However, notice that
the fermionic polarization, which drives the QOBD, is independent of the dynamics of $\P$ and only cares about the direction of the static polarization. Furthermore, as we discussed in Sec.~\ref{sec:quant-order-disord}, the instability of the $t$ mode to QOBD is driven by fluctuations of the $z$ mode, whose dynamics are not affected by the volume preserving symmetry-breaking strain.
% and so the strain does not the unstable term is driven by the soft $z$ mode, not by the $t$ mode \vk{ (Why? Do we neglect QOBD by the $t$ mode completely?)}.
%Thus, the nonanalytic terms we calculated in Sec.~\ref{sec:quant-order-disord} are not modified by the volume-preserving strain.
% instability criterion remains unchanged \vk{ (What's the criterion?)}.

When both $\ve_0$ and $\ve_d$ are present and can be independently controlled, it is possible to drive \emph{both} the $z$ and $t$ modes away from criticality, and then compensate by an appropriate $\ve_d$ to tune the $t$ mode to the critical point while keeping the other one non-critical. 
  %\vk{ (Isn't this whole section about such a scenario? Do we really study the case when both modes are nearly critical? My impression was that we always consider them independently.)}
  % The whole QOBD section is when they ar of the same order
% bring only since $r_t^\ve$ may be driven to the QCP independent of $r_z$, the implication is that we may tune the combined system \emph{away} from criticality, keeping $r_z,r_t \gg \gb \nu_F$, and then drive
In this case, the system is driven through a second-order transition, since when $|r_z - r_t| \gg \gb\nu_F$ the QOBD tendencies are suppressed.
% \vk{ (How do we know it's second-order? And why is it second-order? What about QOBD by the $t$-mode?)}
% The reason is that the nonanalytic terms are strongly suppressed for $\gb/\nu_F/r_i \ll 1$ as can be see by expanding Eqs. \eqref{eq:f-oop-1} and \eqref{eq:f-ip-1} to leading order in $\gb/\nu_F/r_i$. 
We present the schematic phase diagram in the $\ve_0$, $\ve_d$ plane in Fig.~\ref{fig:strain1}, for a situation where $\lambda_{0z},\lambda_{0t}$ and $\lambda_{1t}$ are all non-negligible.
% The schematic phase diagram for this type of behavior is presented in Fig. .

Finally, let us discuss the most common case of in-plane uniaxial strain, parametrized by $\ve_{xx} = \ve_{uni}$, $\ve_{yy} = -\nu \ve_{uni}$ where $\nu$ is the Poisson ratio, which is typically $\nu<0.5$. In that case, both $\ve_0$, $\ve_d$ are nonzero, and changing $\ve_{uni}$ basically traces a straight line through the phase diagram of Fig. \ref{fig:strain1}.

Before completing this section, let us briefly comment on the effect of including a finite lattice anisotropy in the calculation. The lattice anisotropy breaks down the rotationally invariant propagators and the shear strain term $\hat{\ve}_d$ to representations of the discrete $C_4$ rotations of the lattice (in the case of a square lattice). The main effect on the phonon propators is the introduction of diagonal anisotropic terms of the form
\begin{equation}
  \label{eq:anisotropy-props}
  \mathcal{L}_l = \sum_{i} u_i(\q) (D_0k_a^2)^{-1} (c_i^2-c^2)/c^2q_i^2 u_i(-\q),
\end{equation}
where $ c_i$ are modifications of the phonon velocities around the major axes (clearly $c_x = c_y$ for a square lattice). These do not qualitatively change our preceding results, as the bare momentum dependence of the phonons is negligible in our treatment of the QOBD phases, as are non-Fermi liquid effects.
% \vk{ (Is it? What about one-loop level?)}.
Of course, if the anisotropic terms are much larger than the dynamical contribution of the electrons, the correlation effects will be suppressed. From a technical viewpoint, this will take place when the angular integration in the various terms is strongly suppressed by the anisotropy. %The main effect on the coupling to strain is that the $d$-wave representation of the volume-preserving strain breaks into two 1D representations ($B_{1g},B_{2g}$), but since we did not actually utilize rotational invariance to derive the results in this section, we do not expect a strong impact due to the change in representation structure.
% \vk{ How about the effect on fermion propagator?}
% AK Not very big as long as there are no cold spots

\section{Theory of an isotropic 3D {\QFM}}
\label{sec:theory-3d-qfm}

The three-dimensional case differs from the 2D case in two ways. First, in the disordered phase there is no preferred direction, such as $\hz$, which we picked for the out-of-plane direction in our treatment of the 2D system. As a result, there is no $z,t$ splitting in the disordered phase and the response of the system is qualitatively the same as that of a ferromagnet, i.e. the bosonic response is Landau overdamped. Once an ordered state sets in with the polarization, e.g. in the $\hz$ direction again, it itself provides a preferred direction, splitting the response into $z$, $t$ modes.
%, which are now nothing but the longitudinal and transverse sectors in the ordered state \vk{(why does $z/t$ splitting has smth to do with longitudinal/transverse modes? Do you mean longitudinal/transverse with respect to the ordering polarization vector or with respect to $\bf q$?)}\AK{(It's always with respect to polarization.)}. 
The second difference from 2D is the usual weakness of 3D QC fluctuations compared to 2D ones. This gives rise to logarithmic rather than algebraic divergencies, e.g. leading to marginal FL rather than nFL behavior near the QCP \cite{Holstein1973,Son1999,Chubukov2005}.

Except for these issues, the qualitative behavior in 3D is essentially the same as in 2D. The transverse $t$ mode introduces the splitting $\Delta$ as an IR cutoff, giving rise to QOBD. In terms of pairing, the QC fluctuations give rise to enhanced $T_c$. However, all the effects are much weaker than in 2D.

We now summarize the main results for the 3D case. Since the behavior is similar as in 2D, and only the algebra is more complex, we leave all calculation details to the Appendix \ref{sec:deta-calc-sec-iv}.
%\vk{(specify appendix)}.

\subsection{The disordered phase}
\label{sec:disordered-phase}

In the disordered phase, the polarization after projection onto the transverse sector is
\begin{align}
  \label{eq:pi-disordered-3d}
  \hat \Pi_0 (q) = \Pi_0(q)\hat{\mathcal{P}}(\hq)
\end{align} 
where
\begin{equation}
  \label{eq:Pi0-def}
  \Pi_0(q) = \delta r - \delta\Pi(q_0/\vf |\q|), 
\end{equation}
and
\begin{align}
  \label{eq:pi-dis-details-3d}
  \delta r &= \frac{2}{3}\gb \nu_F, \nn\\
  \delta\Pi(x) &= \frac{1}{2}\gb \nu_F x f(x).
\end{align}
 Here $\nu_F = k_F^2/\pi^2 v_F k_a^3$ is the total 3D DOS at the Fermi level, and 
\begin{equation}
  \label{eq:lind-3d}
f(x) =  x - (x^2-1) \arctan(1/x).
\end{equation}
It follows that at the lowest frequencies, $\hat \Pi$ has a Landau overdamped form, $\delta\Pi \sim |q_0|/\vf|\q|$, which is standard for $|\q|=0$ QCPs. Similarly, the electronic self-energy at the QCP is that of a marginal FL
\begin{equation}
  \label{eq:sigma-disordered-3D}
  % \Sg(k) \approx \gb\nu_F\frac{\kf^3}{6 k_a^3}k_0 \log(\vf \Lambda/k_0),
  \Sg(k_0) \approx -i\gb\nu_F\frac{k_a^2}{12 \kf^2}k_0 \log\left|\frac{\omega_\Lambda}{ k_0}\right|,
\end{equation}
where
% \begin{equation}
$\omega_\Lambda = \frac{ v_F \Lambda^3}{\gb\nu_F k_a^2}$ 
% \end{equation}
% where
% \vk{
% \begin{equation}
% \omega_0 \equiv 4 v_F \Lambda^3 / \pi \bar g \nu_F k_a^2.
% \end{equation}}
and $\Lambda$ is an UV momentum cutoff. Since the logarithmic divergence is rather weak, we will neglect it when computing the order-by-disorder mechanism.

% \vk{
% \begin{equation}
%   %\label{eq:sigma-disordered-3D}
%   \Sg(k_0) \approx \gb\nu_F\frac{\kf^2}{12 k_a^2}k_0 \log(\omega_0/k_0),
% \end{equation}}

Before proceeding to discuss pairing and QOBD in 3D systems, it will be convenient to construct a vector basis to decompose the interaction, similar to what we did in two dimensions, see Eqs.~\eqref{eq:trio-def}-\eqref{eq:ff-t-2d}. 
We will construct this basis in a way that is convenient not just in the disordered state, but also in the presence of FE order, which chooses a preferred direction. To
account for fluctuations in the ordered state,
%presence of FE order, 
we assume a FE order parameter polarized along the $\hz$ direction,
\begin{equation}
  \label{eq:Delta-def-3D}
  \bv{\Delta} = \hz \Delta.
\end{equation}
%As discussed above, the order parameter polarization breaks the rotational symmetry. We account for this by 
Accordingly, we define for every vector $\k$ a right-angle trio 
\begin{equation}
  \label{eq:right-trio-3D}
  \hk_u, \,\,\, \hk_t, \,\,\, \hk,
\end{equation}
where
\begin{equation}
  \label{eq:hk-t-u-def}
   \hk_t = \frac{\hz\times\hk}{|\hz\times\hk|}, \qquad \hk_u = \hk_t \times \hk.
\end{equation}

\begin{table}[tp]
    \caption{Table of the first three irreducible representations of a rotationally invariant model with inversion symmetry in three spatial dimensions. The last column denotes the sign of the interaction for each representation compared to the $s$-wave attraction. Thus, negative sign implies a repulsive interaction and zero implies a marginal interaction.}
    \label{tab:SC_decomp-3D}
    \begin{tabular}{| l | c |  c |c |c |}
    \hline
    irrep  &  Matrix form $F_{nj}(\hat \k)$  & Inv. symmetry  & Int. sign   \\ \hline
     $n=0$  &   $1$  &   even  &  + \\ \hline
     $n=1$  &  $ \hk \cdot \boldsymbol \sigma $&  odd & -- \\ \hline
     $n=2 $ &  $\hk \times \bvs$  &  odd  & 0 \\ \hline
    \end{tabular}
\end {table}

Here $\hk_t$ has
\begin{figure}
  \centering
  \includegraphics[width=\hsize,clip,trim=0 100 0 100]
  {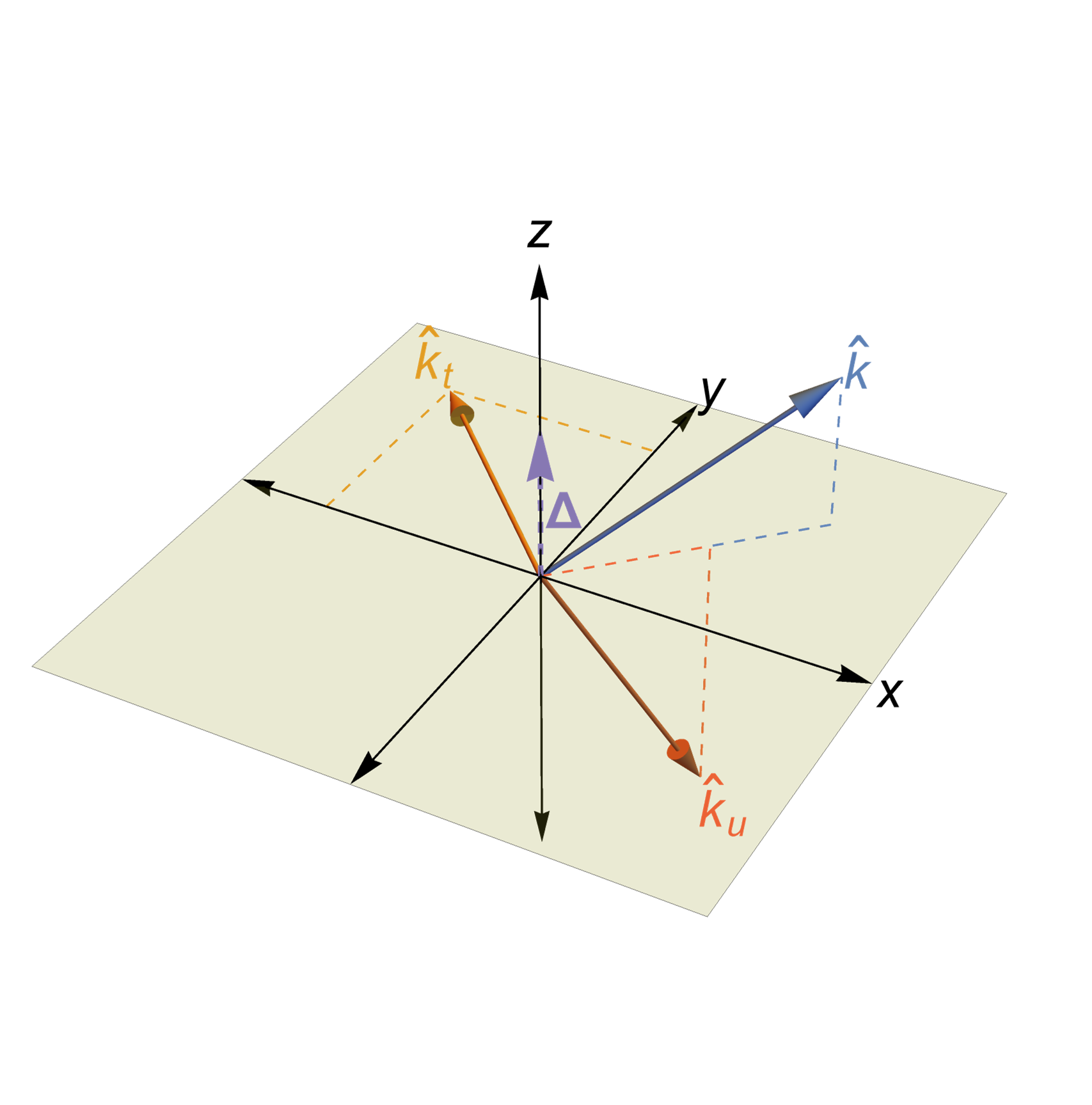}
  \caption{
  %\vk{ (Add $z$-axis.)} 
  The basis (Eq. \eqref{eq:right-trio-3D}) for describing the transverse fluctuations in the 3D system. A given vector $\hk$ (blue) has two transverse vectors: $\hk_t$ in the $xy$ plane  (yellow), and $\hk_u$ , whose projection on the $xy$ plane is parallel to the projection of $\hk$ itself, with a length $\cos\theta_k$, where $\theta_k$ is the polar angle of $\hk$ (orange). The polarization vector of the FE order parameter (dashed, purple) is aligned along the $z$ axis, such that the amplitude of a transverse fluctuation mode is proportional to its projection on the plane perpendicular to the polarization, i.e. the $xy$ plane.}
  \label{fig:basis-3D}
\end{figure}
been chosen to be in the $xy$ plane that is perpendicular to the polarization vector, and $\hk_u$ has the same projection on the $xy$ plane as $\hk$ itself (see Fig.~\ref{fig:basis-3D}). The interaction form factor is now
\begin{equation}
  \label{eq:k-cross-decomp-3d}
  \bv{\gamma}(\hat k) = \hat k \times \bvs = \sigma_u \hk_t - \sigma_t \hk_u,
\end{equation}
where $\sigma_t = \hk_t\cdot\bvs$, $\sigma_u=\hk_u\cdot \bvs$.

\subsection{Pairing in 3D {\QFM}s}
\label{sec:pairing-3d}

Next, we discuss the pairing in 3D. As was done in the 2D case, we focus on the disordered state and neglect the weak effect of the possible first-order transition. The pairing equation has the same structure as in 2D, Eq.~\eqref{eq:pairing-eq-1}.
%, with the pairing vertex given again by Eq. \eqref{eq:pairing-function}. 
In principle, the 3D pairing problem near a QCP is quite different from the 2D case. The reason is that in 2D the pairing interaction mediated by the critical mode is peaked at small momentum transfer, giving rise to an effectively local (in momentum space) pairing potential, stemming from the effective 1D regime for fluctuations transverse to the FS. In 3D, because of the extra dimension, the mode is not really limited to small momentum transfer and is only cut off logarithmically at the UV. Nevertheless, previous work \cite{Chubukov2005} has shown that to a leading approximation the interaction is still effectively local in momentum space, because of the limitation that electrons scatter parallel to the FS. Such arguments can be made rigorous by a large-$N$ expansion for the fermions. Here, we will just assume the validity of the local interaction and proceed.

To identify the relevant pairing channel, we need to compute the summation over spin and momentum indices in the gap equation
\begin{widetext}
  
\begin{equation}
  \label{eq:pairing-eq-1-3D}
  \Phi_{\alpha\beta}(k) \approx \frac{\gb T}{D_0} \sum_{q} G(-k-q/2)G(k+q/2)D(q)\mathcal{P}^{ij}(\hq)\gamma_{\delta\alpha}^i(-\hk)\Phi_{\delta\mu}(k)\gamma_{\mu\beta}^j(\hk).
\end{equation}
\end{widetext}
The equation is the same as Eq. \eqref{eq:pairing-eq-1}, up to a shift in momentum for convenience and an approximation $|\q|\ll |\k|$.
Following the analysis in 2D, we first write the gap function as a sum over irreducible representations
\begin{align}
\label{eq:rep-phi-3d}
\Phi(k) = i\sigma^y\sum_{nj} \phi_{nj}(k_0) F_{n}^j(\hat \k)\,,
\end{align}
and $T_c$ is set by the representation that develops a non-trivial solution at the highest temperature. Equation~\eqref{eq:rep-phi-3d} is again identical in form to Eq. \eqref{eq:rep-phi-2d} from 2D, but the $F_n^j$ should be understood as the representations in 3D (see Table \ref{tab:SC_decomp-3D}). By restricting the momentum transfer $\q$ to the plane that is tangent to the FS and keeping the leading order in $q/k_F$ we obtain that to leading order only the singlet channel ($n=0$) is 
% finite. 
unstable to pairing.
%All other channels cancel out and do not contribute. 
This result should be contrasted with the situation far from criticality, where standard BCS theory can be employed~\cite{Kozii2015,Wang2016}. In that case odd-parity fluctuations induce relatively strong pairing in some non-$s$-wave odd-parity channels.

To understand this result we can simplify the expression for $\Phi$ in Eq.~\eqref{eq:pairing-eq-1-3D}, by recalling that the momentum $\q$ can be factored into components parallel and perpendicular to the FS, 
%just like for the normal state calculations, such that the perpendicular component is small 
and that the pairing fluctuations are largest when $\hq$ is approximately parallel to the FS. Since $\hk_u$ and $\hk_t$ form a basis to this parallel plane, we can approximately expand $\hq$ as
\begin{equation}
  \label{eq:q-perp}
  \hq \approx \cos\phi_q\hk_u+\sin\phi_q\hk_t, 
\end{equation}
from which we immediately obtain,
\begin{widetext}
\begin{equation}
  \label{eq:Phi-3D}
  \mathcal{P}^{ij}(\hq)\gamma_{\delta\alpha}^i(-\hk)\Phi_{\delta\mu}(p_0,\k)\gamma_{\mu\beta}^j(\hk) = -(\cos\phi_q \sigma_t^T - \sin\phi_q\sigma_u^T)\Phi(\cos\phi_q \sigma_t - \sin\phi_q\sigma_u)\,,
\end{equation}
where $(\cdot)^T$ denotes a transpose and we dropped the spin index summation. After integrating over $\phi_q$ only the diagonal terms remain, i.e.
\begin{equation}
  \label{eq:Phi-tilde-3d-2}
 \mathcal{P}^{ij}(\hq)\gamma_{\delta\alpha}^i(-\hk)\Phi_{\delta\mu}(p_0,\k)\gamma_{\mu\beta}^j(\hk)  \to -\frac{1}{2}(\sigma_u^T\Phi\sigma_u+\sigma_t^T\Phi\sigma_t) = \frac{1}{2}i\sigma_y \sigma_t\left[\sum_{nj} \phi_{nj} F_n^j(\hat \k)\right] \sigma_t + (t \leftrightarrow u),
\end{equation}

\iffalse
\begin{align}
  \label{eq:Phi-tilde-3d-2}
 \mathcal{P}^{ij}(\hq)\gamma_{\delta\alpha}^i(-\hk)\Phi_{\delta\mu}(p_0,\k)\gamma_{\mu\beta}^j(\hk)  &\to -\frac{1}{2}(\sigma_u^T\Phi\sigma_u+\sigma_t^T\Phi\sigma_t)\nn\\
  &= \frac{1}{2}i\sigma_y \sigma_t\left[\sum_{nj} \phi_{nj} F_n^j(\hat \k)\right] \sigma_t + (t \leftrightarrow u),
\end{align}
\fi
\end{widetext}
% where $(t\leftrightarrow u)$ denotes replacing instances of $\sigma_t$ with $\sigma_u$.
where we used the fact that $\sigma_{j}^T i\sigma_y = -i\sigma_y\sigma_j$.
%\vk{(what's $\sigma_I$?)}.
% (\sigma_{i})_{\beta\alpha} i(\sigma_{y})_{\beta\gamma} = -i (\sigma_{y})_{\alpha\beta} (\sigma_{i})_{\beta\gamma}$.
% , allowing us to rewrite Eq. \eqref{eq:Phi-tilde-3d} as,
% \begin{align}
%  
%   \tilde\Phi = +i\sigma_y \sigma_t(f + \bv{d} \cdot \bvs) \sigma_t + (t \leftrightarrow u).
% \end{align}
Let us perform the spin matrix products explicitly. Expanding $\bvs$ in our basis we find
\begin{align}
  \label{eq:spin-sums-3d}
  \sigma_t \bvs \sigma_t &= \sigma_t \hk_t - \sigma_u \hk_u - \sigma_k \hk, \\
  \sigma_u \bvs \sigma_u &= -\sigma_t \hk_t + \sigma_u \hk_u - \sigma_k \hk, \\
  \sigma_{t,u} \sigma_0 \sigma_{t,u} &= \sigma_0.
\end{align}
Plugging them back into Eq.~\eqref{eq:Phi-tilde-3d-2} and keeping only the lowest order in $k$ representations $n=0$, 1, and 2 (corresponding to the scalar, pseudoscalar, and vector) we obtain
\begin{equation}
  \label{eq:Phi-tilde-3d-3}
  \Phi = + 2i\sigma_y \left[\phi_0 F_0 - \phi_1 F_1 (\hk)\right].
  % -\phi_{2j}F_2^j(\hat \k)\right].
  % AK I thought we agreed the F2 term is absent?
\end{equation}
Thus, as discussed above, \emph{only} the singlet ($n=0$) channel is attractive, and the triplet $n=2$ channel is \emph{marginal} - neither attractive nor repulsive, as evidenced by its absence from Eq.~\eqref{eq:Phi-tilde-3d-3}. The intuition for this is that the effect of the $u$, $t$ modes on the triplet pairing channels are opposite: each promotes pairing in its own spin polarization, and suppresses pairing in the other polarizations, due to the ``double-repulsive'' nature of the pairing interaction, as we discussed for the 2D case.

It is fairly straightforward to estimate $T_c$ for the 3D {\QFM}.
% Considering for simplicity only the spin-singlet state in the unstrained system, and plugging
Plugging Eq.~\eqref{eq:spin-sums-3d} back into Eq.~\eqref{eq:pairing-eq-1}, we see that the pairing equation just describes an isotropic system coupled to a 3D QCP via an isotropic interaction. We can therefore just take the result from the literature \cite{Chubukov2005},
\begin{equation}
  \label{eq:Tc-3D}
  T_c \approx \w_\Lambda e^{-\pi^2\sqrt{3\frac{\vf k_a}{\gb}}}
\end{equation}
where $\w_\Lambda$ was defined after Eq.~\eqref{eq:sigma-disordered-3D}.
%and $g_0 = 12\pi^2$.
%$\w_0 = \vf k_a \Lambda/(\gb\nu_F)^{1/3}, g_0 = \frac{9\pi^3}{16}\frac{k_F^2}{k_a^2}$. 
Here we have the well-known BCS-like result, but with a dependence on $\sqrt{\gb/(k_a\vf)}$ rather than $\gb/(k_a\vf)$ resulting from the logarithmic divergence of the pairing interaction (see Appendix~\ref{sec:deta-calc-sec-iv}). Interestingly, we also found a square-root-BCS $T_c$ for the $t$ mode in the $2D$ system. However, the physical mechanism there was completely different and arose from the singular strength of the bosonic interaction with no Cooper instability. For finite $r$ we recover a BCS-like transition temperature, $T_c \approx \w_r \exp(-\vf k_a/g_r)$, where $g_r = \gb \log(\Lambda^2/(k_a^2 r))/(8\pi^2)$ and  $\w_r=\mbox{min}\left(\frac{c k_a r^{1/2}}{2\pi}, \frac{2 r^{3/2}\vf k_a}{\pi^2 \gb\nu_F}\right)$ is a UV cutoff set either by Landau damping or by the bare bosonic speed of sound.

\subsection{Quantum Order by Disorder in 3D}
\label{sec:quant-order-disord-1}

Next, we consider whether a first-order transition or FDW phase preempt the second-order FE transition. Without loss of generality, we take the FE polarization in the ordered state to be along the $z$ axis, see Eq.~\eqref{eq:Delta-def-3D}.

We compute the polarization in the presence of FE order. It has the form
\begin{align}
  \label{eq:Pi-Delta-3D}
  \hat \Pi_\Delta (q) &= \left(\Pi_t(q,\Delta)\right)\hq_t\hq_t \nn\\
  &+\left(\sin^2\theta_q\Pi_0(q) + \cos^2\theta_q \Pi_u(q,\Delta)\right)\hq_u\hq_u,
\end{align}
where $\theta_q$ is the polar angle of $\hq$. $S_t$ and $S_u$ have a complicated form (see Appendix \ref{sec:deta-calc-sec-iv}).
%\vk{(specify appendix)}, h
However, as long as the polar angle is not too big, they can be approximated by,
\begin{align}
  \label{eq:Su-St-approx}
  \Pi_u &\approx \Pi_t \approx \frac{1}{2}\delta r - \frac{1}{2}\gb\nu_F s_0\left(\frac{q_0+2i\tilde{\Delta}}{\vf|\q|}\right) + \cc
  % - s_0\left(\frac{q_0}{\vf|\q|}\right) \right]
          ,\nn\\
          &\qquad
              \qquad\tilde\Delta = \sqrt{2/3}\Delta,
\end{align}
where $s_0(x) = \arctan(1/x)$.
%\vk{(what's $s_0$?)} 
The meaning of Eqs.~\eqref{eq:Pi-Delta-3D} and~\eqref{eq:Su-St-approx} is as follows:  $\Pi_t$ and $\Pi_u$ encode the transverse response, which comes from processes with $\hq$ perpendicular to the ordering vector, i.e. in the $xy$ plane. By construction $\hq_t$ is in the $xy$  plane, but the projection of $\hq_u$ on the $xy$  plane is $|\hq|\cos\theta_q$ (see Fig. \ref{fig:basis-3D}), which is the source of the $\cos^2\theta_q$ prefactor to $\Pi_u$ in Eq.~\eqref{eq:Pi-Delta-3D}. 
%The $\sqrt{2/3}=\frac{1}{2}\int \sin^3\theta d\theta$ factor is just the result of angular averaging of the Zeeman splitting for a fluctuation vector over the polar angles.

From the above discussion, it is already clear that there will be a QOBD effect, since $\Pi_t$ and $\Pi_u$ are nonanalytic functions and $\Delta$ is an infrared cutoff. Performing the calculations numerically for the exact $\Pi_t$ and $\Pi_u$ we find
\begin{equation}
  \label{eq:deltaF-3D}
  \delta F(\Delta) \approx - 0.19 \frac{\Delta^4}{\vf^3k_a^3}\log\left|\frac{\vf \Lambda}{\Delta}\right|,
\end{equation}
where $\Lambda$ is a UV momentum cutoff. Up to a numerical prefactor, this is the same result one finds for QOBD in 3D ferromagnets. Since the 3D system is isotropic, there is no splitting between the $t$ and $u$ sectors. Thus, the free energy has the form
\begin{equation}
  \label{eq:F-3d-1st}
  F_{\Delta} = \frac{1}{\gb}\left[r \Delta^2 +\left(- 1.88 \frac{\gb\nu_F}{\kf^2\vf^2}\log\left|\frac{\vf \Lambda}{\Delta}\right| + \frac{u_jD_0}{2\gb}\right)\Delta^4\right].
\end{equation}
A similar logarithmic dependence is found for the FDW (finite-$q$) transition, whose expressions we omit for simplicity.

\subsection{The phase diagram in 3D and coupling to strain}
\label{sec:strain-effects}

Based on the results of the preceding sections, 
the phase diagram of the 3D system is similar to the 2D one, consisting of a normal paraelectric and ferroelectric state, and a superconducting dome. Moreover, the second-order FE transition may be preempted by a first-order one, or by a transition to a finite-$q$ FDW state, which for simplicity we ignore like we did in 2D (see the discussion in Sec.~\ref{sec:phase-diagram-finite-T}).

The free energy for the 3D case was given in Eq. \eqref{eq:F-3d-1st}. As the discussion above has made clear, as far as the normal state and QOBD go, the {\QFM} in 3D is very similar to its ferromagnetic counterpart. In particular, the first-order transition is extremely weak, characterized by a jump $\Delta^* \sim e^{-a_{3D}/\gb^2\nu_F^2}$, where $a_{3D}$ is some constant. Moreover, as is known for the ferromagnetic case, the nFL state near the QCP should further weaken the nonanalyticity that gives rise to the QOBD ~\cite{Maslov2009}. Hence, the SC phase, which at the critical point has a $T_c$
%a maximum 
at a much higher temperatures than the energy scale set by $\Delta^*$, see Eq.~\eqref{eq:Tc-3D}, should rise above the first-order transition. This implies that the phase diagram in 3D is somewhat similar to the case of the 2D  $z$ mode, as shown in Fig. \ref{fig:phaseDiag2d}a, with a far narrower first-order region.

Next, we consider the effect of external strain. In the 3D system, uniaxial strain along one of the coordinate axes splits the $t$ and $u$ modes. As a result, the transition will remain second-order as QOBD is frozen out. To see this, it is enough to consider the  expression for strain in an isotropic medium, Eq. \eqref{eq:L-strain}. It is readily checked that strain $\ve_0$ shifts the mass of both $t$ and $u$ modes by a constant shift $\lambda_0\ve_0$. In the case of uniaxial strain that is volume-preserving, we have:
\begin{equation}
  \label{eq:uni-strain-3d}
  \ve_{uni} = \ve_{xx} = \ve_{yy} = -2\ve_{zz}, \qquad \ve_{i\neq j} = 0.
\end{equation}
Then, the strain contribution to the action, after removing purely longitudinal components,  is
\begin{align}
  \label{eq:L-strain-uni}
 &\mathcal{L}_{\ve} = u_i(\q)D_0^{-1}\lambda_1\ve_{uni}\left(\hat{I}-3\hz\hz\right)_{ij}u_j(\q) \\
                    &\approx \eta_i(\q)D_0^{-1}\lambda_1\ve_{uni}\left(\hq_t\hq_t + (1-3\sin^2\theta_q)\hq_u\hq_u\right)_{ij}\eta_j(\q), \nn
\end{align}
where in the first line $\hat I$ is the identity matrix, and in the second line we dropped all longitudinal components.

The impact of the strain depends on the sign of $\lambda_1\ve_{uni}$. For $\lambda_1\ve_{uni} > 0$, the energy can be minimized by selecting $\theta_q = \pi/2$. Then, the $t$ mode is pushed away from the QCP but the $u$ mode is pushed towards it,
\begin{equation}
  \label{eq:delta-r-strain-3d-1}
  r_t = r + \lambda_1\ve_{uni}, \qquad r_u = r - 2\lambda_1\ve_{uni}.
\end{equation}
Since $\hq$ is in-the-plane, the polarization of the mode is out-of-plane $\hat{\eta} = \pm \hz$, such that the order parameter is Ising-like. If the strain-induced splitting is large enough, the system will be truly Ising-like and display a second-order transition, whereas if the splitting is small, QOBD can still render the transition weakly first-order.
On the other hand, for $\lambda_1\ve_{uni} < 0$ the energy is minimized when $\theta_q = 0,\pi$, in which case the two transverse modes remain degenerate and are both pushed towards the QCP. Thus, the system orders in an easy-plane XY-like configuration, $\hat{\eta} = \cos\phi\hat x + \sin\phi \hat y$, and QOBD still renders the transition weakly first-order.

Strain dramatically modifies the picture for pairing. As we saw in Sec. \ref{sec:pairing-3d}, only spin-singlet pairing is attractive in 3D at the critical point, and the spin-triplet pairing is neither attractive nor repulsive. In the presence of uniaxial strain, with $\lambda_1\ve_{uni} > 0$, the $t$ and $u$ modes split and only the $u$ mode remains relevant. Furthermore, it is softest for fluctuation wavevectors in the $xy$ plane, see Eq. \eqref{eq:L-strain-uni}. The strain breaks the isotropy of the FS, so solving the pairing equation for the entire FS is challenging. Fortunately, since the pairing attraction is effectively local in momentum space (i.e. it does not couple distant FS momenta), we can obtain information about $T_c$ just by considering the specific points on the FS where pairing is maximal, which turn out to be the points on the circle where the FS cuts the $xy$ plane. This is because $\hq$ should be in the $xy$ plane as discussed above, and also parallel to the FS as usual, which implies $\hq_u = -\hz$. Therefore, the projection of the interaction on the $u$ mode $\hq_u \cdot \hk \times \bvs= \hk_y\sigma_x-\hk_x\sigma_y$ is maximal if $\hk$ is also in the plane.
%These are the FS points with $\hk = \pm \hz$, since at those points the FE modes that contribute to pairing, i.e. with momenta that are parallel to the FS, are also in the plane. 
Going back to Eq.~\eqref{eq:spin-sums-3d} and neglecting the $\hq_t$ contributions, we find that now both spin-singlet and 
%$m=0$ 
a single spin-triplet mode are degenerate, with
\begin{align}
  \label{eq:Phi-tilde-split}
  \Phi 
  %= \tilde\Phi_u 
  &= i\sigma_y \left( \phi_0 F_0 + \phi_{2z}F^{2z}\left(\hk\right)\right).
  %\bv{d}\cdot \hk_u \sigma_u) \nn \\
  %&= i\sigma_y( f + \bv{d}\cdot \hz \sigma_z).
\end{align}
%We conclude, that by applying uniaxial strain the {\QFM} can be tuned from a spin-singlet to an almost degenerate singlet-triplet state.

\section{Discussion}
\label{sec:discussion}

%{\jr{***[The paragraphs in the conclusions are way too long. They need to be broken up in to smaller paragraphs, with a single point for each.]***}}

In this work, we constructed a theory of 
%the normal state of 
a ferroelectric metal in the vicinity of a QCP,
%\vk{(I wouldn't specify ``normal state'' since we said smath about the SC state as well.)}, 
starting from a minimal theory of a FE transverse polar phonon interacting with low-energy electrons via a dynamical Rashba spin-orbit coupling.
We found three properties that determine the qualitative behavior of the coupled system: the spin-charge mixing arising from the coupling term; the nonlinear coupling between FE modes and strain; and the geometric constraint imposed by the transverse nature of the phonon, the coupling term, and the prevalence of forward scattering for fermions on the FS. In particular, the spin-charge mixing gives rise to attraction in both spin-singlet and spin-triplet pairing channels
%rather than the more ``conventional'' scenario where a quantum critical metal is attractive to \emph{either} singlet or triplet pairing but not both. We 
and also
%found 
a tendency to QOBD.
%rather than charge fluctuations. 
Strain, on the other hand, acts as a convenient tuning parameter for the phase diagram. As for the geometric constraints, they have a profound impact on the phase diagram, giving rise to a qualitatively different behavior for 2D and 3D systems. 

In 2D, there are two distinct FE modes, the $z$ and $t$ modes corresponding to out-of-plane and in-plane  polarizations, respectively. While the former is Landau overdamped and creates a nFL, the latter remains underdamped even at the QCP, rendering the fermionic system a marginal FL. Both $z$ and $t$ modes are unstable to pairing in both spin-singlet and spin-triplet channels, which are degenerate to leading order. However, the pairing mediated by the $t$ mode is much weaker due to the geometric constraint. Finally, each mode by itself does \emph{not} give rise to QOBD. Instead, QOBD arises due to the interaction between the two modes when they are close in energy, and therefore will typically appear only in the for appropriate values of external strain. In contrast, in a 3D system, the two transverse modes are degenerate 
%in the absence of strain 
and QOBD is always present (albeit weakly), unless the externally applied strain is too strong. On the other hand,
%, again due to the geometric constraints, 
spin-triplet pairing only appears in the presence of strain in 3D, which however can be used to make the singlet and triplet pairing channels almost degenerate.
%to leading order.

Many properties of our theory rely on the splitting between the LO and TO polar modes that is a hallmark of 3D FE materials. However, in our analysis, we also considered 2D {\QFM}s, both because their behavior is more straightforward to determine and because conventional wisdom tells us that the effects of quantum fluctuations are stronger in reduced dimensions (as we found). We therefore need to comment on the relevance of our model to ``real'' 2D materials where there is no LO-TO gap. 

In 2D materials, the LO-TO gap is replaced by a square-root singularity in the dispersion, so that the inverse propagator for the LO mode is given by $D^{-1}_{LO} \propto q^2 + c^{-2}q_0^2 + (\w_T/c)^2 + q_{LO} q$, where $q_{LO} \approx 2\pi Q^2$ and $Q$ is the ionic charge per site \cite{Sohier2017}. The additional linear term in the inverse propagator is enough to render the LO mode irrelevant. The reason is that fluctuations are not confined to small momenta and are therefore weak at weak coupling, as can be seen from  a straightforward dimensional analysis. For example, even though the LO mode is Landau overdamped, by itself it does not cause the electrons to form a nFL.
%it remains a marginal FL and does not become a nFL like the $z$ mode 
Moreover, it can be verified that the prefactor for the self-energy is parametrically smaller than that generated by the $t$ mode, because the $t$ mode fluctuations \emph{are} dominated by the IR limit, and the marginal FL is a result of the lack of a FS singularity, as discussed in detail in Sec.~\ref{sec:self-energies-one}. Finally, we can formally take the limit $q_{LO}/k_a \gg 1$ as a control parameter and remove the effects of the LO mode entirely from the theory. We do note that  including the LO mode in 2D will modify the QOBD effect in the $t$ channel, since the dispersion of the phonons is irrelevant for the generation of the nonanalytic terms in the free energy. Thus, we expect QOBD to be somewhat stronger for the $t$ mode than predicted in our work (but still small).

In our work, we have not concentrated on specific material realizations, despite mentioning several different {\QFM} compounds. Indeed, many {\QFM} candidates have not been sufficiently well characterized for us to attempt a quantitative comparison of our theory with experiment. Rather, we expect our theory to provide insights into the search for and engineering of materials that display {\QFM} properties. As a specific example, we now briefly discuss the relevance of our results to the 3D quantum paraelectric SrTiO$_3$ (STO). 

Recent calculations have shown that the dynamic Rashba coupling in STO is significant, of the order of several meV. However,
it is not clear that this Rashba coupling can explain the normal state transport and SC properties of STO, at least at low doping levels \cite{Gastiasoro2021,yu2022theory}. It is similarly unsettled whether two-phonon processes may be a more important mechanism for SC, despite being formally irrelevant at the QCP \cite{kiseliov2021theory,Zyuzin2022}.  At the same time, strain significantly enhances the superconducting $T_c$ and also may promote a tendency to triplet pairing \cite{Ahadi2019,Schumann2020,Hameed2020}. This is consistent with our results. A rough estimate for the distance of STO from the QCP yields a typical wavevector  $|\q|a \sim \sqrt{r_0} = \w_Ta/c \sim 0.4$~\cite{Yamada1969,Courtens1993}.
%, which is not particularly small, especially considering the extremely small FS of STO at low doping. 
This should be contrasted with the typical momentum transfer for fluctuations near the QCP, which in 2D is of the order of $\gb\nu_F$ and, in 3D, is bounded from above by a cutoff of the order of $\kf a$. For a carrier density of $n\sim 10^{18}\mbox{cm}^{-3}$, below which a single band is occupied in SrTiO$_3$, Ref.~\onlinecite{Gastiasoro2021} estimated $\kf a \sim 0.15$ and $\gb \nu_F \sim 0.01$, suggesting that the electronic system should not evince QC behavior.
On the other hand, our theory predicts that strain drives the system closer to the QCP and that a tendency to triplet pairing exists only in 2D and in strained 3D systems. We cannot directly compare our theory to the experiments in Refs. \cite{Ahadi2019,Schumann2020,Hameed2020} as the strain in those experiments was enough to drive the material into the FE state. Naively, in the ordered state one expects FE to compete with SC, but this may not be  the case for the FE mode \cite{Kozii2021}. We expect our theory to be easier to compare with experiment in very thin films or heterostructures, provided that the issues of epitaxial strain and breaking of inversion symmetry by the substrate are avoided -- e.g. by appropriately capping the film. We note that the current theory does not apply to Dirac fermions at the charge neutrality point, which was studied in Refs. \cite{Kozii2019,Kozii2021}, but it does apply to doped Dirac systems provided the FS is well established (i.e. $E_F \gg T$).
Overall, our work establishes a solid framework to elucidate the fascinating  properties of {\QFM}s.

\begin{acknowledgments}
We thank A. V. Chubukov, D. M. Maslov, A. Kumar, P. Volkov, J. Schmalian, M. H. Christensen,  M. Feigel'man, A. Kundu, M. Navarro-Gastiasoro, D. Pelc and D. van der Marel for many helpful discussions. A.K. and J.R. acknowledge support by the Israel Science Foundation (ISF), and the Israeli Directorate for Defense Research and Development (DDR\&D) under grant No. 3467/21. V.K. was supported by  the Quantum Materials program at LBNL, funded by the U.S. Department of Energy under contract number DE-AC02-05CH11231. A part of the work by V.K. was performed at Aspen Center for Physics, which is supported by National Science Foundation grant PHY-1607611 and by a grant from the Simons Foundation. R.M.F. was supported by the U.S. Department of Energy through the University of Minnesota Center for Quantum Materials, under Grant No. DE-SC-0016371
\end{acknowledgments}
% \cite{Kozii2019}, and computed the normal state properties as well as the tendencies to pairing instabilities.
% % We found that the linear coupling, which mixes spin and charge degrees of freedom, gives rise to a rich phase diagram, 
% We found, that for a {\QFM}, the geometric constraints enforced by the coupling, the transverse property of the phonon, and the existence of a Fermi surface, strongly influence the dynamics of both the phonons and the electrons. As a result, we found that the behavior in 2D and in 3D are qualitatively different. In 2D, there are two modes: an out of plane (z) mode and an in-plane (t) mode, and they have completely different properties in terms of bosonic and fermionic scattering as well as pairing. The z mode resembles an Ising spin or Ising nematic mode, while the t mode is 

% Subjects for discussion:
% \begin{enumerate}
% \item Relevance of the 2D limit
% \item Applicability to STO
% \item Resistivity ?
% \end{enumerate}
\onecolumngrid
\appendix
\section{Detailed calculations for Sec. \ref{sec:low-energy-theory}}
\label{sec:deta-calc-sec-2d}

\subsection{One-loop self-energies in the disordered phase}
\label{sec:one-loop-self}

In this Appendix we present details of the calculations for the one-loop self-energies within the Eliashberg approximation. The bosonic self-energy is given by Eqs. \eqref{eq:Pi-1loop-0} and \eqref{eq:Pi0-dis-final}, the latter of which we reproduce here:

  \begin{align}
  \label{eq:bubble-bare-app}
  \Pi_0^{lk}(q) &=\frac{\gb T}{k_F^2} \Trc  \sum_{p}  (\p \times \bvs)_l   G(p-q/2) G(p+q/2) (\p\times \bvs)_k\nn\\
  &\approx 2\gb a^2 \int \frac{d^3p}{\tpp^3}  \frac{1}{i\Sgt(p_0-q_0/2)-\epsilon(\p-\q/2)}\frac{1}{i\Sgt(p_0+q_0/2)-\epsilon(\p+\q/2)}\frac{|\p|^2}{\kf^2}\hat P^{lk}(\hat p). \nn\\
  % &\approx \gb \int \frac{d^3}{\tpp^3}\Trc\left[\left(|\p|\hat p_t^k\sigma_z + \hat z^k \sigma_{p;t}\right)\left(|\p|\hat p_t^l\sigma_z + \hat z^l \sigma_{p;t}\right)\right]
\end{align}
Here $\gb = \lambda^2D_0$, $\kf$, and  $a$ are respectively the effective fermion-boson coupling, the Fermi wavenumber which we assume constant for simplicity, and the lattice constant, all as defined in the main text. We have also  defined the shorthand notation for the generalized electron
self-energy,
\begin{equation}
  \label{eq:Sgt-def}
  i\Sgt(k) = i k_0  -\Sg(k_0,\k) \approx i k_0 - \Sg(k_0).
\end{equation}
Henceforth we implicitly assume that the self-energy does not depend on the momentum, which is justified if we treat $\vf$ as the renormalized Fermi velocity. In principle, Eq. \eqref{eq:bubble-bare-app} should be evaluated self-consistently with the fermionic self-energy and also with vertex corrections (which are not shown here). However, we will assume (and later verify), that both these modifications can be neglected. In that case we can integrate over $p_0$, and $\hat \Pi_0$ has the well-known form of the Lindhard function:
\begin{align}
  \label{eq:Pi0-FS}
  \Pi_0^{lk}(q)
  % &\approx i\gb \int \frac{d^3p}{\tpp^3} \frac{G(p+q)-G(p)}{\epsilon(\p+\q) - \epsilon(\p)-i q_0}\frac{|\p|^2}{\kf^2}\hat P^{lk}(\hat p) \nn\\
    &\approx
    2
      \gb
      % k_a^{-2}
      a^2
      \int\frac{d^2p}{\tpp^2}\frac{\Theta(-\epsilon(\p+\q))-\Theta(-\epsilon(\p))}{\epsilon(\p+\q) - \epsilon(\p)-i q_0}\frac{|\p|^2}{\kf^2}\hat P^{lk}(\hat p), 
\end{align}
where
% $k_a=2\pi/L$ is the inverse lattice spacing \vk{(there's no $k_a$ in this expression. Besides, it was defined previously as $k_a = 1/a$)}, \vk {and
$\Theta(x)$ is the Heaviside step function. Linearizing near the FS and performing the energy integral we obtain Eq. \eqref{eq:Pi0-dis-final}, where for our model 
\begin{equation}
  \label{eq:nuF-def}
  \nu_F =
  \frac{m^*a^2}{\pi} \equiv
  % \frac{1}{\pi \vf \kf}\frac{\kf^2}{k_a^2},
  \frac{1}{\pi \vf \kf}(\kf a)^2,
\end{equation}
with $m^*$ denoting the FL effective mass. Adding and subtracting $iq_0$ in the numerator of Eq. \eqref{eq:Pi0-dis-form-1}, we find
\begin{equation}
  \label{eq:pi0-app-1}
  \hat\Pi_0(z) \approx \gb \nu_F\int \frac{d\theta_p}{\tp} \left[1  + i z \frac{1}{\cos(\theta_p-\theta_q) - i z}\right]\hat{\mathcal P}(\theta_p),
\end{equation}
where
\begin{equation}
  \label{eq:z-def}
  z=\frac{q_0}{\vf |\q|}.
\end{equation}
We change variables to  $\theta = \theta_p - \theta_q$ and decompose $\hat p$ onto $\hat q$, $\hat q_t$, and $\hat z$ directly, so that
\begin{equation}
  \label{eq:proj-vec-not}
  \mathcal{P} = \hat I -\hat p \hat p =  \hat q\hat q \sin^2\theta + \hat q_t\hat q_t \cos^2\theta + \hat z\hat z - \sin\theta\cos\theta (\hq \hq_t + \hq_t\hq).
\end{equation}
Performing the integral in Eq. \eqref{eq:pi0-app-1} then gives Eqs. \eqref{eq:Pi0-dis-form-1}-\eqref{eq:delta-pi-defs}.

Next we calculate the fermionic self-energy, given in Eq. \eqref{eq:Sig-dis-1loop}:
\begin{align}
  \label{eq:Sig-dis-1loop-app}
  \Sg_{\alpha\beta}(k) &\approx \frac{\gb a^2}{D_0\kf^2} \int \frac{d^3p}{\tpp^3} (\k \times \bvs_{\alpha\gamma})_iG(k-p)D(p)(\k \times \bvs_{\gamma\beta})_j \nn \\
                       &= (\sigma_0)_{\alpha\beta}
                       % \frac{\gb L^2}{k_a^2}
                         \gb a^2
                         \int \frac{d^3p}{\tpp^3} \frac{D_z(p) + \cos^2(\theta_k-\theta_p)D_t(p)}{i\Sgt(k_0-p_0) - \vf p \cos(\theta_k-\theta_p)}
\end{align}
Unsurprisingly, we see that the self-energy depends on both modes, even though only one of them is at the QCP. For simplicity, we will from now on drop the $\sigma_0$. Since $D_z$ and $D_t$ depend on $|\p|$ only, we can perform the angular integral exactly, yielding
\begin{align}
  \label{eq:Sg-decom-app}
  \Sg(k) &= \Sg_z(k) + \Sg_t(k), \\
  \Sg_z(k) &= -i\frac{\gb a^2}{\vf \tpp^2} \int_{-\infty}^{\infty}dp_0\int_0^\infty dp \frac{\sgn(Z)}{\sqrt{1+Z^2}}D_z(p), \nn \\
  \Sg_t(k) &= -i
  % \frac{\gb}{\vf k_a^2}
             \frac{\gb a^2}{\vf \tpp^2}
             \int_{-\infty}^{\infty}dp_0\int_0^\infty dp~ Z\left(1-\frac{|Z|}{\sqrt{1+Z^2}}\right)D_t(p), \nn
\end{align}
where
\begin{equation}
  \label{eq:Z-def-app}
  Z = \Sgt(k_0-p_0)/\vf |\p|.
\end{equation}
The self-energy contribution from $D_z$ is well known from the context of e.g. ferromagnetic or nematic QCPs, while the contribution from $D_t$ is distinct. The conventional way to solve the integral over $D_z$ is to note that the $\sgn(Z)$ term limits the $p_0$ integral to a scale of $k_0$, which in turn implies that $|Z| \ll 1$, allowing one to neglect the $\sqrt{1+Z^2}$ contribution from the fermionic sector. Thus, the $p$ integral  only involves $D_z$, yielding an effective local frequency-dependent interaction. The separation of scales given by the sign function is the root of the Eliashberg approximation. Here, we proceed a bit differently, so as to treat both $\Sg_z$ and $\Sg_t$ on an equal footing, solving first the integral with the more familiar $D_z$. We assume and then verify that we may neglect self-energy corrections to Eqs. \eqref{eq:Sg-decom-app}, i.e. that a self-consistent treatment yields the same result as a non-self-consistent treatment with $Z \approx |k_0-p_0|/\vf|\p|$. Then we  shift the frequency integral and change variables, such that,
\begin{align}
  \label{eq:Sg-z-app-1}
  \Sg_z(k) &= i\frac{\gb a^2}{\tpp^2} \int_{-\infty}^{\infty}dZ\int_0^\infty p dp \frac{\sgn(Z)}{\sqrt{1+Z^2}}D_z(p,|Z+k_0/\vf p|) \nn\\
           &= -i\frac{\gb a^2}{\tpp^2} \int_{0}^{\infty}p dp\frac{dZ}{\sqrt{1+Z^2}}\left[\frac{1}{p^2a^2 + r_z + \gb\nu_F \left|Z - \frac{k_0}{\vf p}\right|}-\frac{1}{p^2a^2 + r_z + \gb\nu_F \left|Z + \frac{k_0}{\vf p}\right|}\right] \nn\\
           &= -i\frac{\gb}{\tpp^2
             \vf k_a} k_0\int_{0}^{\infty}dp\frac{dZ}{\sqrt{1+\frac{k_0^2}{\vf^2k_a^2 p^2}Z^2}}\left[\frac{p}{p^3 + p r_z + \frac{\gb\nu_Fk_0}{\vf k_a} \left|Z - 1\right|}-\frac{p}{p^3 + p r_z + \frac{\gb\nu_Fk_0}{\vf k_a} \left|Z + 1\right|}\right].
\end{align}
In Eq. \eqref{eq:Sg-z-app-1} we dropped for simplicity the $c^{-2}(k_0-p_0)^2$ frequency term from the bosonic bare propagator. This term does not contribute to the low-energy theory for the $z$ mode and only appears as a logarithmic cutoff for the $t$ mode, as we show below. Here, we also defined, as in the main text,
\begin{equation}
  \label{eq:ka-def}
  k_a = a^{-1}
\end{equation}
to make the dimensional analysis more transparent.
For simplicity, consider $r_z = 0$. Then, it follows that $p\sim k_0^{1/3}$ and $Z\sim 1$, justifying neglecting the square root term, and also justifying our neglect of the self-energy in the fermionic propagator and of the bare bosonic frequency dependence in the bosonic propagator. Rescaling momentum gives
\begin{align}
  \label{eq:Sg-z-app-2}
  \Sg_z(k) &=-i\frac{\gb}{\tpp^2
             \vf k_a} k_0\left(\frac{\gb\nu_Fk_0}{\vf k_a}\right)^{-1/3}\int_{0}^{\infty} dp~dZ \left[\frac{p}{p^3 +\left|Z - 1\right|}-\frac{p}{p^3+  \left|Z + 1\right|}\right] \nn\\
  &= -i\w_z^{1/3}k_0^{2/3}, % Check sign
\end{align}
where
\begin{equation}
  \label{eq:w-z-def}
  \w_z^{1/3} = \frac{1}{2\sqrt{3}}\left(\frac{\gb^{2}}{\pi^2\vf\kf}\right)^{1/3}.
\end{equation}
For finite $r_z$, when the frequency $k_0$ is small enough, the integral over the bosonic momentum in Eq. \eqref{eq:Sg-z-app-1} is dominated by $r_z$, leading to  a linear self-energy as shown in the main text, Eq. \eqref{eq:sig-z-fl}.
% the self energy has a scaling form
% \begin{equation}
%   \label{eq:Sg-z-app-3}
%   \Sg_z(k) = -i\w_z^{1/3}k_0^{2/3} \mathcal{U}_z\left(r_z\left(\frac{\vf k_a}{\gb\nu_Fk_0}\right)^{2/3}\right),
% \end{equation}
% where $\mathcal{U}_z(x) = 1 + \cdots$ is a scaling function that can be readily obtained from Eq. \eqref{eq:Sg-z-app-1}, and yields the linear self-energy at the smallest frequencies (where $r_z$ dominates the other terms in the boson propagator) \cite{Rech2006}as shown in Eq.  \vk{ (Write explicit expression for $\mathcal{U}_z(x)$.)}. 
Now, we repeat the treatment for $\Sg_t(k)$. Going through the same steps yields
\begin{equation}
  \label{eq:Sg-t-app-1}
  \Sg_t(k)=-i\frac{\gb}{\tpp^2k_a^2}\int_{0}^{\infty} p dp dZ Z\left(1-\frac{|Z|}{\sqrt{1+Z^2}}\right) \left[\frac{1}{p^2/k_a^2 + r_t + \gb\nu_F \left(Z - \frac{k_0}{\vf p}\right)^2}-\frac{1}{p^2/k_a^2 + r_t +\gb\nu_F \left(Z + \frac{k_0}{\vf p}\right)^2}\right].
\end{equation}
Upon setting $r_t = 0$ and rescaling, as before, $Z\to Z k_0/(\vf k_a p) $, we find that the $Z$ integral diverges logarithmically. Thus, the main contribution to $\Sg_t$ is from the region $Z\gg (k_0/\vf p)$, which allows us to just expand the contribution in the square brackets to leading order as 
%\vk{ (smth is wrong with the units here)}
\begin{align}
  \label{eq:Sg-t-app-2}
  \Sg_t(k) &\approx -i\frac{\gb}{\tpp^2}\int_{Z_{IR}}^{\infty} p dp dZ Z\left(1-\frac{|Z|}{\sqrt{1+Z^2}}\right) \frac{1}{(p^2 + \gb\nu_F Z^2)^2}\frac{4 \gb \nu_F k_0 Z}{\vf k_a p}
  \\
           &=  -i  \frac{\gb}{4\pi(\gb\nu_F)^{1/2}\vf k_a} k_0 \int_{Z_{IR}}^{Z_{UV}} \frac{dZ}{Z}\left(1-\frac{|Z|}{\sqrt{1+Z^2}}\right). \nn\\
           % &\approx i\frac{\sqrt{\gb\nu_F}k_a}{8 k_F} k_0 \log\left(\frac{\sqrt{\gb \nu_F}\vf k_a}{k_0}\right).
               \label{eq:Sg-t-app-3}
\end{align}
In Eq. \eqref{eq:Sg-t-app-3} the lower  cutoff is determined by the lower cutoff of the momentum and given by $Z_{IR}^2 = k_0 / (\vf  k_a \sqrt{\gb \nu_F})$. The upper cutoff $Z_{UV}$ is of order one if the bare frequency dependence of the bosonic propagator is neglected, as in the previous section. If the latter is taken into account, the upper cutoff is
\begin{equation}
  \label{eq:Zuv-def}
  Z_{UV} = \mbox{min}\left(1, \frac{c}{\vf}\right),
\end{equation}
and the final result is
\begin{equation}
  \label{eq:Sg-t-app-4}
  \Sg_t(k) \approx -i
  %\frac{\sqrt{\gb\nu_F}k_a}{4 k_F} k_0 \log\left(Z_{UV}\frac{\sqrt{\gb \nu_F}\vf k_a}{k_0}\right),
  \frac{\sqrt{\gb\nu_F}k_a}{8 k_F} k_0 \log\left(Z_{UV}^2\frac{\sqrt{\gb \nu_F}\vf k_a}{|k_0|}\right),
\end{equation}
in agreement with Eq. \eqref{eq:Sg-t-qcp} of the main text.
%\vk{ (This expression agrees with my expression (apart from $i$) and disagrees with the main text.)}.
Note that the logarithm is IR divergent so that the theory is fully self-consistent. For finite $r_t$, we notice from Eq. \eqref{eq:Sg-t-app-2} that the lower cutoff is simply replaced by $Z_{IR}^2 = \mbox{max}(k_0 / (\vf k_a\sqrt{\gb \nu_F}),r_t/\gb\nu_F)$, so that the system is in a FL regime at the lowest frequencies.

\subsection{Vertex corrections}
\label{sec:vertex-corrections}

In order to assess whether the expressions for $\hat \Pi$ remain valid near the QCP, we evaluate the particle-hole vertex correction. The reason for this is that in magnetic systems, the form of the Landau damping term changes when the system goes into the nFL phase, so that e.g. $\q_0/\sqrt{\vf^2|\q|^2+q_0^2}$ becomes $\q_0/\sqrt{\vf^2|\q|^2+\Sigma(q_0)^2}$. On the other hand, the Landau damping term in the charge channel remains unchanged due to the Ward identity reflecting charge conservation. The form of the Landau damping term is important in determining the nonanalytic terms in the free energy, and as a result vertex corrections can change the power-law behavior of these terms near the QCP.
% thus the scaling of the QOBD transitions \vk{(smth's wrong with the sentence)}.
In this section, we show that the {\QFM} polarization operator is almost unaffected by vertex corrections.
%behaves more like a charge channel \vk{(rephrase here)}, in the sense that the form of the polarization \vk{operator} is almost unchanged.

The fully dressed vertex
% vertex equation has the form
has the form,
\begin{equation}
  \label{eq:gamma-1}
  \hat \Gamma(k,q) = \hz \Gamma_z(k,q) \sigma_t - \hk_t \Gamma_t(k)\sigma_z, 
\end{equation}
where we recall that $\hat \gamma =   \hz \sigma_t - \hk_t \sigma_z$ is the bare vertex, see Eq. \eqref{eq:k-cross-decomp}.
% , and we assume $\q \ll \k$ and neglect the $\q$ dependence in the bare vertex.
The polarization bubble has the form
\begin{align}
  \label{eq:Pi-gamma}
  \Pi_0^{lk}(q) &=\frac{\gb T}{k_F^2} \Trc  \sum_{p}  \Gamma_l(p,q)   G(p-q/2) \gamma_k(\p) G(p+q/2) \nn\\
                &= \frac{\gb T}{k_F^2} \Trc  \sum_{p}  \left(\hz\hz \Gamma_z(p,q) + \hp_t\hp_t \Gamma_t(p,q)\right)_{kl} G(p-q/2) G(p+q/2) \nn\\
                &=\frac{\gb T}{k_F^2} \Trc  \sum_{p}
                \frac{\left(\hz\hz \Gamma_z(p,q) + \hp_t\hp_t \Gamma_t(p,q)\right)_{kl}}{i p_0 - \left(\Sg(p_0+q_0/2)-\Sg(p_0-q_0/2)\right) - \vf \hat p \cdot \q}
                  \left[G(p-q/2)- G(p+q/2)\right].
\end{align}
Upon integrating over fermionic energies, the Green's functions on the right just give Fermi distribution functions (becoming Heaviside step functions at $T=0$). The implication is that if the vertex functions obey
\begin{equation}
  \label{eq:Ward-1}
  \frac{\Gamma_i(p,q)}{i p_0 -\left(\Sg(p_0+q_0/2)-\Sg(p_0-q_0/2)\right) - \vf \hat p \cdot \q} = \frac{1}{i p_0 - \vf \hat p \cdot \q},
\end{equation}
then the polarization bubble is not changed by a finite or even divergent self-energy. For the case $\gamma = \sigma_0$, i.e. just the charge vertex, this is well established \cite{Chubukov2005a,Chubukov2018}. For the {\QFM} case, we note that the vertex equation has the form
\begin{equation}
  \label{eq:deltaGamma-def}
  \hat\Gamma(k,q) = \hat\gamma(k) + \delta\hat\Gamma(k,q),
\end{equation}
where
\begin{equation}
  \label{eq:gamma-2}
  \delta\Gamma_{\alpha\beta} (k,q) = \frac{\gb T}{D_0\kf^2}\sum_p\gamma_{\alpha\delta}^{i}(\k)\Gamma^{l}_{\delta\eta}(k,q)\gamma^{j}_{\eta\beta}(\k)D^{ij}(p)G(k+p-q/2)G(k+p+q/2).
\end{equation}
Here, we assumed $\q \ll \k$ and neglected the $\q$ dependence in the bare vertex. After performing the various summations we find

\begin{align}
  \label{eq:gamma-3}
  \delta\hat{\Gamma}(k) =  \frac{\gb T}{D_0\kf^2}\sum_p\left[D_z(p)(\hz \sigma_t \Gamma_z - \hk_t \sigma_z \Gamma_t)+\cos^2(\theta_k-\theta_p)D_t(p)(\hk_t \sigma_z \Gamma_t - \hz \sigma_t \Gamma_z) \right]G(k+p-q/2)G(k+p+q/2).
\end{align}
A careful analysis of the term inside the brackets reveals that, for each channel, the correction arising from its own channel comes with a positive sign whereas the corrections arising from the other channel comes with a negative sign. Thus, assuming that only one channel is critical, this implies that we may neglect the contribution of the non-critical channel and obtain

\begin{align}
  \label{eq:gamma-4}
  \delta\Gamma_{z} (k) &=  \frac{\gb T}{D_0\kf^2}\sum_p\frac{ \Gamma_z(p,q)}{i p_0  -\left(\Sg(p_0+q_0/2)-\Sg(p_0-q_0/2)\right) - \vf \hat p \cdot \q} [G(k+p-q/2)- G(k+p+q/2)]D_z(p), \\
  \delta\Gamma_{t} (k) &=  \frac{\gb T}{D_0\kf^2}\sum_p\frac{ \Gamma_t(p,q) \cdot \cos^2(\theta_p-\theta_q)}{i p_0  -\left(\Sg(p_0+q_0/2)-\Sg(p_0-q_0/2)\right) - \vf \hat p \cdot \q} [G(k+p-q/2)- G(k+p+q/2)]D_t(p).
\end{align}
Using Eq.~\eqref{eq:Ward-1} as an ansatz,  expanding $\hp_t\hp_t = \cos^2(\theta_p=\theta_q)\hq_t\hq_t + \cdots$, and going back to the definition of the self-energies, we find
\begin{equation}
  \label{eq:Ward-2}
  \delta\Gamma_j(p,q) = \frac{-\left(\Sg_j(p_0+q_0/2)-\Sg_j(p_0-q_0/2)\right) - \vf |\q| \cos(\theta_p-\theta_q)}{i p_0  - \vf |\q| \cos(\theta_p-\theta_q)},
\end{equation}
which proves the ansatz. Thus, the polarization bubble retains its bare form.

\section{Details of the calculations in Sec. \ref{sec:pairing-2d-qfms}}
\label{sec:deta-calc-sec}

In this Appendix we provide a detailed calculation of the solution of the pairing equations.
% , shown in Fig. (...) of the main text.
In Sec.~\ref{sec:pairing-2d-qfms}, we found that both the $z$ and $t$ modes have two nearly degenerate pairing channels, a singlet and a triplet. In this section, we calculate the degenerate pairing $T_c$ in the leading order. In practice, the two channels are typically split by subleading terms in the pairing equation \cite{Klein2018c,Klein2019}, but we shall not deal with that issue here.

\subsection{Pairing near the $z$ mode QCP}
\label{sec:pairing-near-z}

The pairing equation is given by,
%Eq. \eqref{eq:pairing-z-1}, which we reproduce here,
\begin{equation}
  \label{eq:pairing-z-1-app}
  \phi_{nj}(k) = \gb T\sum_p D_z(k-p)\phi_{nj}(p)G(p)G(-p),
\end{equation}
where $\phi_{nz}$ was defined in Eq. \eqref{eq:pairing-eq-1} and Table \ref{tab:SC_decomp}.
%$\Phi_z = i\sigma_y(f(k) + \bv{d}(k)\cdot \bvs)$, where $d_z = 0, d_x+i d_y = e^{i\theta_p} d_{xy}(k)$. 
We shift the momentum integration, $\p \to \p + \k$ and integrate over $\theta_p$ to obtain
\begin{align}
  \phi_{nj}(k_0,\theta_k) &= \frac{\pi \gb T \nu_F}{k_F} \sum_{k_0\neq p_0}\int_0^\infty \frac{dp}{2\pi}\frac{1}{|p_0|+|\Sg(p_0)|} \frac{\phi_{nj}(p_0,\theta_k)}{r_z + p^2/k_a^2 + \gb\nu_F|p_0-k_0|/(\vf p)}.
  \label{eq:pairing-z-2-app}
\end{align}
As we did for the normal state properties, we neglected for simplicity the bare $(p_0-k_0)^2/c^2$ bosonic frequency term. Away from the QCP, we may neglect the Landau damping component of $D_z$ and the self-energy, so the gap equation reads,
\begin{align}
  \phi_{nj}(k_0,\theta_k) &= \frac{\pi \gb T \nu_F k_a}{4k_F\sqrt{r_z}} \sum_{k_0\neq p_0}\frac{\phi_{nj}(p_0,\theta_k)}{|p_0|},
  \label{eq:pairing-z-3-app}
\end{align}
which is a standard BCS-type equation. The $p_0$ sum yields a logarithm that is cut off by the Landau damping at $p_0 \sim \frac{\vf k_a}{\gb\nu_F} r_z^{3/2}$, which yields the FL limit in Eq. \eqref{eq:Tc-estimates}. At the QCP, $r_z \approx 0$, the integral over $dp$ yields
\begin{align}
  \phi_{nj}(k_0,\theta_k) &= \frac{\pi \gb T \nu_F k_a}{3\sqrt{3}k_F} \sum_{k_0\neq p_0}\frac{\phi_{nj}(p_0,\theta_k)}{(|p_0|+|\Sg(p_0)|)\left(\frac{\gb\nu_F |k_0-p_0|}{\vf k_a}\right)^{1/3}}.
  \label{eq:pairing-z-4-app}
\end{align}
Similar equations appear in a class of QC pairing models called $\gamma-$ models, with $\gamma =1/3$ characterizing the power law of the pairing interaction, and $2\gamma$ the nFL self-energy. Similar results hold for  pairing from nematic fluctuations. The result for $T_c$ is known and we write it explicitly in Eq. \eqref{eq:Tc-estimates} (with $a_z,b_z = O(1)$)  \cite{chubukov2020interplay}.

\subsection{Pairing near the $t$ mode QCP}
\label{sec:pairing-near-t}

For the $t$ mode, the gap equation is given by
%(see Eq.~\eqref{eq:pairing-t-1}) \vk{ (what equation?)},
\begin{flalign}
  \label{eq:pairing-t-1-app}
  \phi_{nj}(k) = \gb T\sum_{p}\left(\hk \cdot \frac{\k-\p}{|\k-\p|}\right)^2D_t(k-p)
 \phi_{nj}(p) G(p) G(-p).
\end{flalign}
Again shifting momenta and integrating over angles yields
\begin{align}
  \phi_{nj}(k_0,\theta_k) = \frac{\pi \gb T \nu_F}{\vf \kf} \sum_{p_0 \neq k_0}\int_0^\infty \frac{dp}{2\pi p }
  % \left(1 - \frac{p_0}{\sqrt{\vf^2p^2+p_0^2}}\right)
  ~l_1\left(\frac{p_0}{\vf p}\right)\frac{\phi_{nj}(p_0,\theta_k)}{r_t + p^2/k_a^2 + \gb\nu_F(p_0-k_0)^2/(\vf p)^2}.\label{eq:pairing-t-2-app}
\end{align}
As discussed in the main text, there is no logarithmic $1/p_0$ term from the fermions, neither in the FL nor in the marginal FL regime. Since the bosonic propagator has a dynamic critical exponent $z=2$, $p\sim \sqrt{p_0-k_0}$, we may safely assume that $\vf p \gg p_0$ and approximate $l_1(p_0/(\vf p)) \approx 1$. Then, integrating over $p$, we find
\begin{align}
  \phi_{nj}(k_0,\theta_k) = \frac{\pi \gb T \nu_F}{\vf \kf} \sum_{p_0 \neq k_0} \frac{\phi_{nj}(p_0,\theta_k)}{\frac{\sqrt{\gb\nu_F}}{\vf k_a}|p_0-k_0|} \mathcal{Y}\left(\frac{r_t}{\frac{\sqrt{\gb\nu_F}}{\vf k_a}|p_0-k_0|}\right),\label{eq:pairing-t-3-app}
\end{align}
where $\mathcal{Y}$ is the function,
\begin{align}
  \label{eq:scale-phi-t}
  \mathcal{Y}(x) &= \int_0^\infty \frac{dy}{\tp y} (x + y^2 + y^{-2})^{-1} = \frac{1}{2\sqrt{4-x^2}}-\frac{\tan^{-1}\left(\frac{2+x}{\sqrt{4-x^2}}\right)}{\pi\sqrt{4-x^2}}
\end{align}
with asymptotic behaviors
\begin{align}
    \mathcal{Y}(x)&=\left\{\begin{array}{ll}
    1/8 & x\to 0\\
    \log(x)/(\tp x) &x \to \infty
  \end{array}\right..
\end{align}
% $\mathcal{Y}$ is a scaling function with the limits $\mathcal{Y}(x \to 0) = \pi / 4, \mathcal{Y}(x \to \infty) = \log(x)/(2x)$.
In the FL regime, the gap equation then reads

\begin{align}
  \phi_{nj}(k_0,\theta_k) = \frac{\pi \gb T \nu_F}{\vf \kf} \sum_{p_0 \neq k_0} \frac{1}{2r_t}\phi_{nj}(p_0,\theta_k)\log\left(\frac{r_t}{\frac{\sqrt{\gb\nu_F}}{\vf k_a}|p_0-k_0|}\right)\label{eq:pairing-t-4-app},
\end{align}
and the upper limit for the frequency sum is just $\vf k_a r_t / \sqrt{\gb\nu_F}$. The frequency sum is not divergent, so there is no solution for the gap equation unless $\gb \nu_F \sim 1$. On the other hand, for $r_t\to 0$, the gap equation reads

\begin{align}
  \phi_{nj}(k_0,\theta_k) = \frac{\pi \sqrt{\gb \nu_F}  T k_a}{8\kf} \sum_{p_0 \neq k_0} \frac{\phi_{nj}(p_0,\theta_k)}{|p_0-k_0|}.\label{eq:pairing-t-5-app}
\end{align}
As written here, the upper cutoff for the frequency sum is  $p_0 \sim \vf k_a \sqrt{\gb\nu_F}$, which is obtained from the $l_1$ function in Eq. \eqref{eq:pairing-t-2-app}. A more careful calculation including the previously neglected bare bosonic frequency term yields a modified cutoff $p_0 \sim Z_{UV}^2 \vf k_a \sqrt{\gb\nu_F}$, where $Z_{UV}$ was defined in Eq. \eqref{eq:Zuv-def}. This leads directly to  Eq. \eqref{eq:Tt-p}, with $a_t =O(1)$. To see how $T_c$ vanishes with finite $r_t$, we expand $\mathcal{Y}(x) \approx \frac{1}{8}(1-x/\pi)$ and convert to a frequency integration with $k_0\approx 0$ to obtain
\begin{equation}
  \label{eq:Phi-t-finite-rt}
  1 \approx \frac{\sqrt{\gb \nu_F}k_a}{8 \kf}\left(\log \frac{\vf k_a \sqrt{\gb \nu_F}}{2\pi T_c}-\frac{1}{\pi}\frac{r_t}{2\pi T_c}\right).
\end{equation}
For $r_t = 0$ we obtain
\begin{equation}
  \label{eq:Tc-t-rt-0}
  2\pi T_{c,t} \approx \vf k_a \sqrt{\gb \nu_F} e^{-\frac{8\kf}{\sqrt{\gb \nu_F} k_a}}
\end{equation}
in accordance with  Eq. \eqref{eq:Tt-p}. Expanding $T_c = T_{c,t}-\delta T_{c,t}$ we obtain
\begin{equation}
  \label{eq:deltaT-t-rt}
  2\pi\delta T_{c,t} \approx \frac{r_t}{\pi\sqrt{\gb \nu_F}}\vf k_a
\end{equation}
leading to the expressions following Eq. \eqref{eq:Tt-p}.

\section{Detailed calculations for Sec. \ref{sec:quant-order-disord}}
\label{sec:deta-calc-sec-quant}

In this Appendix we give a detailed derivation of the results in Sec. \ref{sec:quant-order-disord} on the order-by-disorder induced phases. 
\subsection{The first-order transition}
\label{sec:1st-ord-transition}

In the main text, we noted that to obtain the nonanalytic terms in the free energy we must (a) assume that the system has spontaneously formed static  uniform order, (b) compute the polarization bubble in the presence of that order, and  (c) calculate the correction to the free energy from that polarization. Let us proceed step by  step.

The polarization bubble is given by Eq. \eqref{eq:bubble-Delta}, which upon performing the rotation in Eq. \eqref{eq:U-def} has the form, for $T \to 0$,
\begin{align}
  \label{eq:bubble-Delta-app}
  \Pi_\Delta^{kl} &= \frac{\gb}{k_a^2} \Trc \int \frac{d^3p}{\tpp^3} (\hp_t \sigma_x + \hz \sigma_z)^k \left[\left(E_p + \frac{1}{2}\varepsilon_q\right)\sigma_0 - \Delta \sigma_z\right]^{-1}(\hp_t \sigma_x + \hz \sigma_z)^l\left[\left(E_p - \frac{1}{2}\varepsilon_q\right)\sigma_0 - \Delta \sigma_z\right]^{-1},
                  % &= \gb \nu_F \left\{
                    % \hp\hp 
                    % \right\}
\end{align}
where
\begin{equation}
  \label{eq:epq-def}
  E_p = ip_0 - \vf (|\p| - \kf), \qquad  \varepsilon_q = i q_0 - \vf\hp\cdot{\bf q}
\end{equation}
give the fermionic dispersions near the FS. As the Green's functions are diagonal in spin space the crossterms between $\sigma_{x}$ and  $\sigma_z$ vanish upon tracing over the spin indices, yielding a two-block diagonal polarization. Performing the frequency and then momentum integrals yields
\begin{align}
  \label{eq:Pi-delta-FS}
  \hat{\Pi}_\Delta(q) &= \frac{\gb \nu_F}{2} \int \frac{d\theta_p}{\tp}\left[ \hp_t\hp_t \left(\frac{\vf|\q|\hp\cdot\hq-2\Delta}{\vf|\q|\hp\cdot\hq-2\Delta- iq_0} + \frac{\vf|\q|\hp\cdot\hq+2\Delta}{\vf|\q|\hp\cdot\hq+2\Delta- iq_0}\right)
                       + 2\hz\hz \frac{\vf|\q|\hp\cdot\hq}{\vf|\q|\hp\cdot\hq- iq_0}\right].
\end{align}
The global factor $1/2$ in Eq. \eqref{eq:Pi-delta-FS} is due to the definition of $\nu_F$ in Eq. \eqref{eq:nuF-def} to include spin summation. Changing variables and projecting onto $\hq$, $\hq_t$, and $\hat z$ as we did in the disordered case we find,
\begin{equation}
  \label{eq:Pi-delta-final}
  \hat{\Pi}_{\Delta}(q) = \frac{\gb\nu_F}{2}\left(\hq_t\hq_t\Pi_{\Delta;t} + \hq\hq \Pi_{\Delta;l} + 2 \hz\hz  \Pi_{\Delta;z}\right),
\end{equation}
where
\begin{align}
  \label{eq:Pi-delta-forms}
  \Pi_{\Delta;t} &= 1 + \delta\Pi_{\Delta;t}, \nn\\
  \Pi_{\Delta,l} &= 1 + \delta\Pi_{\Delta;z} - \delta\Pi_{\Delta;t}, \nn\\
  \Pi_{\Delta;z} &= 1 + \delta\Pi_{\Delta;z},  
\end{align}
and $\delta\Pi_{\Delta}$ given in Eq. \eqref{eq:delta-pi-Deltas} of the main text.

For a $t$ transition, the polarization bubble has the form (see Eqs. \eqref{eq:k-cross-decomp} and \eqref{eq:zeeman-ip})
\begin{align}
  \label{eq:bubble-Delta-ip-app}
  \Pi_\Delta^{kl} &= \frac{\gb}{k_a^2} \Trc \int \frac{d^3p}{\tpp^3} (\hz \sigma_{p;t} -\hp_t \sigma_z ) \left[\left(E_p + \frac{1}{2}\varepsilon_q\right)\sigma_0 + \Delta \sin\theta_p\sigma_z\right]^{-1}(\hz \sigma_{p;t} -\hp_t \sigma_z)\left[\left(E_p - \frac{1}{2}\varepsilon_q\right)\sigma_0 +\Delta \sin\theta_p\sigma_z\right]^{-1} \nn\\
                  &=  \frac{\gb \nu_F}{2} \int \frac{d\theta_p}{\tp}\left[ \hz\hz \left(\frac{\vf|\q|\hp\cdot\hq-2\Delta\sin\theta_p}{\vf|\q|\hp\cdot\hq-2\Delta\sin\theta_p- iq_0} + \frac{\vf|\q|\hp\cdot\hq+2\Delta\sin\theta_p}{\vf|\q|\hp\cdot\hq+2\Delta\sin\theta_p- iq_0}\right)
                    + 2\hp_t\hp_t \frac{\vf|\q|\hp\cdot\hq}{\vf|\q|\hp\cdot\hq- iq_0}\right].
                    % \hp\hp 
                    % \right\}
\end{align}
Obviously, the roles of $t$ and $z$ modes are simply reversed, with the $z$ mode acting as a transverse fluctuation to the $t$ mode order.
% and longitudinal/transverse \vk{(modes?)} are simply reversed
In addition, the integral for the $\hz\hz$ component is dominated by the region $\cos(\theta_p-\theta_q) \sim \mbox{max}(\Delta/(\vf|\q|),q_0/(\vf|\q|)$.
% \vk{(maybe add max or min?)}.
In all the computations of the free energy, these quantities are $\lesssim 1$. Therefore we may safely assume $\cos(\theta_p-\theta_q)$ is small, and replace $\Delta \sin\theta_p \to \Delta\cos\theta_q$, effectively neglecting some small quantitative corrections.  Projecting $\hp_t$ onto $\hq$, $\hq_t$, and $\hat z$ and performing the angular integrals yields Eqs.~\eqref{eq:Pi-delta-final} and \eqref{eq:Pi-delta-forms}, where this time $\delta\Pi_\Delta$ is given in Eq.~\eqref{eq:Pi-delta-ip}. Note that we are using the same notations for $\delta\Pi_\Delta$ in both $t$ and $z$ cases to minimize the notation burden.

To compute the RPA free energy, we plug Eq. \eqref{eq:Pi-delta-final} into the relevant expression, Eq. \eqref{eq:free-1st-order-def}, neglecting the longitudinal $\delta\Pi_l$ component, which contributes to the gapped-out longitudinal mode. Only one of the sectors ($z$ or $t$) depends on $\Delta$. For the $z$ case, the free energy is given by Eq. \eqref{eq:f-oop-1}, which we reproduce here for clarity,
\begin{align}
  \label{eq:f-oop-1-app}
  \delta F &= T\sum_q \left(\log D_{t}^{-1}(q,\Delta\hz)- \log D_t^{-1}(q,0) \right) \nn\\
           % &= k_a^{-2} \int \frac{d^3q}{\tpp^3} \log \left[\frac{r_t + |\q|^2 + \delta\Pi_{\Delta;t}(q)}
             % {r_t + |\q|^2 +\delta\Pi_{0;t}(q)}\right]\nn\\
           &=k_a^{-2} \int \frac{d^3q}{\tpp^3} \log \left[\frac{r_t + |\q|^2 + \frac{z_q}{2}\left(l_1(z_q+2i\Delta_q)+\cc\right)}
             {r_t + |\q|^2 + z_ql_1(z_q)}\right],
\end{align}
where
\begin{equation}
  \label{eq:zq-dq}
  z_q = \frac{q_0}{\vf|\q|},\quad \Delta_q = \frac{\Delta}{\vf|\q|}.
\end{equation}
The integrand in Eq. \eqref{eq:f-oop-1-app} is even in $\Delta$, so that a power expansion yields only even powers $\Delta^2, \Delta^4, \cdots$. However, $l_1(x)$ is not analytic, which dramatically affects the result. To see this, we change variables to $z=z_q$. Then, the integral in Eq. \eqref{eq:f-oop-1-app} is of the form
\begin{equation}
  \label{eq:delta-F-2-app}
  \delta F \propto \int_0^\infty dz \int_0^\Lambda q^2 dq \left(f_1(z) \frac{\Delta^2}{\vf^2q^2} + f_2(z) \frac{\Delta^4}{\vf^4q^4} + \cdots \right).
\end{equation}
Here, $f_1,f_2,\ldots$ are convergent functions of $z$ and $\Lambda$ is some UV cutoff which will not play a role in the final result. We see that the first term in the expansion is quadratic in $\Delta$ and UV divergent. Thus, it merely generates some correction to $r_z$ which we ignore. The second term, however, is IR divergent, and since the integrand is \emph{not} analytic we cannot extend the $q$ contour over the complex plane. Instead, we may estimate it by introducing an IR cutoff $q_\Delta \sim \Delta/\vf$, which immediately gives us the nonanalytic $\Delta^3$ correction to the free energy. Successive terms in the expansion all diverge in the same manner, so they all give the same $\Delta^3$ contribution. To compute Eq. \eqref{eq:f-oop-1-app} exactly, we rescaled it by $x = \vf q/\Delta$ and explicitly subtracted the second-order expansion term in $x^{-1}$. The resulting integral is convergent, yielding the final line of Eq. \eqref{eq:delta-f-final}.
% \subsection{in-plane transition}
% \label{sec:plane-transition}

For the in-plane ($t$) transition, the treatment is analogous. The rescaled summation is given by
\begin{align}
  \label{eq:f-ip-1-app}
  \delta F &= T\sum_q \left(\log D_{\bv\Delta,z}^{-1}- \log D_z^{-1} \right) \nn\\
           &\approx k_a^{-2} \int \frac{d^3q}{\tpp^3} \log \left[\frac{1 + \frac{z_q}{2}\left(l_0(z_q-2i\Delta_q\cos\theta_q)+\cc\right)}
             {1 + z_ql_0(z_q)}\right]\nn\\
           &= \frac{k_a^{-2}}{2\pi^2}\int_0^\infty dz \int_0^\infty x^2 dx \int_0^{2\pi} \frac{d\theta_q}{\tp}  \left|\cos\theta_q\right|^3 \log \left[
             \frac{1 + \frac{z}{2}\left(l_0(z+2ix^{-1})+\cc\right)}
             {1 + zl_0(z)}\right].          
\end{align}
We subtracted the second-order term and computed the numerical prefactor exactly, obtaining Eq.~\eqref{eq:f-ip-1}.

\subsubsection*{Finite-temperature phase diagram}
\label{sec:finite-temp-phase}

\begin{figure}
  \centering
  \includegraphics[width=0.5\hsize]{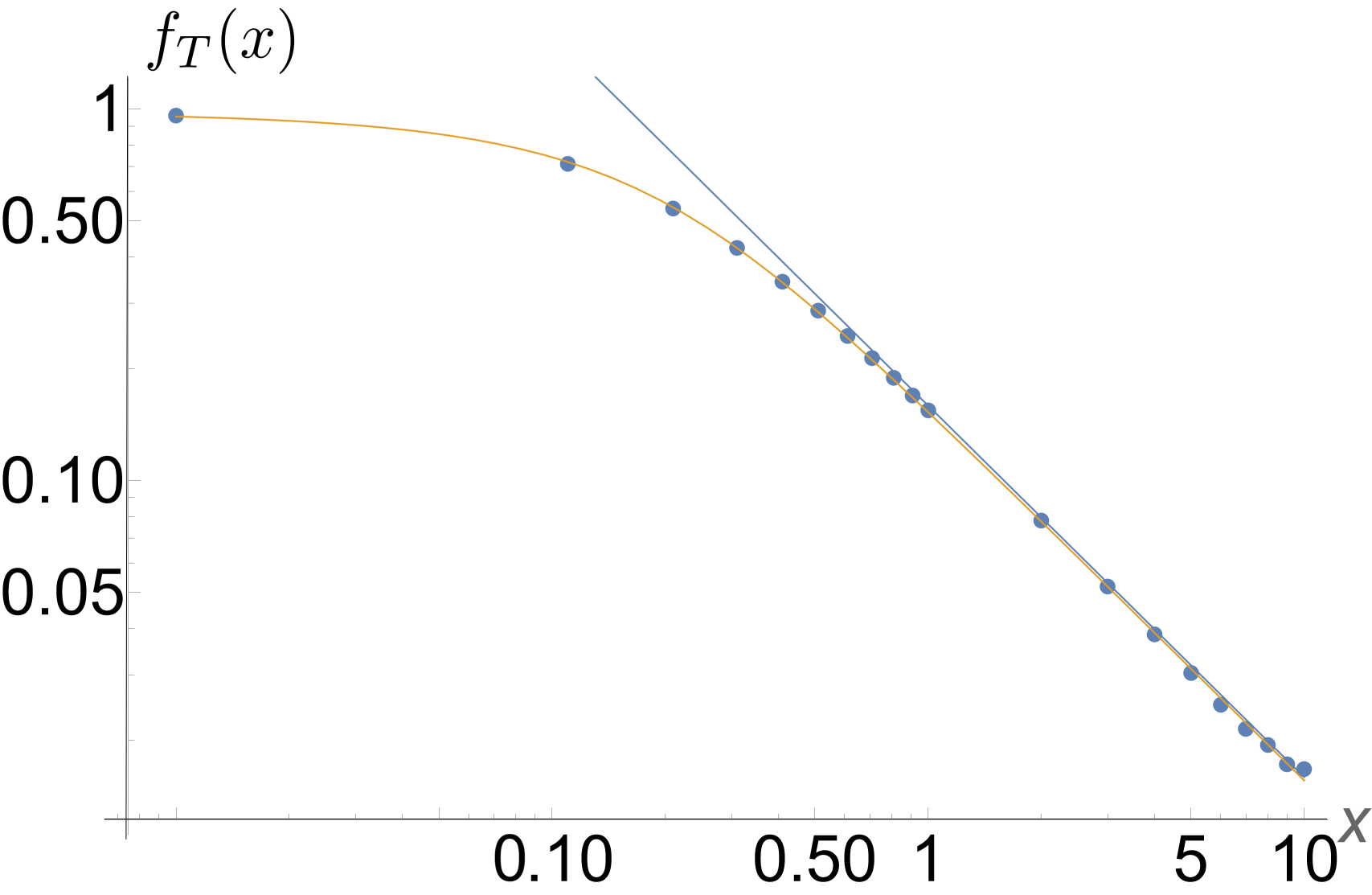}
  \caption{Computation of $f_T(x)$. The dots are obtained from the numerical evaluation of the nonanalytic term at finite $T$. The solid blue line is the asymptotic expression with the exact numerical prefactor, and the solid yellow line is Eq.~\eqref{eq:fT-found}. 
  }
  \label{fig:ftfound}
\end{figure}

To create the schematic finite-temperature phase diagram, we also solved Eq. \eqref{eq:f-oop-1-app} in the finite-temperature regime. This can be done numerically by rescaling both $\q$ and $\Delta$ with $T$.
% and performing the sum numerically.
Clearly, the result will be that
\begin{equation}
  \label{eq:delta-F-finite-T}
  \delta F \propto -\Delta^3 f_T(T/\Delta),
\end{equation}
where $f_T(0) = 1$. Expanding in large $T/\Delta$, one readily finds that $f_T(x) = a_z/x$ for $x\gg 1$, showing that, as expected the nonanalytic term vanishes at high temperature. The pre-factor $a_z$ can be evaluated numerically, giving $a_z \approx 0.16$. Furthermore, to excellent numerical accuracy we found that
\begin{equation}
  \label{eq:fT-found}
  f_T(x) = \frac{2}{\pi}\tan^{-1}\left(\frac{b_z}{x}\right).
\end{equation}
with $b_z \approx 1/4$. In Fig. \ref{fig:ftfound} we depict the numerical evaluation of $f_T$ along with the exact asymptotic expression and the fitted expression given by Eq. \eqref{eq:fT-found}. The result is similar for the $t$ mode, with $a_t \approx 0.19$ and  $b_t \approx 0.28$.

\begin{figure}
  \centering
  \includegraphics[width=0.5\hsize]{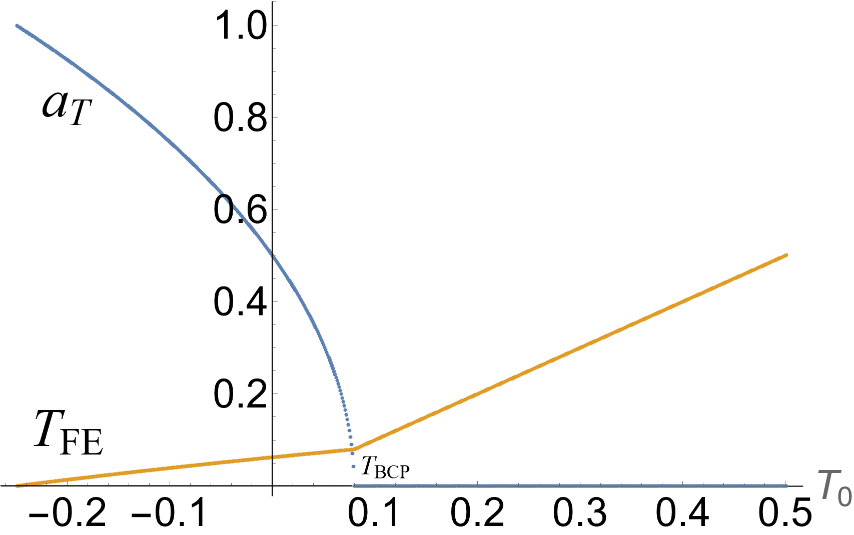}
  \caption{Numerical evaluation of the self-consistency equation, Eq.~\eqref{eq:aT-sc}. The parameters used were $a=1$, $b=1$, $\alpha=1$. }
  \label{fig:firstComp}
\end{figure}

Using the expression for $f_T(x)$, we are able to fully determine the phase diagram. The free energy has at $T=0$ the following form
\begin{equation}
  \label{eq:free-energy-app}
  F = r \Delta^2 - a \Delta^3 + b \Delta^4,
\end{equation}
where we suppress all $z/t$ subscripts. The prefactors (for the $z$ mode) are given in Eq. \eqref{eq:F-oop-final}, and the first-order phase transition is given by
\begin{align}
  \label{eq:phase-trans-eq}
  0 &= \pd_\Delta F \propto 2r - 3a\Delta+4b\Delta^2, \nn\\
  0 &= F \propto r - a \Delta + b\Delta^2.
\end{align}
The solution is
\begin{equation}
  \label{eq:trans-app}
  \Delta^* = \frac{a}{2b}, \qquad r^* = \frac{a^2}{4b}.
\end{equation}
The finite-temperature phase diagram can be obtained by introducing a temperature-dependent coefficient for the $\Delta^3$ term in Eq. \eqref{eq:free-energy-app},
\begin{equation}
  \label{eq:aT-def}
  a \to a_T = a f_T(T/\Delta).
\end{equation}
For small $a$, the correction to Eq. \eqref{eq:phase-trans-eq} is negligible, and Eq. \eqref{eq:trans-app} still represents an approximate solution provided $a\to a_T$. Plugging the solution back into Eq. \eqref{eq:aT-def} and using our exact expression for $f_T$, Eq. \eqref{eq:fT-found}, results in a self-consistency equation for the transition temperature $T=T_{FE}$,
\begin{equation}
  \label{eq:aT-sc}
  a_T = a \frac{2}{\pi}\tan^{-1}\left(\frac{1}{4}\frac{\frac{a_T}{2b}}{\frac{a_T^2}{4b\alpha} +  T_0}\right),
\end{equation}
where we used Eq. \eqref{eq:T-j-def} to express the temperature in terms of $r^*$. Taking the denominator to zero recovers the correct $T=0$ result $\alpha T_0 = r^*$. Taking the large $T_0$ limit, we obtain the tricritical temperature in Eq.~\eqref{eq:T-TCP}. In Fig.~\ref{fig:firstComp}, we show how the self-consistency equation can be used to compute the entire phase transition line. The phase diagram for the $t$ mode can be obtained using exactly the same treatment.

\subsection{The FDW (finite-$q$) transition}
\label{sec:finite-transition-1}

The three diagrams that we computed were presented in Fig. \ref{fig:finiteQOBD} of the main text. Their values are given by Eq. \eqref{eq:delta-D-def}, which we reproduce here,

\begin{align}
  \label{eq:delta-D-def-app}
  \Pi_q^{ij}(\q,0) &= d^{ij}_1(\q,0) + d^{ij}_2(\q,0),\\
  d^{ij}_1(q=(\q,0)) &= 2\frac{\gb^2T^2}{\kf^4D_0} \sum_{p,k} G(k-q)G(k-p)G^2(k) D^{mn}(p)\mathcal{F}^{imnj}_0(\k), \\
  d^{ij}_2(q=(\q,0)) &= \frac{\gb^2T^2}{\kf^4D_0} \sum_{p,k} G(k-q)G(k-p)G(k)G(k-p-q) D^{mn}(p)\mathcal{F}_0^{imjn}(\k),
\end{align}
where
\begin{align}
  \label{eq:form-factors-app}
  \mathcal{F}_0^{imnj} = \Trc(\k\times\bvs)^i(\k\times\bvs)^m(\k\times\bvs)^n(\k\times\bvs)^j.
\end{align}
Up to a form factor, $d_{ij}^{(2)}$ can be cast into the same form as $d_{ij}^{(1)}$. To see this, we utilize the identity, which holds in the FL regime,
\begin{equation}
  \label{eq:GG-iden}
  G(k)G(k-q) = G(q)\left(G(k)-G(k-q)\right).
\end{equation}
Then we find
\begin{align}
  \label{eq:GG-d2-switch}
  &\sum_k G(k) G(k-p) G(k-q) G(k-p-q) = \sum_k G(k)G(k-p)G(p)[G(k-p-q)-G(k-q)] =\nn\\
  &\sum_k G(k)G(k-q)G(p)[G(k+p)-G(k-p)] = \sum_k G(k)G(k-q)G(p)[(G(k+p)-G(k))-(G(k-p)-G(k))] \nn\\
  &= -\sum_k G(k)^2G(k-q)[G(k+p)+G(k-p)].
\end{align}
If we assume further that $|\p|\ll|\k|$, so that the momentum shifts in Eq. \eqref{eq:GG-d2-switch} do not change $\mathcal F_0$, we find Eq. \eqref{eq:Pi-q-form-2} of the main text, which we reproduce here,
\begin{equation}
  \label{eq:Pi-q-form-2-app}
  \hat \Pi_q = 8\gb^2T^2\sum_{p,k}G(k-q)G(k-p)G^2(k) \left[\hz\hz \cos^2(\theta_p-\theta_k) D_t(p) + \hk_t\hk_t D_z(p)\right].
\end{equation}
As discussed in the main text, we see that the $z$ mode polarization is renormalized only by the $t$ mode
% bubble
and vice versa.
%\vk{(why bubble? it's $D$ that renormalizes $\Pi.$)}.

To solve the integrals, we specialize to the case where $r_z \ll r_t$, thus dropping the contribution proportional to $\hk_t\hk_t$. Since $q_0 =0$, the fermionic poles are split only when $p_0$ and $k_0$ are in separate half-planes. We assume that $\k$ is near the FS and split the integral as $2k_a^{-2}\int d^2k/\tpp^2 = \nu_F \int d\varepsilon_{k}d\theta_k/\tpp$. Integrating over $\varepsilon_{k}$ and then $k_0$ yields
\begin{align}
  \label{eq:d-1-calc-1}
  \int d\varepsilon_k dk_0 G(k)^2G(k-q)G(k-p)
  % = \frac{ \tp i p_0}{(\vf\hat k \cdot(\p-\q) + i p_0)(\vf \hk \cdot \p + i p_0)^2}
  = \frac{ \tp i p_0}{(\vf |\p| \cos\theta_p + i p_0 - \vf |\q| \cos\theta_k)(\vf|\p| \cos\theta_p + i p_0)^2},
\end{align}
where without loss of generality we picked $\hq = \hat x$ and shifted $\theta_p \to \theta_p + \theta_k$. Integrating over $\theta_k$ we find
% Integrating over $\theta_k$ will average out the $\hk_t\hk_t$ term in Eq. \eqref{eq:Pi-q-form-2-app}, leaving a longitudinal component we neglect and a transverse component, which we keep,
\begin{align}
\label{eq:d-1-calc-2}
  \Pi_{q,z} =
  % \frac{2\gb^2\nu_F}{\pi\vf k_a^2 |\q|} \int \frac{d^3p|p_0|}{\tpp^3}\frac{\hz\hz\cos^2\theta_pD_t(p)l_0(z_{pq}) + \hq_t\hq_t D_z(p)l_1(z_{pq})}{(\vf|\p|\cos\theta_p + i p_0)^2}.
  \frac{2\gb^2\nu_F}{\pi\vf k_a^2 |\q|} \int \frac{d^3p|p_0|}{\tpp^3}\frac{\cos^2\theta_pD_t(p)l_0\left((\vf|\p| \cos\theta_p + i p_0)/(\vf|\q|)\right)}{(\vf|\p|\cos\theta_p + i p_0)^2},
\end{align}
where $\Pi_{q,z}$ denotes the $z$ mode component and $z_{pq} = (\vf|\p| \cos\theta_p + i p_0)/(\vf|\q|)$.
Equation~\eqref{eq:d-1-calc-2} shows that $\Pi_{q,z}$ has a constant term which is cut off in the UV, which we incorporate into $r_z$. By dimensional analysis, its convergent part is linear in $\vf |\q|$, which can be obtained by differentiating the integrand,
\begin{align}
\label{eq:d-1-calc-3}
% \delta\hat \Pi_q = -\vf |\q| \frac{2\gb^2\nu_F}{\pi k_a^2} \int \frac{d^3p|p_0|}{\tpp^3(\vf |\q|)^4}\frac{\hz\hz\cos^2\theta_pD_t(p)\left( l_0(z_{pq})+ z_{pq}l_0'(z_{pq})\right) + \hq_t\hq_t D_z(p)\left( l_1(z_{pq})+  z_{pq}l_1'(z_{pq})\right)}{z_{pq}^2}.
  \delta\Pi_{q,z} = -\vf |\q| \frac{2\gb^2\nu_F}{\pi k_a^2} \int \frac{d^3p|p_0|}{\tpp^3(\vf |\q|)^4}\frac{\cos^2\theta_p\left( l_0(z_{pq})+ z_{pq}l_0'(z_{pq})\right)}{z_{pq}^2(r_t+\delta\Pi_t(p))},
\end{align}
where we wrote out $D_t$ explicitly and neglected the analytic $|\p|^2$ term. The integrand in Eq. \eqref{eq:d-1-calc-2} is convergent and dimensionless. Furthermore, because $r_t = \gb\nu_F/2$, the two contributions to $D_t^{-1}$ are of the same order. After a change of variables we find
\begin{align}
\label{eq:d-1-calc-4}
  \delta \Pi_{q,z} = \frac{|\q|}{k_a} \frac{16\gb}{\pi \vf k_a} \int_0^\infty dr \int_0^\pi d\theta \int_0^{\pi/2}d\phi \frac{r^2 \sin\phi\cos\phi}{\tpp^3}
                     % \nn\\
                   % &\qquad\qquad\qquad
                     \frac{\cos^2\theta\left( l_0(\zeta)+ \zeta l_0'(\zeta)\right)}{\zeta^2 (1 + 2\tan^2\phi l_1(\tan\phi))},
                     % \frac{\hz\hz\cos^2\theta D_t(\tan\phi)\left( l_0(\zeta)+ \zeta l_0'(\zeta)\right) + \hq_t\hq_t D_z(\tan\phi)\left( l_1(\zeta)+  \zeta l_1'(\zeta)\right)}{\zeta^2} + \cc,
\end{align}
where $\zeta = r(\sin\phi+i \cos\phi\cos\theta)$. Performing the numerical integral yields the $z$ mode contribution in Eq. \eqref{eq:Pi-q-final}.

% \subsection{The in-plane transition}
% \label{sec:plane-transition-1}

The case of the $t$ mode is analogous to the $z$ mode case. We start with Eq. \eqref{eq:Pi-q-form-2-app} and keep only the $t$ mode contribution, proportional to $D_z(p)$. The integrals over $\varepsilon_k$ and $k_0$ are identical to the $z$ case and yield the same result as Eq. \eqref{eq:d-1-calc-1}. The only difference is the integral over $\theta_k$, which now averages over the $\hk_t\hk_t$ term. It yields a component proportional to $\hq\hq$ and a component proportional to $\hq_t\hq_t$. We drop the first one and are left with
\begin{align}
  \label{eq:d1-calc-5}
  \delta\Pi_{q,t} &= -\vf |\q| \frac{2\gb^2\nu_F}{\pi k_a^2} \int \frac{d^3p|p_0|}{\tpp^3(\vf |\q|)^4}\frac{\left( l_1(z_{pq})+ z_{pq}l_1'(z_{pq})\right)}{z_{pq}^2(r_z+\delta\Pi_z(p))} \nn\\
  &=  \frac{|\q|}{k_a} \frac{4\gb}{\pi \vf k_a} \int_0^\infty dr \int_0^\pi d\theta \int_0^{\pi/2}d\phi \frac{r^2 \sin\phi\cos\phi}{\tpp^3}
                     % \nn\\
                   % &\qquad\qquad\qquad
                     \frac{\left( l_1(\zeta)+ \zeta l_1'(\zeta)\right)}{\zeta^2 (1 + \tan\phi l_0(\tan\phi))}.
\end{align}
Performing the integral we obtain the $t$ mode contribution of Eq. \eqref{eq:Pi-q-final}.
\section{Detailed calculations for Sec. \ref{sec:theory-3d-qfm}}
\label{sec:deta-calc-sec-iv}

In this Appendix we present the detailed calculations for a 3D {\QFM}. First we compute the polarization bubble, assuming right away the presence of order (along the $z$ axis, e.g. $\bv{\Delta} = \hz \Delta$ for the homogeneous transition). Then we compute both normal and pairing self-energies simultaneously.

The 3D polarization bubble in the presence of $\bv\Delta$ is given by a similar expression to that of the 2D case, Eq.  \eqref{eq:bubble-Delta}, and the diagonalization transformation is identical to the one for the $z$ mode in the 2D case, Eqs. \eqref{eq:U-def} and \eqref{eq:G-delta-def}. The interaction changes to
\begin{equation}
  \label{eq:ff-change-3d}
  (\hat p \times \bvs) \to U^\dagger (\hat p \times \bvs) U = \hp_t (\sin\theta_p\sigma_x-\cos\theta_p\sigma_y) - \hp_u \sigma_z,
\end{equation}
where we defined the azimuthal and polar variables $\phi_p$, $\theta_p$. For convenience, we write down explicit expressions for the basis vectors
\begin{equation}
  \label{eq:pt-pu-def}
  \hp_t = -\sin\phi_p\hat x +\cos\phi_p\hat y,\quad \hp_u = \cos\theta_p\cos\phi_p\hat x +\cos\theta_p\sin\phi_p \hat y - \sin\theta_p \hz.
\end{equation}
Note that $\hp_u$ is perpendicular  to $\hp$ but its projection on the $xy$ plane is parallel to the projection of $\hp$ on the plane, hence it does not cause spin-flip processes.  The polarization bubble is
\begin{flalign}
  \label{eq:bubble-Delta-app-3d}
  \Pi_\Delta^{kl} = \frac{\gb}{k_a^3} \Trc \int \frac{d^4p}{\tpp^4} ( \hp_t (\sin\theta\sigma_x-\cos\theta\sigma_y) - \hp_u \sigma_z) \left[\left(E_p + \frac{1}{2}\varepsilon_q\right)\sigma_0 + \Delta_p \sigma_z\right]^{-1}\nn\\
 \times ( \hp_t (\sin\theta\sigma_x-\cos\theta\sigma_y) - \hp_u \sigma_z)\left[\left(E_p - \frac{1}{2}\varepsilon_q\right)\sigma_0 + \Delta_p \sigma_z\right]^{-1},
                  % &= \gb \nu_F \left\{
                    % \hp\hp 
                    % \right\}
\end{flalign}
where
\begin{equation}
  \label{eq:Delta-theta-def}
  \Delta_p = \Delta \sin\theta_p
\end{equation}
is the Zeeman-like splitting for $\hp$. After tracing out and integrating over frequency and energy we obtain
\begin{align}
  \label{eq:Pi-delta-FS-3d-app}
  \hat{\Pi}_\Delta(q) &= \frac{\gb \nu_F}{2} \int \frac{d\cos\theta_pd\phi_p}{4\pi}\left[ \hp_t\hp_t \left(\frac{\vf|\q|\hp\cdot\hq-2\Delta_p}{\vf|\q|\hp\cdot\hq-2\Delta_p- iq_0} + \frac{\vf|\q|\hp\cdot\hq+2\Delta_p}{\vf|\q|\hp\cdot\hq+2\Delta_p- iq_0}\right)
                       + 2\hp_u\hp_u \frac{\vf|\q|\hp\cdot\hq}{\vf|\q|\hp\cdot\hq- iq_0}\right],
\end{align}
where
\begin{equation}
  \label{eq:nu_F-3d}
  % \nu_F = \frac{m^* k_F}{2\pi^2k_a^3}
  \nu_F = \frac{m^* k_F}{\pi^2k_a^3}
\end{equation}
is the 3D density of states. To evaluate Eq. \eqref{eq:Pi-delta-FS-3d-app} it is convenient to add and subtract a $\hp_t\hp_t$ term next to the $\hp_u\hp_u$ term the right-hand side. Then, the polarization is seen to consist of two contributions,
\begin{equation}
  \label{eq:Pi-delta-2-3d}
  \hat{\Pi}_\Delta(q) = \hat \Pi_0(q) + \delta  \hat\Pi_\Delta(q).
\end{equation}
Here, $\hat \Pi_0$ has the same form as the polarization in the disordered phase, namely
%is the disordered-phase polarization,
\begin{align}
  \label{eq:hat-pi-0-3d}
   \hat \Pi_0(q) &= \gb \nu_F \int \frac{d\cos\theta_pd\phi_p}{4\pi}(\hp_t\hp_t + \hp_u\hp_u) \frac{\vf|\q|\hp\cdot\hq}{\vf|\q|\hp\cdot\hq- iq_0}.
\end{align}
By rotating $\hq$ to the $\hz$ axis and back, we find
\begin{equation}
  \label{eq:pi-0-full-3d}
  \hat \Pi_0(q) = \gb\nu_F\left[\frac{2}{3} + z_q\left(\frac{1}{2}(s_0(z_q) + s_1(z_q))(\hq_t\hq_t+\hq_u\hq_u) + (s_0(z_q) - s_1(z_q)) \hq\hq\right)\right],
\end{equation}
where
\begin{align}
  \label{eq:lind-3d-app}
  s_0(x) &=  \arctan(1/x)\\
           % \frac{i}{2}\log\frac{1-i x}{1+i x}\\
  s_1(x) &= x - x^2\arctan(1/x)
           % i \log \frac{1-i x}{1+i x}\right)
\end{align}
and
% $s_0$ and $s_1$ were defined in Eq.~\eqref{eq:lind-3d}, and
$z_q$ defined in Eq. \eqref{eq:zq-dq}. This result gives Eq. \eqref{eq:pi-disordered-3d} of the main text.

The $\Delta$-dependent contribution is obtained from the spin-flip $\hp_t\hp_t$ part of the polarization in Eq. \eqref{eq:Pi-delta-FS-3d}. To compute it, we expand $\hp_t$ in the $\hq$ basis,
\begin{align}
  \label{eq:hp-t-expand-3d}
  \hp_t = \cos(\phi) \hq_t -\sin(\phi)(\cos\theta_q \hq_u + \sin\theta_q \hq),
  % \Rightarrow \hp_t\hp_t &= \cos^2\phi
\end{align}
where $\phi = \phi_p-\phi_q$, such that,
\begin{equation}
  \label{eq:hphp-proj}
  \hp_t\hp_t = \cos^2\phi\hq_t\hq_t + \sin^2\phi\cos^2\theta_q\hq_u\hq_u + \cdots,
\end{equation}
where the ellipsis denotes terms that either vanish upon angular integration or belong to the longitudinal sector. The meaning of Eq. \eqref{eq:hphp-proj} is that spin-flip contributions in the $\hq_u\hq_u$ sector depend on the polar alignment of $\hq$, and are maximal when $\hq$ itself is in the $\hz$ direction. Thus, plugging the above into Eq. \eqref{eq:hat-pi-0-3d} we obtain
\begin{align}
  \label{eq:Pi-delta-FS-3d}
  \delta\hat{\Pi}_\Delta(q) = \frac{\gb \nu_F}{2} z_q (S_t(|\q|,\Delta) \hq_t\hq_t + \cos^2\theta_qS_u(|\q|,\Delta)\hq_u\hq_u + \cc).
\end{align}
$S_t$ and $S_u$ are given by rather cumbersome expressions,
\begin{align}
  \label{eq:St-Su-def}
  S_t &= \frac{1}{2}\int \frac{d\cos\theta}{\sin\theta_q\sin\theta} \left[l_1\left(Z_{q,\Delta}\right) - l_1\left(Z_{q,0}\right) \right], \nn \\
  S_u &= \frac{1}{2}\int \frac{d\cos\theta}{\sin\theta_q\sin\theta} \left[l_0\left(Z_{q,\Delta}\right) - l_1\left(Z_{q,\Delta}\right) -l_0\left(Z_{q,0}\right)+l_1(Z_{q,0})\right],
\end{align}

where
\begin{equation}
  \label{eq:Z-def}
  Z_{q,\Delta}(\theta) = \frac{q_0 - 2i\Delta\sin\theta-\vf|\q|\cos\theta_q\cos\theta}{\vf|\q|\sin\theta_q\sin\theta},
\end{equation}
and $l_0$ and $l_1$ are just the Lindhard functions we obtained for the 2D problem, see Eq. \eqref{eq:delta-pi-defs}. While it is possible to work out the asymptotics of $S_{t,u}$ in detail, it will not be necessary for our calculations and so we omit them. Instead, we note that in the regime of small angles, $\cos\theta_q \gg \sin\theta_q$, both functions have the approximate form 
\begin{equation}
  \label{eq:St-Su-simple}
  S_u \approx S_t \approx \frac{1}{2}s_0\left(\frac{q_0-2i\tilde{\Delta}}{\vf|\q|}\right) - \frac{1}{2}s_0\left(\frac{q_0}{\vf|\q|}\right).
\end{equation}
Here we replaced $\Delta_p$ from Eq.~\eqref{eq:Delta-theta-def} with $\Delta_p \approx \tilde{\Delta} = \sqrt{2/3} \Delta$,
% \vk{(what is $\Delta p$?)},
see Eq.~\eqref{eq:Pi-Delta-3D} in the main text, by taking the angular average of the sin term. Equation~\eqref{eq:St-Su-simple} shows that the system exhibits both Landau damping and a nonalytic dependence that generates the QOBD terms, as discussed in the main text. Equations~\eqref{eq:Pi-delta-2-3d} and~\eqref{eq:Pi-delta-FS-3d} are equivalent to Eq. \eqref{eq:Pi-Delta-3D} of the main text, and are obtained by defining $\Pi_t = S_t(q,\Delta) + \hq_t \cdot \hat \Pi_0(q) \cdot \hq_t$, $\Pi_u =  S_u(q,\Delta) + \hq_u \cdot \hat \Pi_0(q) \cdot \hq_u$.

The free energy in the presence of finite $\Delta$ is obtained just as in the 2D case, by tracing over the action, see Eq.~\eqref{eq:free-1st-order-def}. As we saw previously, typical scales for the trace are $\vf |\q| \sim q_0 \sim \Delta$. Thus, at the critical point and for small enough $\Delta$, the analytic $r + q^2/k_a^2$ terms in the propagators are negligible, and the free energy correction is
\begin{align}
  \label{eq:delta-F-3d}
  \delta F &\approx \sum_q \left(\log\frac{\Pi_t(q,\Delta)}{\Pi_t(q,0)} + \log\frac{\Pi_u(q,\Delta)}{\Pi_u(q,0)}\right) \nn \\
  &= \frac{1}{\tpp^4k_a^3} \int_0^\infty dq_0 \int_0^\Lambda q^2 dq d\Omega_q \left(\log\frac{\Pi_t((q,\Delta)}{\Pi_t(q,0)} + \log\frac{\Pi_u(q,\Delta)}{\Pi_u(q,0)}\right).
\end{align}
Similarly to the 2D case, both $\Pi_t$ and $\Pi_u$ are functions  of $q_0/(\vf q)$ and $\Delta/(\vf q)$ only. Hence, by dimensional analysis, the logarithmic terms have an expansion of the form $A/q^2  + B/q^4 + \cdots$, where $A,B,\ldots$ are functions of $\omega$ and $\Delta$. The quadratic terms are UV divergent, and the quartic terms are logarithmically divergent. To evaluate them, we rescale $q_0 \to \vf q z$, $q \to \Delta y / \vf$, which renders the polarization functions dimensionless, e.g. $\Pi_t(q,\Delta) = \Pi_t(z, \theta_q,1/y)$, such that
\begin{align}
  \label{eq:F-3d-2}
  \delta F \approx  \frac{\Delta^4}{\tpp^3(\vf k_a)^3} \int_0^{\frac{\Lambda \vf}{\Delta}} y^3 dy \int_0^\infty dz  \int_0^\pi\sin\theta d\theta \left(\log\frac{\Pi_t((z,\theta,y^{-1})}{\Pi_t(z,\theta,0)} + \log\frac{\Pi_u(z,\theta,y^{-1})}{\Pi_u(z,\theta,0)}\right).
\end{align}
We evaluated the integrand numerically as a function of $y$ by integrating over $z$ and $\theta$. Then, we fitted the result to a series of power laws and obtained at $y \gg 1$ that the integrand has the form
\begin{equation}
  \label{eq:num-fit}
  -\frac{A}{y^2}-\frac{B}{y^4} + \cdots, \qquad  A = 0.70 \pm 0.01, \qquad B = 46.7 \pm 0.9.
\end{equation}
This yields the expression in Eq. \eqref{eq:deltaF-3D}.

The pairing equation is given by Eq. \eqref{eq:pairing-eq-1}. In order to compute it, we first write down the (linearized) equations for both the normal self-energy and the pairing vertex,
\begin{equation}
  \label{eq:sigma-phi}
  \Sigma_{\alpha\beta}(k) = \Sigma(k) \delta_{\alpha\beta}, \qquad \Phi_{\alpha\beta}(k) = f(k) (i \sigma_y)_{\alpha\beta},
\end{equation}
where
\begin{align}
  \label{eq:SE-3d}
  % \Sigma(k) &= \frac{\gb}{k_a^3 D_0} T\sum_{p_0}\int \frac{d^3p}{\tpp^3}G(p)D(p-k), \\
  % f(k) &= \frac{\gb}{k_a^3 D_0}T\sum_{p_0}\int \frac{d^3p}{\tpp^3}f(p)G(p)G(-p)D(p-k)
           \Sigma(k) &= \frac{\gb}{k_a^3 D_0} T\sum_{p_0}\int \frac{d^3p}{\tpp^3}G(p)D(p-k), \\
  f(k) &= \frac{\gb}{k_a^3 D_0}T\sum_{p_0}\int \frac{d^3p}{\tpp^3}f(p)G(p)G(-p)D(p-k).
\end{align}
Here, we already performed spin summations and projected onto the transverse component. Furthermore, we used the facts that only the spin-singlet channel is attractive and that
\begin{equation}
  \label{eq:D-3d-def}
  D(p) = D_0(r + |\p|^2/k_a^2 +\delta \Pi_0(p))^{-1}
\end{equation}
is the boson propagator in the disordered phase. We assume that the momentum integration for both $\Sigma$ and $f$ factorizes to components parallel and transverse to the FS, and obtain the effective frequency-dependent propagator
\begin{equation}
  \label{eq:d-def}
  d(q_0) = \int_0^\Lambda p dp \frac{1}{p^2 + \frac{\pi\gb\nu_F|q_0|}{4\vf k_a p}} = \log \left( 1 + (\Lambda/k_a)^3\frac{4\vf k_a}{\pi\gb\nu_F|q_0|}\right)^{1/3}.
\end{equation}
The zero-temperature normal-state self-energy is then found to be
\begin{equation}
  \label{eq:Sg-3d}
  \Sigma(k_0) \approx -i\frac{\gb}{\vf k_a} \frac{k_0}{4\pi^2}\log\frac{\Lambda/k_a}{(\frac{\gb\nu_F|k_0|}{\vf k_a})^{1/3}},
  % \Sigma(k_0) \approx \frac{\gb}{\vf k_a} \frac{k_0}{4\pi^3}\log\frac{\Lambda}{(\frac{\gb\nu_F|k_0|}{\vf k_a})^{1/3}},
\end{equation}
which gives Eq. \eqref{eq:sigma-disordered-3D} of the main text. At finite temperatures, both $f$ and $\Sg$ have similar forms,
\begin{align}
  \label{eq:gen-form}
  \Sg(k_0) &= -i\frac{\gb/\vf k_a}{4\pi} T \sum_{p_0} d(p_0-k_0)\sgn(p_0), \\
  f(k_0) &=  \frac{\gb/\vf k_a}{4\pi} T \sum_{p_0} d(p_0-k_0) \frac{f(p_0)}{|p_0 + \Sg(p_0)|}.
\end{align}
It may be verified that this is precisely the form of Eq. 7 in Ref. \onlinecite{Chubukov2005}. To connect the two problems, one may define
\begin{align}
  \label{eq:pairing-defs}
  g = \frac{\gb}{12\pi^2 \vf k_a },  \qquad \w_\Lambda = \frac{\vf\Lambda^3 }{k_a^2\gb\nu_F}.
\end{align}
Then $T_c$ is given by Eq. (27) in that paper, namely
\begin{equation}
  \label{eq:TC-CH-SCH}
  T_c \approx \w_\Lambda \exp\left(-\frac{\pi}{2\sqrt{g}}\right).
\end{equation}
For finite $r$ that is large enough to neglect the Landau damping term, the logarithmic frequency-dependent enhancement of the interaction is replaced by a constant logarithm, $d(q_0)\approx d_0 = \log\left(1+\frac{\Lambda^2}{k_a^2 r}\right)^{1/2}$, and $T_c$ goes back to  a BCS-like form,
\begin{equation}
\label{eq:tc-3d-r-app}
    T_c \approx \w_{r} e^{-\frac{1}{3g d_0}}, \qquad \w_r = \mbox{min}\left(\frac{c k_a r^{1/2}}{2\pi}, \frac{2 r^{3/2}\vf k_a}{\pi^2 \gb\nu_F}\right).
\end{equation}
 %\vk{ \approx  (Can we fix the reference list?)}

\section{Numerical parameters for Fig. \ref{fig:strain1}}
\label{sec:app-num}

The qualitative shape of the phase diagram in the presence of strain depends on a variety of parameters. For clarity, we present here the numerical parameters used in constructing the phase diagram of Fig. \ref{fig:strain1}. 

To create the figure, we used the following dimensionless parameters. The second order gaps were $r_t = 0.0, r_z = -0.05$, and the first order transitions were given by $r_j^* = r_j - \delta r$, where $\delta r$ = 0.05. The elastic couplings were $\lambda_{0z} = 0.8, \lambda_{0t} = 1, \lambda_{1t} = 0.2$. We picked the parameters for visual clarity rather than physical significance. Finally, for simplicity, we did not account for the complicated dependence of the border between the first-order and second-order regions of the t mode on $\Delta r, \overline{r}$. Instead we used a simple linear relation (which is justified very near the critical point at small strains), $r_j^* = \rho r_j - \delta r$, with $\rho = 0.9$. The size of the SC phase in the figure is not to scale.

\twocolumngrid
\bibliography{qfm}
\end{document}